\documentclass[natbib]{svjour3}
\usepackage{graphicx,color}
\usepackage{amsfonts,amsmath,amssymb,amsxtra}
\def\vec#1{{\mathbf{#1}}}
\def\idm#1{{\mbox{\small{#1}}}}
\newcommand{\LL}{\mathcal{L}}
\newcommand{\Tcl}{T_{\idm{p-p}}}
\newcommand{\Vcl}{V_{\idm{p-p}}}
\newcommand{\Trot}{T_{\idm{rot}}}
\newcommand{\Vrot}{V_{\idm{rot}}}
\newcommand{\Vtid}{V_{\idm{tid}}}
\newcommand{\LLrel}{\mathcal{L}_{\idm{rel}}}
\newcommand{\Irot}{I^{(\idm{rot})}}
\newcommand{\Izero}{I}
\newcommand{\IIrot}{\mathbb{I}^{(\idm{rot})}}
\newcommand{\IIzero}{\mathbb{I}^{(0)}}
\newcommand{\tildeIIrot}{\widetilde{\mathbb{I}}^{(\idm{rot})}}
\newcommand{\Ff}{\pmb{\mathcal{F}}}
\newcommand{\Gg}{\pmb{\mathcal{G}}}

\newcommand{\fE}{\mathcal{E}}
\newcommand{\msun}{m_{\odot}}
\newcommand{\mJ}{m_{\idm{J}}}
\newcommand{\au}{\mbox{au}}
\newcommand{\RS}{R_{\odot}}
\newcommand{\RJ}{R_{\idm{J}}}

\newcommand{\imut}{i_{\idm{mut}}}
\newcommand{\Icrit}{\theta_{0}^{(\idm{crit})}}
\newcommand{\thetacrit}{\theta_{1}^{(\idm{crit})}}
\def\corr#1{{{#1}}}
\def\Corr#1{{{#1}}}

\bibstyle{kluwer}

\begin{document}

\title{The generalized non-conservative model of a 1-planet system - revisited}



\author{Cezary Migaszewski}


\institute{CM \at
      Toru\'n Centre for Astronomy, Gagarin Str. 11, 87-100 Toru\'n, Poland \\
              \email{c.migaszewski@astri.umk.pl} \\
}

\date{Received: date / Accepted: date}

\maketitle

\begin{abstract}
We study the long-term dynamics of a planetary system composed of a star and 
a planet. Both bodies are considered as extended, non-spherical, rotating 
objects. There are no assumptions made on the relative angles between the 
orbital angular momentum and the spin vectors of the bodies. Thus, we 
analyze full, spatial model of the planetary system. Both objects are 
assumed to be deformed due to their own rotations, as well as due to the 
mutual tidal interactions. The general relativity corrections are 
considered in terms of the post-Newtonian approximation. Besides the 
conservative contributions to the perturbing forces, there are also taken 
into account non-conservative effects, i.e., the dissipation of the 
mechanical energy. This dissipation is a result of the tidal perturbation 
on the velocity field in the internal zones with non-zero turbulent 
viscosity (convective zones). Our main goal is to derive the equations of 
the orbital motion as well as the equations governing time-evolution of the 
spin vectors (angular velocities). We derive the Lagrangian equations of 
the second kind for systems which do not conserve the mechanical energy. 
Next, the equations of motion are averaged out over all fast angles with 
respect to time-scales characteristic for conservative perturbations.  The 
final equations of motion are then used to study the dynamics of the 
non-conservative model over time scales of the order of the age of the 
star. We analyze the final state of the system as a function of the initial 
conditions. Equilibria states of the averaged system are finally discussed.
\end{abstract}

\section{Introduction}
\label{intro}
Since the first discovery of an exoplanet revolving around the main sequence 
star \citep{Mayor1995}, many planetary companions were detected in a few 
day orbits . The dynamics of such systems are strongly affected by 
relativistic as well as non-point and non-Newtonian effects. These 
perturbations cause the periapses rotation and the precession of the 
orbital nodes as well as spins of the rotating bodies. Considering  the 
longest time scale which is comparable to the age of the parent star, one has to 
take into account also a dissipation of the mechanical energy. It is 
believed, that the most important physical mechanism dissipating the energy 
is the tidal perturbation of the velocity field in parts of bodies 
possessing non-zero turbulent viscosity \citep[e.g.,][]{Zahn1977}. This 
takes place in the convective zones of stars and planets. The mechanisms of 
the energy dissipation, particularly active in stars and Jupiter-like 
planets, were studied by many authors \citep[e.g.,][]{Goldreich1966,Ogilvie2004, Wu2005a, Wu2005b, Ogilvie2007, 
Miller2009, Gu2009, Arras2010}. An open problem is to estimate values of 
physical parameters characterising the strength and time-scales of these 
processes, and in turn, the time-scale of the dissipative evolution of 
planetary systems.

It is well known, that the energy loss leads to a variation (a decrease in 
general) both the semi-major axis and eccentricity, as well as to evolution 
of spins, their directions and magnitudes. The planetary dynamics of 
systems with energy loss were considered both in cases with one and two 
planets \citep[e.g.,][]{Mardling2002, Witte2002, Dobbs-Dixon2004, 
Mardling2007, Barker2009, Leconte2010, Rodriguez2010, Michtchenko2011, Rodriguez2011, Correia2011, Laskar2012}. The 
equations of motion of such systems are usually derived through introducing 
a dissipative force. For instance, the well known model of \cite{Hut1981}, 
assumes that this force emerges due to a time delay in forming the tidal 
bulge and the orbital motion of a perturber. That implies a non-zero angle 
between the radius vector of the deformable body with respect to its 
companion, and the axis of symmetry of the tidally deformed object. In more 
elegant way, the tidal force was derived from the energy loss function 
$\dot{E}$ defined by \cite{Eggleton1998}.

In this paper, we found a more straightforward derivation of the dissipative 
model that relies on the Lagrangian equations of the second kind. As we 
will show, this approach makes it possible to obtain quite simply the 
dissipative forces acting in the $N$-body system; however, we limit here 
the derivation to $N=2$. Moreover, to derive the equations governing the 
evolution of angular velocities, both in  conservative, as well as in 
non-conservative models, we {\em should not} apply the Euler equations, 
which hold only for a specific form of the potential energy $V$ that has to 
be then a function of Euler angles $\phi, \theta, \psi$, and should not 
depend on their time derivatives $\dot{\phi}, \dot{\theta}, \dot{\psi}$. 
This is only true in the case of {\em the rigid} body. For deformable 
objects, $V$ is a function of the angular velocity $\pmb{\Omega} = 
\pmb{\Omega}(\phi, \theta, \psi, \dot{\phi}, \dot{\theta}, \dot{\psi})$ and 
thus it does not fulfill these assumptions. Hence, the Euler equations 
stating that the time derivative of the rotational angular momentum equals 
to the torque acting on the rigid body do not hold in general. However, we 
will show  here that it is still possible to obtain the equations of the 
evolution of the angular velocities in vectorial form, which is reminiscent 
of the classic Eulerian equations.

The plan of this paper is the following one. In Section 2 we derive the 
equations of motion for a general form of the Lagrangian $\LL = 
T_0(\pmb{\Omega}_0) + T_1(\pmb{\Omega}_1) - V_0(\vec{r}, \pmb{\Omega}_0) - 
V_1(\vec{r}, \pmb{\Omega}_1) + \LL_1(\vec{r}, \dot{\vec{r}})$, and a 
dissipative function $\dot{E} = \dot{E}_0(\vec{r}, \dot{\vec{r}}, 
\pmb{\Omega}_0) + \dot{E}_1(\vec{r}, \dot{\vec{r}}, \pmb{\Omega}_1)$. Here, 
symbols $\vec{r}, \dot{\vec{r}}$ denote the planetary position and the 
orbital velocity vectors relative to the star, and $\pmb{\Omega}_0, 
\pmb{\Omega}_1$ stand for the angular velocity vectors of the star and the 
planet, respectively.

In the next Section 3, we find expressions for these functions in a 
particular model considered in this paper. We use the polytropic model of 
\cite{Chandrasekhar1933a,Chandrasekhar1933b} to calculate the internal 
structure and a deformation of figures of both objects. Relativistic 
correction to the Lagrangian is taken from \cite{Brumberg2007}. To 
determine the energy loss function, we use a simple model by \cite{Eggleton1998}.

In Section 4, the derived equations of motion are averaged out over all 
angles that vary in time-scales related to the conservative evolution. 
These angles, ordered from the fastest to the slowest one are the following: 
the mean anomaly, the horizontal angle of the precessing planetary spin in 
the orbital frame, and the argument of pericenter. As the result we obtain 
the equations of motion describing the dissipative dynamics only, which are 
then studied in Section 5. We analyze the final state of the planetary 
system in terms of the initial conditions. As a particular solution, we 
discuss the equilibrium permitted in the system.

\section{The equations of motion}
%
We shall consider a planetary system in terms of the mechanical system with
the Lagrangian $\LL = \LL(q_1, \dots, q_n, \dot{q}_1, \dots, \dot{q}_n)$ and
energy loss function $\dot{E} = \dot{E}(q_1, \dots, q_n, \dot{q}_1, \dots,
\dot{q}_n)$, where $q_i$ are the generalized coordinates, and $\dot{q}_i$
are the generalized velocities. Index $i$ spans $1$ to $n$, where $n$ is
the number of the degrees of freedom. The dynamical evolution of the system
is governed by solutions to the Lagrangian equations of the second kind (e.g.,
\cite{Greiner2003}, p.~328):
\begin{equation}
\frac{d}{dt} \left( \frac{\partial \LL}{\partial \dot{q}_i} \right) -
\frac{\partial \LL}{\partial q_i} =
\frac{1}{2} \frac{\partial \dot{E}}{\partial \dot{q}_i}, \quad i = 1, \dots, n,
\end{equation}
\corr{The above equations are correct when $\dot{E}$ fulfills the following condition \citep[e.g.,][p.~63]{Goldstein2002}
\begin{equation}
\sum_{i=1}^n \dot{q}_i \, \frac{\partial \, \dot{E}}{\partial \, \dot{q}_i} = 2 \, \dot{E}.
\label{eq:condition_Edot}
\end{equation}
Particularly, it takes place when $\dot{E}$ is a quadratic form of the generalized velocities. The explicit form of $\dot{E}$ is given further in the text. Although it is not a quadratic form in $\dot{q}$, it fulfills the above condition as well (it is shown in Appendix~\ref{appB}).
}

In the problem considered here, the generalized coordinates are the following:
\[
\lbrace q_i \rbrace_{i=1}^{i=9} =
\lbrace x, y, z, s_x, s_y, s_z, p_x, p_y, p_z \rbrace.
\]
The first three coordinates are components of the position vector of the
planet relative to the star, $\vec{r} = [x, y, z]^T$. The last six
coordinates are vectors of three independent components in two unit
quaternions, $\vec{s} = [s_x, s_y, s_z]^T$, $\vec{p} = [p_x, p_y, p_z]^T$:
\[
\mathfrak{s} = s_0 + \mathfrak{i} \, s_1 + \mathfrak{j} \,
s_2 + \mathfrak{k} \, s_3, \quad s_k \in \mathbb{R},
\]
\[
\mathfrak{p} = p_0 + \mathfrak{i} \, p_1 + \mathfrak{j} \,
p_2 + \mathfrak{k} \, p_3, \quad p_k \in \mathbb{R}.
\]
The components of the quaternions as well as their time derivatives are
related as follows:
\begin{equation}
s_0^2 + s_1^2 + s_2^2 + s_3^2 = 1, \quad
s_0 \, \dot{s}_0 + s_1 \, \dot{s}_1 + s_2 \, \dot{s}_2 + s_3 \, \dot{s}_3 = 0
\label{eq:unit_quaternions}
\end{equation}
and similarly, for $\mathfrak{p}$. Therefore, only three components of each
quantity are independent. We choose these independent components as follows:
$s_x = s_1, s_y = s_2, s_z = s_3$ and $p_x = p_1, p_y = p_2, p_z = p_3$.
Thus $s_0, \dot{s}_0, p_0, \dot{p}_0$ are functions of $\vec{s},
\dot{\vec{s}}, \vec{p}, \dot{\vec{p}}$, according to Eq.~(\ref
{eq:unit_quaternions}).

The Lagrangian of the system, as well as $\dot{E}$, are assumed to be 
functions of planetary position, the velocity and spin vectors of both 
bodies, i.e., $\LL = \LL(\vec{r}, \dot{\vec{r}}, \pmb{\Omega}_0, 
\pmb{\Omega}_1)$, $\dot{E} = \dot{E}(\vec{r}, \dot{\vec{r}}, 
\pmb{\Omega}_0, \pmb{\Omega}_1)$. It is well known, that the angular 
velocity may be expressed with the help of quaternions in the following 
form, e.g., \cite{Heard2006}, p.~49:
\begin{equation}
\left( \begin{array}{c} 0 \\ \pmb{\Omega}_0 \end{array} \right) = 
2 \left( \begin{array}{c c c c} 
s_0 & s_1 & s_2 & s_3\\
-s_1 & s_0 & -s_3 & s_2\\
-s_2 & s_3 & s_0 & -s_1\\
-s_3 & -s_2 & s_1 & s_0
\end{array} \right) \,
\left( \begin{array}{c} 
\dot{s}_0 \\ \dot{s}_1 \\ \dot{s}_2 \\ \dot{s}_3
\end{array}\right)
\label{eq:Omega_quaternions}
\end{equation}
and similarly for $\pmb{\Omega}_1$ with quaternion $\mathfrak{p}$. Thus
$\pmb{\Omega}_0 = \pmb{\Omega}_0 (\mathfrak{s}, \dot{\mathfrak{s}})$ and
$\pmb{\Omega}_1 = \pmb{\Omega}_1 (\mathfrak{p}, \dot{\mathfrak{p}})$. There
exists also an inverse relation:
\begin{equation}
\left( \begin{array}{c} 
\dot{s}_0 \\ \dot{s}_1 \\ \dot{s}_2 \\ \dot{s}_3 \end{array}\right) = 
\frac{1}{2} \left( \begin{array}{c c c c} 
s_0 & -s_1 & -s_2 & -s_3\\
s_1 & s_0 & s_3 & -s_2\\
s_2 & -s_3 & s_0 & s_1\\
s_3 & s_2 & -s_1 & s_0
\end{array} \right) \,
\left( \begin{array}{c} 0 \\ \pmb{\Omega}_0 \end{array} \right).
\end{equation}
The angular velocities are then $\pmb{\Omega}_0 = \pmb{\Omega}_0(\vec{s},
\dot{\vec{s}})$ and $\pmb{\Omega}_1 = \pmb{\Omega}_1(\vec{p}, \dot{\vec{p}})$.

The full set of Lagrange equations are then the following:
\begin{equation}
\left\{
\begin{array}{l}
\displaystyle{\frac{d}{dt} \left( \frac{\partial \, \LL}
{\partial\, \dot{\vec{r}}} \right) - \frac{\partial \, \LL}
{\partial \, \vec{r}} = \frac{1}{2} \, \frac{\partial \, \dot{E}}
{\partial \, \dot{\vec{r}}}},\\
\displaystyle{\frac{d}{dt} \left( \frac{\partial \, \LL}
{\partial \, \dot{\vec{s}}} \right) - \frac{\partial \, \LL}
{\partial \, \vec{s}} = \frac{1}{2} \, \frac{\partial \, \dot{E}}
{\partial \, \dot{\vec{s}}}},\\
\label{eq:Lagrange_vec_s}
\displaystyle{\frac{d}{dt} \left( \frac{\partial \, \LL}
{\partial \, \dot{\vec{p}}} \right) - \frac{\partial \, \LL}
{\partial \, \vec{p}} = \frac{1}{2} \, \frac{\partial \, \dot{E}}
{\partial \, \dot{\vec{p}}}}.
\end{array}
\right.
\end{equation}
It may be shown \corr{(see Appendix \ref{appA})}, that the second equation in the above set may be
transformed into the matrix equation
\begin{equation}
\mathbb{A}_1 \vec{X} = \mathbb{A}_2 \vec{Y}, \quad
\vec{X} \equiv \frac{d}{dt} \left( \frac{\partial \, \LL}
{\partial \, \pmb{\Omega}_0} \right) - \frac{1}{2} \frac{\partial \, \dot{E}}
{\partial \, \pmb{\Omega}_0}, \quad \vec{Y} \equiv \frac{\partial \, \LL}
{\partial \, \pmb{\Omega}_0},
\label{eq:Lagrange_matrix}
\end{equation}
\begin{equation}
\mathbb{A}_1 = \left( \begin{array}{c c c}
s_0^2 + s_1^2 & s_0 s_3 + s_1 s_2 & -s_0 s_2 + s_1 s_3 \\
-s_0 s_3 + s_1 s_2~~ & s_0^2 + s_2^2 & s_1 s_0 + s_2 s_3 \\
s_0 s_2 + s_1 s_3 & -s_0 s_1 + s_2 s_3~~ & s_0^2 + s_3^2
\end{array} \right),
\end{equation}
\begin{equation}
\mathbb{A}_2 = -2 \left( \begin{array}{c c c}
s_0 \dot{s}_0 + s_1 \dot{s}_1 & s_0 \dot{s}_3 + s_1 \dot{s}_2 & -s_0 \dot{s}_2 + s_1 \dot{s}_3 \\
-s_0 \dot{s}_3 + s_2 \dot{s}_1~~ & s_0 \dot{s}_0 + s_2 \dot{s}_2~~ & s_0 \dot{s}_1 + s_2 \dot{s}_3 \\
s_0 \dot{s}_2 + s_3 \dot{s}_1 & -s_0 \dot{s}_1 + s_3 \dot{s}_2 & s_0 \dot{s}_0 + s_3 \dot{s}_3
\end{array} \right).
\end{equation}
This equation may be solved with respect to $\vec{X}$, leading to the
following solution:
\begin{equation}
\vec{X} = \corr{\left(\mathbb{A}_1^{-1} \, \mathbb{A}_2 \right) \vec{Y}} = \left( \begin{array}{c c c}
0 & -\Omega_{0,z} & \Omega_{0,y}\\
\Omega_{0,z} & 0 & -\Omega_{0,x}\\
-\Omega_{0,y} & \Omega_{0,x} & 0
\end{array} \right) \vec{Y}.
\end{equation}
where the angular velocity vectors have components $\pmb{\Omega}_0 =
[\Omega_{0,x}, \Omega_{0,y}, \Omega_{0,z}]^T$ \corr{for the star. Similar expression may be obtained for the angular velocity of the planet, $\pmb{\Omega}_1 =
[\Omega_{1,x}, \Omega_{1,y}, \Omega_{1,z}]^T$.} The product in the right-hand side of the equation is simply
the vector product $\pmb{\Omega}_0 \times \vec{Y}$. The final equations of
the evolution of $\pmb{\Omega}_l$ have the following form:
\begin{equation}
\frac{d}{dt} \left( \frac{\partial \LL}{\partial \, \pmb{\Omega}_l} \right)
=  \pmb{\Omega}_l \times \frac{\partial \LL}{\partial \, \pmb{\Omega}_l}
+ \frac{1}{2} \, \frac{\partial \dot{E}}{\partial \, \pmb{\Omega}_l}, \quad l=0,1.
\label{eq:Lagrange_Omega}
\end{equation}
The above equations are valid for systems, which Lagrangian depends on the
space orientation of extended objects only through the angular velocities.
As we will show, for the case considered here, these equations have similar
explicit form as the Euler equations. Nevertheless, they were derived under
assumption of a particular form of the potential function $V = V(\vec{r},
\pmb{\Omega}_0, \pmb{\Omega}_1)$, suitable to our model.
%
\section{The Lagrangian and the dissipative function}

The Lagrangian of the system is given by the following expression:
\begin{equation}
\LL = \Tcl - \Vcl - \Vrot - \Vtid + \Trot  + \LLrel,
\end{equation}
where
\begin{equation}
\Tcl = \frac{1}{2} \, \beta \, \dot{\vec{r}}^2, \quad
\Vcl = -\frac{k^2 \, m_0 \, m_1}{r}
\end{equation}
are the kinetic and potential energies of two point masses interacting with
accord to the Newtonian gravity. Masses of the star and the planet are
denoted with $m_0$ and $m_1$, respectively, $\beta \equiv (1/m_0 +
1/m_1)^{-1}$ is the reduced mass, $r \equiv ||\vec{r}||$ and $k$ is the
Gauss constant. Terms $\Vrot$ and $\Vtid$ are for the perturbing potential
energy of two extended, non-spherical objects. We assume, that each object
is deformed due to its own rotation, as well as to the mutual tidal
interaction with the other body. These two effects are considered
separately. Such a simplification is correct to the first order. Moreover,
we assume that the planet deforms the shape of the star due to the point
mass interaction, and the star is a point-mass perturber deforming the figure
of the planet. The rotational kinetic energy of a deformable object is
given by $\Trot$, while the relativistic term of the Lagrangian is denoted
by $\LLrel$.
%
\subsection{Potential of the axially symmetric object}
%
Both perturbing terms $\Vrot$ and $\Vtid$ have an axial symmetry. The axis
of symmetry of $\Vrot$ coincides with a direction of the angular velocity vector of a
particular object, while the axis of symmetry of $\Vtid$ coincides with a
direction of a vector $\vec{r}$ joining the mass centers of the
bodies.
It is well known, that the potential energy of a system that consists of a
point mass $m$ and extended axially symmetric object of mass $M$ and
characteristic radius $R$ may be developed in Taylor series with respect to
$(R/r)$:
\begin{equation}
V_{\idm{axial}} = \frac{k^2 \, M \, m}{r} \sum_{l=1}^{\infty}
J_{2l} \left(\frac{R}{r}\right)^{2l} \, P_{2l}\,(\hat{\vec{r}} \cdot \hat{\vec{z}})
\approx \frac{k^2 M m}{r^3} \, J_2\, R^2\, P_2\,(\hat{\vec{r}} \cdot \hat{\vec{z}}),
\label{eq:V_axial}
\end{equation}
where $J_{2l}$ are the Stokes coefficients, and $P_{2l}\,(x)$ are the
Legendre polynomials \corr{[see, e.g., \cite{Schaub2003}, p.~480 or \cite{Murray2000}, p.~133]}. Here, these series are truncated at terms proportional
to $J_2$. In the case of the rotational deformation, $\hat{\vec{z}} =
\pmb{\Omega}/\Omega$, where $\Omega \equiv \|\pmb{\Omega}\|$. For the tidal
perturbations induced by a body at $\vec{r}$, $\hat{\vec{z}} = \hat{\vec{r}}
\equiv \vec{r}/r$. The zonal Stokes coefficients $J_2$ are expressed by the
following integrals \corr{\citep[e.g.,][p.~477]{Schaub2003}}:
\begin{eqnarray}
J_2 &=& -\frac{2\pi}{M R^2} \, \int_{0}^{\pi} \int_{0}^{R'(\vartheta)}
\rho\,(\mathcal{R}, \vartheta) \, \mathcal{R}^4 \, \sin\vartheta \, P_2
\,(\cos\vartheta) \, d\vartheta \, d\mathcal{R} \nonumber \\
&\approx& -\frac{2\pi}{M R^2} \, \int_{0}^{\pi} \int_{0}^{R} \rho\,(\mathcal{R},
\vartheta) \, \mathcal{R}^4 \, \sin\vartheta \,
P_2\,(\cos\vartheta) \, d\vartheta\, d\mathcal{R},
\label{eq:J2_def}
\end{eqnarray}
where the density $\rho\,(\mathcal{R}, \vartheta)$ depends on the magnitude
of perturbations due to the rotation and tides. That function may be
determined in a coordinate system which has its $z$-axis along
$\hat{\vec{z}}$. Angle $\vartheta$ is measured from this axis, between $0$
(the ''north pole'') to $\pi$ (the ''south pole''), and the radial variable
is $\mathcal{R} \in [0,R'(\vartheta)]$. The shape of a body is described in
terms of $R'(\vartheta)$, which may be approximated by $R$ when the
perturbation is small.
%
\subsubsection{Calculating the Love numbers}
%
To determine and describe tidal perturbations, we shall need to calculate
the Love numbers\corr{\footnote{\corr{In stellar astrophysics, the so called apsidal motion constant (instead of the Love number, which is twice as much) is used as a physical parameter giving the rate of the rotation of apsidal line of the binary orbit in space. On the other hand, in planetary astrophysics, the Love number is usually used (not only when a planet-moon system is considered, but also for a star-planet system). In this work we use this planetary convention.}}}. 
To accomplish this task, we apply remarkable results of
\cite{Chandrasekhar1933a}, who considered uniformly rotating polytropes\corr{\footnote{\corr{A polytrope is understood here as a gaseous object being in hydrostatic equilibrium under its own gravity. The pressure-density relation is given by the formulae $P = K \rho^{(n+1)/n}$ (where pressure is denoted by $P$, density by $\rho$, $K$ is a constant factor and $n$ is a polytropic index). A rotating polytrope or/and being attracted by some external force is then called {\it a perturbed polytrope}.}}} and
obtained equations for the density function $\rho\,(\mathcal{R},
\vartheta)$. He postulated the following form of $\rho$:
\begin{equation}
\rho\,(z, \vartheta) = \rho_{\idm{c}} \, \mathcal{W}^n\,(z, \vartheta),
\quad z = A\,\mathcal{R},
\end{equation}
where $\rho_{\idm{c}}$ is the central density and $n$ is the polytropic
index \corr{and $A$ is a function, which is usually introduced to define dimensionless variable $z$ \citep[see, e.g.,][p.~176 for the explicit formulae]{Kippenhahn1994}}. He found that for slowly rotating body:
\begin{equation}
\mathcal{W}\,(z, \vartheta) = w(z) + \frac{\Omega^2}{2 \pi k^2 \rho_{\idm{c}}}
\left[ \mathcal{U}_0\,(z) - a_n \, \mathcal{U}_2\,(z) \, P_2\,(\cos\vartheta) \right].
\end{equation}
Function $w(z)$ may be obtained by solving the Lane--Emden equation
(Eq.~[12] in the cited paper). Functions $\mathcal{U}_0\,(z)$ and
$\mathcal{U}_2\,(z)$ may be found by solving equations [37$_1$] and
[37 $_2$] in the cited paper, respectively (let us note that the notation
used here is different from the original one). Coefficients $a_n$ depend on
the polytropic index $n$ and may be derived as solutions to the problem
considered by Chandrasekhar. Using the resulting $\rho\,(\mathcal{R},
\vartheta)$, it is relatively easy to show that the Stokes coefficient $J_2$
defined by Eq.~(\ref{eq:J2_def}) has the following form
\begin{eqnarray}
&&J_2 = J_2^{\idm{(rot)}} = \frac{1}{3} \, k_{\idm{L,r}} \, \frac{R^3 \,
\Omega^2}{k^2 \, M}, \\
&&k_{\idm{L,r}} = k_{\idm{L,r}}(n) =
\frac{6 \, a_n}{5 \, z_n^5} \int_0^{z_n} n \, \left[ w\,(z) \right]^{n-1}
\mathcal{U}_2\,(z) \, z^4 \, dz,
\end{eqnarray}
where $z_n$ is dimensionless size of undisturbed body, i.e., $w\,(z_n) = 0$
and $k_{\idm{L,r}}$ may be attributed to the fluid Love number of the object
deformed due to the rotation (\cite{Munk1975}, p.~26), which is a function
of index $n$. In the range $n \in [0,4]$, this function is very well
approximated by the formulae:
\begin{eqnarray}
&&\log_{10} k_{\idm{L,r}} \approx f(n) = \log_{10} \frac{3}{2} + \alpha_1 \,
n + \alpha_2 \, n^2 + \alpha_3 \, n^3, \nonumber\\
&&\alpha_1 = -0.4872, \quad
\alpha_2 = +0.0424, \quad
\alpha_3 = -0.0238.
\label{eq:Love_number_formulae}
\end{eqnarray}
In the next paper, \cite{Chandrasekhar1933b} considered the deformation of
the polytropic body having mass $M$ by the tidal force emerged due to the
outer point-mass perturber of mass $m$. He found the density function, which
may be then used to calculate $J_2$ coefficient of tidally distorted object. It
may be shown that for small perturbations, the linear approximation is valid,
and then:
\begin{equation}
J_2 = J_2^{(\idm{tidal})} = - \left(\frac{R}{r}\right)^3 \,
\frac{m}{M} \,\, k_{\idm{L,t}},
\end{equation}
where $k_{\idm{L,t}}$ is the tidal Love number (\cite{Munk1975}, p.~27) and
is given by the formulae:
\begin{equation}
k_{\idm{L,t}} = k_{\idm{L,t}}(n) = \frac{5 \, z_n^2}{6 \, a_n} \,
\left[ 3 \, \mathcal{U}_2\,(z_n) + z_n \, \mathcal{U}_2\,'\,(z_n) \right]^{-1}
\, k_{\idm{L,r}}^{(n)} = k_{\idm{L,r}}^{(n)}.
\end{equation}
Thus the fluid Love numbers of rotationally and tidally deformed object are
numerically equal. Let's denote them by $k_{\idm{L}}$. \corr{It is not surprising, because $k_{\idm{L}}$ is a physical measure of deformability of an object under the attraction of some force (here, centrifugal and tidal forces are considered).} The Love number relates
to the quantity $Q$ introduced in (\cite{Eggleton1998}, Eq.~15c), i.e., $k_L
= Q/(1-Q)$. \corr{The numerical results obtained above are in agreement with the results stated in the cited work. However, the computation of $k_L$ (or the apsidal motion constant) were performed by many authors before \citep[e.g.,][]{Brooker1955}, to make this work self-contained, we present a brief overview of one of the way which leads to numerical values of $k_{\idm{L}}$. We choose to make use of Chandrasekhar's results, nevertheless this is not the only approach possible \citep[see, e.g.,][]{Eggleton1998}. Equation~15c from this last paper gives a formulae for $Q$ for a more general model of mass distribution. For a polytropic model they find a polynomial expression, Eq.~19. Our numerical results arisen from Eq.~(\ref{eq:Love_number_formulae}) are in agreement with the results in the literature \citep[e.g.,][]{Brooker1955,Eggleton1998}.}

\corr{However, the model of a polytrope being deformed by centrifugal as well as tidal forces is expected to work well for stars and gaseous planets, it is not expected so, when considering rocky planets \citep[see, e.g.,][]{Bursa1984}.}

To conclude, we write down the following forms of $\Vrot$ and $\Vtid$
\begin{equation}
V_{\idm{rot}} = \frac{m_1 \, k_{\idm{L},0} \, R_0^5 \, \Omega_0^2}{3 \, r^3} \,
P_2\,(\hat{\vec{r}} \cdot \hat{\pmb{\Omega}}_0) + \frac{m_0 \, k_{\idm{L},1} \,
R_1^5 \, \Omega_1^2}{3 \, r^3} \, P_2\,(\hat{\vec{r}} \cdot \hat{\pmb{\Omega}}_1),
\end{equation}
\begin{equation}
V_{\idm{tid}} = -\frac{k^2 \, m_1^2 \, R_0^5 \, k_{\idm{L},0}}{r^6}
- \frac{k^2 \, m_0^2 \, R_1^5 \, k_{\idm{L},1}}{r^6},
\end{equation}
where $k_{\idm{L,0}}$ and $k_{\idm{L,1}}$ are the Love numbers of the star
and the planet, respectively. Their numeric values are given by Eq.~(\ref
{eq:Love_number_formulae}). It is believed, that a mass distribution in
Sun-like stars is well approximated by a polytropic model of index $n \in
[3,4]$. In a case of Jupiter-like planets, the appropriate value of $n \in
[1,2]$.

\subsection{The kinetic energy of rotating, extended bodies}
%
The kinetic energy of rotations of extended objects is given by a sum:
\begin{equation}
\Trot = \frac{1}{2} \, \pmb{\Omega}_0^T \, \mathbb{I}_0 \,
\pmb{\Omega}_0 + \frac{1}{2} \, \pmb{\Omega}_1^T \, \mathbb{I}_1 \, \pmb{\Omega}_1,
\end{equation}
where $\mathbb{I}_0$ and $\mathbb{I}_1$ are the moments of inertia of the
star and the planet, respectively. The moment of inertia is defined in a
coordinate system with the origin that coincides with the mass center of the
body as the following integral (e.g., \cite{Schaub2003}, p.~129):
\begin{equation}
\mathbb{I}_l = \int_{\idm{body~l}} \rho_l\,(\tilde{\vec{r}}) \,
\left( - \big[ \tilde{\vec{r}} \big] \, \big[ \tilde{\vec{r}} \big] \right)
\, d^3\tilde{\vec{r}},
\label{eq:inertia_moment}
\end{equation}
where $\tilde{\vec{r}}$ points towards the mass element in the body, and
$[\tilde{\vec{r}}]$ is the so called {\em tilde matrix} of a vector
$\tilde{\vec{r}}$. For a vector $\vec{a} = [a_x, a_y, a_z]^T$, it has the form
of:
\begin{equation}
\big[ \vec{a} \big] \equiv \left( \begin{array}{c c c} 
0 & -a_z & a_y \\
a_z & 0 & -a_x \\
-a_y & a_x & 0
\end{array} \right).
\end{equation}
The matrix multiplication of $[\vec{a}]$ and a vector $\vec{b}$ is
equivalent to the the vector product $\vec{a} \times \vec{b}$. A
multiplication in Eq.~(\ref {eq:inertia_moment}) gives:
\begin{equation}
- \big[ \tilde{\vec{r}} \big] \, \big[ \tilde{\vec{r}} \big] = 
\left(\begin{array}{c c c}
\tilde{z}^2 + \tilde{y}^2 & - \tilde{x} \, \tilde{y} & - \tilde{x} \, \tilde{z}\\
- \tilde{x} \, \tilde{y} & \tilde{x}^2 + \tilde{z}^2 & - \tilde{y} \, \tilde{z}\\
- \tilde{x} \, \tilde{z} & - \tilde{y} \, \tilde{z} & \tilde{x}^2 + \tilde{y}^2
\end{array}\right).
\end{equation}
The density function of the $l$-th object ($l=0,1$) may be written as a sum:
\begin{equation}
\rho_l\,(\tilde{\vec{r}}) = \rho_l^{(0)} (\tilde{r}) + \Delta\rho_l^{(\idm{rot})}
(\tilde{\vec{r}}),
\end{equation}
where $\rho_l^{(0)} (\tilde{r})$ is for the density function of undisturbed
body, $\tilde{r} \equiv || \tilde{\vec{r}} || $, $\Delta\rho_l^{(\idm{rot})}
(\tilde{\vec{r}})$ is a correction stemming from the rotational deformation.
Thus the moment of inertia may be expressed as the following sum:
\begin{equation}
\mathbb{I}_l = \IIzero_l + \Delta\IIrot_l.
\end{equation}
Each term in that equation represents the moment of inertia for some density
function. The first term is the moment of inertia of undistorted body
possessing spherically symmetric distribution of its mass. Using a simple
polytropic model of that object, we obtain:
\begin{eqnarray}
&&\IIzero_l = \Izero_l \mathbb{E}, \quad \Izero_l =
\frac{2}{5} \, m_l \, R_l^2 \, \kappa_n, \\ 
&&\kappa = \kappa(n) = \frac{5}{3 \, z_n^4 \, |w'(z_n)|}
\int_{0}^{z_n} w^n(z) \, z^4 \, dz,
\end{eqnarray}
\corr{where $\mathbb{E}$ is a unit $3 \times 3$ matrix, i.e., $\mathbb{E} = \mbox{diag} (1,1,1)$.} 
For the homogeneous mass distribution ($n=0$), coefficient $\kappa = 1$. In
the range of $n \in [0,4]$, the coefficient is well approximated by the
formulae similar to that ones derived for the Love number:
\begin{eqnarray}
&&\log_{10} \kappa \approx f(n) = \alpha_1 \, n + \alpha_2 \, n^2 + \alpha_3 \, n^3, \\
&&\alpha_1 = -0.2061, \quad
\alpha_2 = +0.0297, \quad
\alpha_3 = -0.0140.
\end{eqnarray}
The rotational contribution to the moment of inertia  in the principal axes
frame (with the $z$ axis determined by $\hat{\pmb{\Omega}}$), has the following
form:
\begin{eqnarray}
&&\Delta\tildeIIrot_l = I_l^{(\idm{rot,1})} \,
\mathbb{E} - \, I_l^{(\idm{rot,2})} \,
\mbox{diag} \, (1,1,-1), \\
&&I_l^{(\idm{rot,1})} = \frac{4 \, \Omega_l^2 \, R_l^5 \, \tau_l}{3 \, k^2}, \quad
I_l^{(\idm{rot,2})} = \frac{\Omega_l^2 \, R_l^5 \, k_{\idm{L},l}}{9 \, k^2}.
\end{eqnarray}
The coefficient $\tau$ is given in the polytropic model by the integral:
\begin{equation}
\tau = \tau(n) = \frac{1}{z_n^5} \int_0^{z_n} n \, w^{n-1}(z) \,
\mathcal{U}_0(z) \, z^4 \, dz.
\end{equation}
Similarly to the coefficients $k_{\idm{L}}$ and $\kappa$, $\tau$ may be also
approximated by the polynomial
\begin{eqnarray}
&&\log_{10} \tau \approx f(n) = \alpha_1 \, n + \alpha_2 \,
n^2 + \alpha_3 \, n^3, \\
&&\alpha_1 = -0.3818, \quad
\alpha_2 = +0.0220, \quad
\alpha_3 = -0.0125.
\end{eqnarray}
The resulting form of $\Trot$ is then:
\begin{equation}
\Trot = \frac{1}{2} \left( \Izero_0 + \Irot_0 \right) \Omega_0^2 +
\frac{1}{2} \left( \Izero_1 + \Irot_1 \right) \Omega_1^2.
\end{equation}
%
\subsection{The relativistic perturbing Lagrangian}
%
The model considered here includes also the relativistic perturbations. We do not
include terms containing $\pmb{\Omega}_0, \pmb{\Omega}_1$, because their
contributions to the equations of motion are of the second order. The
relativistic contribution to the Lagrangian of the two point masses is given
by the formulae  \citep[e.g.,][]{Brumberg2007}:
\begin{equation}
\LL_{\idm{rel}} = \frac{1}{c^2} \bigg\lbrace \gamma_1 \left( \dot{\vec{r}}^2 \right)^2{}
 + \gamma_2 \frac{\dot{\vec{r}}^2}{r} +
 \gamma_3 \frac{\left(\vec{r} \cdot \dot{\vec{r}}\right)^2}{r^3}
 - \gamma_4 \frac{1}{r^2} \bigg\rbrace,
\end{equation}
where the mass parameters have the following form:
\begin{eqnarray}
&&\gamma_1 = \frac{\beta}{8} \frac{m_0^3 + m_1^3}{\left(m_0 + m_1\right)^3}, \quad
\gamma_2 = \frac{k^2}{2} \left( 3 \, m_0 \, m_1 + \beta^2 \right), \\
&&\gamma_3 = \frac{k^2}{2} \beta^2, \quad
\gamma_4 = \frac{1}{2} \, \beta \, \mu^2.
\end{eqnarray}
%

\subsection{The dissipative function}
%
A function describing dissipation of the mechanical energy of the system
is the following sum:
\begin{equation}
\dot{E} = \dot{E}_0 + \dot{E}_1,
\end{equation}
where the first term comes from the energy dissipation in the star due to
tidal interaction with a planet, while the second term expresses the energy
dissipation in the planet due to the interaction with the star. The
particular term of $\dot{E}$ are given by a simple model proposed by \citep[][\corr{Eq.~43}]{Eggleton1998}:
\begin{equation}
\dot{E}_l = -\frac{9 \, \sigma_l \, m_l^2 \, R_l^{10} \, k_{\idm{L},l}^2}{2 \, r^8}
\bigg[ 2 \, \frac{\left( \vec{r} \cdot \dot{\vec{r}} \right)^2}{r^2}
+ \dot{\vec{r}}^2 - 2 \, \dot{\vec{r}}
\cdot \left( \pmb{\Omega}_l \times \vec{r} \right)
+ \left( \pmb{\Omega}_l \times \vec{r} \right)^2 \bigg],
\label{eq:Edot}
\end{equation}
for $l=0,1$. Parameters $\sigma_0, \sigma_1$ are dissipation constants
for the star and the planet, respectively\corr{\footnote{\corr{We assume that $\sigma_l$ are constant, which is physically wrong, while these quantities depend on the tidal frequency (which is a function of orbital elements and time). Nevertheless, because of poor knowledge of their values, we make this simplification of the model. For an overview on the more realistic tidal models see, e.g., \citep{FerrazMello2008, Efroimsky2009, Efroimsky2011}.} \Corr{The dependence of $\sigma_l$ on the tidal frequency is also discussed in subsection~4.3.}}}. They may be expressed in the
following form \citep{Kiseleva1998}:
\begin{equation}
\sigma_l = \frac{\lambda_l}{m_l \, R_l^2} \left( \frac{1 +
k_{\idm{L},l}}{k_{\idm{L},l}} \right)^2 \sqrt{\frac{k^2 \, m_l}{R_l^3}},
\end{equation}
where $\lambda_0$ and $\lambda_1$ are non-dimensional coefficients of the
energy loss rates in the star and the planet, respectively. A calculation of
these values is complex, and we postpone the derivation to other work,
actually, that is basically beyond the scope of the present paper. Following
the literature \cite[e.g.,][]{Ogilvie2004, Wu2005a, Wu2005b, Ogilvie2007, Hansen2010}, we may assume these coefficients in the range of
$\lambda_0 \in [10^{-7}, 10^{-4}]$ and $\lambda_1 \in [10^{-7}, 10^{-4}]$.
Particular values of the dissipative coefficients are not well known. In the
paper by \cite{Hansen2010}, he calibrated the equilibrium tide model
comparing the results of calculations with the observed relation of the
orbital period and eccentricity of close binaries for which their age is
known. He obtained $\bar{\sigma}_* \approx 5.3 \times 10^{-5}$, which may be
``translated'' to $\lambda_0$, through relation $\lambda_0 = \bar{\sigma}_*
(1 + k_{\idm{L},0})^{-2}$, which then gives $\lambda_0 \approx 5 \times
10^{-5}$. The dissipation constants for Jupiter-like planets were derived by
analysis the period-eccentricity relations of known systems with these
planets. The resulting $\bar{\sigma}_p \approx 6.8 \times 10^{-7}$ gives
$\lambda_1 \approx 5 \times 10^{-7}$.
%
\section{The explicit form of the equations of motion}
%
Furthermore, we consider a particular form of the Lagrangian and the dissipative
function:
\begin{eqnarray}
&&\LL = T_0(\pmb{\Omega}_0) + T_1(\pmb{\Omega}_1) - V_0(\vec{r},
\pmb{\Omega}_0) - V_1(\vec{r}, \pmb{\Omega}_1) + \LL_1(\vec{r}, \dot{\vec{r}}), \\
&&\dot{E} = \dot{E}_0(\vec{r}, \dot{\vec{r}}, \pmb{\Omega}_0)
+ \dot{E}_1(\vec{r}, \dot{\vec{r}}, \pmb{\Omega}_1).
\end{eqnarray}
The derivative of $\LL$ over $\dot{\vec{r}}$ does not depend on
$\pmb{\Omega}_l$ and similarly, the derivative of $\LL$ over particular
angular velocity does not depend on the other $\pmb{\Omega}$, nor on the
linear velocity. Hence, the vector equations for $\vec{r}$, $\pmb{\Omega}_0$
and $\pmb{\Omega}_1$ are separable. Equations in the top row of (\ref
{eq:Lagrange_vec_s}) describe the orbital  motion of the planet. The
equations~(\ref {eq:Lagrange_Omega}) with $l=0,1$ are for the evolution of
the angular velocities of the star and the planet, respectively.
%
\subsection{The equations of translational motion}
%
The top-row vector equation of~(\ref{eq:Lagrange_vec_s}) has to be solved
with respect to $\ddot{\vec{r}}$ to obtain the explicit form of the
equations of translational motion. The dependence of $\LL$ on $\dot{\vec{r}}$
appears in the $\Tcl$ and $\LLrel$ terms. Only the first one is a uniform
quadratic function of $\dot{\vec{r}}$. Nevertheless, we may solve the problem.
The solution has the following form in the first order approximation:
\begin{eqnarray}
\ddot{\vec{r}} & \simeq & -\frac{\mu}{r^3} \, \vec{r} 
- \frac{6 \, \mu}{r^8} \left( \frac{A_0}{m_0} + \frac{A_1}{m_1} \right) \vec{r} \nonumber \\ 
&&- \frac{1}{2 \, \beta} \frac{1}{r^5} \sum_{l=0,1} A_l \bigg\lbrace \Big[ \Omega_l^2
- 5 \frac{\left( \vec{r} \cdot \pmb{\Omega}_l \right)^2}{r^2} \Big] \vec{r} + 2 \left( \vec{r} \cdot \pmb{\Omega}_l \right) \pmb{\Omega}_l \bigg\rbrace \nonumber\\
&&+ \frac{1}{c^2} \bigg\lbrace \Gamma_1 \frac{\dot{\vec{r}}^2}{r^3} \vec{r} + \Gamma_2 \frac{\vec{r} \cdot \dot{\vec{r}}}{r^3} \dot{\vec{r}} + \Gamma_3 \frac{\left( \vec{r} \cdot \dot{\vec{r}} \right)^2}{r^5} \vec{r} + \Gamma_4 \frac{\vec{r}}{r^4} \bigg\rbrace \nonumber\\
&& - \frac{9}{2 \, \beta} \frac{1}{r^8} \sum_{l=0,1} \sigma_l \, A_l^2 \bigg\lbrace 2 \frac{\vec{r} \cdot \dot{\vec{r}}}{r^2} \, \vec{r} + \dot{\vec{r}} - \pmb{\Omega}_l \times \vec{r} \bigg\rbrace,
\label{eq:final_translational}
\end{eqnarray}
where the first term is the Keplerian acceleration, the second term has its
origin in the perturbing potential of tidally deformed bodies. The term in
the second row emerges due to the rotational deformation of both objects.
The perturbing acceleration in the third row is for the relativistic
contribution, while the last term comes from the dissipation of a mechanical
energy. \Corr{In spite of very different approaches used in \citep{Eggleton1998} and in this work, the final equations of motion, especially their dissipative part, are the same (see their Eq.~34 with $\vec{f}_{\idm{TF}}$ given by Eq.~45).}
According with the above notation $A_0 \equiv m_1 \, R_0^5 \,
k_{\idm{L,0}}$, $A_1 \equiv m_0 \, R_1^5 \, k_{\idm{L,1}}$ and
\begin{eqnarray}
\Gamma_1 \equiv - \mu - 3 \, k^2 \beta, \quad
\Gamma_2 \equiv 4 \, \mu - 2 \, k^2 \beta, \quad
\Gamma_3 \equiv \frac{3}{2} \, k^2 \beta, \quad
\Gamma_4 \equiv 2 \, \mu \left( 2 \, \mu + k^2 \beta \right).\nonumber
\end{eqnarray}
%

\subsection{The equations of the evolution of angular velocities}
%
The equations of the evolution of angular velocities of the star and the
planet, denoted here by $\pmb{\Omega}_0$ and $\pmb{\Omega}_1$ are obtained
through solving Eq.~(\ref{eq:Lagrange_Omega}) with respect to
$\dot{\vec{\Omega}}_l$ ($l=0,1$). To the first order, one derives the
following form of these equations:
\begin{eqnarray}
\dot{\pmb{\Omega}}_l & \simeq & -\frac{A_l}{\Izero_l} \frac{1}{r^5} \, \left( \vec{r} \cdot \pmb{\Omega}_l \right) \left( \pmb{\Omega}_l \times \vec{r} \right) \nonumber \\ 
&&+ \frac{A_l}{\Izero_l} \frac{1}{r^5} \bigg\lbrace \left( \dot{\vec{r}} \cdot \pmb{\Omega}_l \right) \vec{r} + \left( \vec{r} \cdot \pmb{\Omega}_l \right) \dot{\vec{r}} + \left( \vec{r} \cdot \dot{\vec{r}} \right) \pmb{\Omega}_l - 5\, \left( \vec{r} \cdot \pmb{\Omega}_l \right) \left( \vec{r} \cdot \dot{\vec{r}} \right) \frac{\vec{r}}{r^2} \bigg\rbrace \nonumber \\
&& - \frac{9}{2} \, \frac{\sigma_l \, A_l^2}{\Izero_l} \, \frac{1}{r^8} \, \vec{r} \times \bigg\lbrace \left( \pmb{\Omega}_l \times \vec{r} \right) - \dot{\vec{r}} \bigg\rbrace,
\label{eq:final_spins}
\end{eqnarray}
where the first-row term is the same as would be derived directly from
the Euler equations. The second-row terms emerge due to dependency of  the
potential energy on $\pmb{\Omega}_l$. We note that the non-rigid-body
contribution to the equations are the same order as the rigid-body term. The
dissipative term stands in the last row of~(\ref{eq:final_spins}) and it is
exactly the same as in \citep{Eggleton1998}. \Corr{To compare the results, see their Eq.~36 with $\vec{f}_{\idm{TF}}$ given by Eq.~45.}
%
\subsection{\Corr{Effects} not included in the model}
%
The equations derived so far do not take into account a few physical and
perturbing phenomena, which we would like to list here explicitly. First of
all, we assumed that the star does not evolve in time keeping its internal
structure constant. However, if to consider the evolutionary time scale, the
stellar radius as well as the internal structure change: the  time evolution
of $R_0$ and the mass distribution alter the inertia moment, which then may
provide an additional term in Eq.~(\ref {eq:final_spins}). Hence, also the
Love number is not constant over time. Evolution  of the internal structure,
for instance, the size of the convective zone, implies that none of
coefficients $\lambda_0$ remains constant. Also a significant stellar wind
would be responsible for yet additional term in Eq.~(\ref {eq:final_spins}),
\citep[e.g.,][]{Dobbs-Dixon2004, Barker2009}.

The same discussion applies to the planet. Its radius and internal structure
also may vary over time, which is mainly attributed to the tidal and thermal
influence by the parent star \citep[e.g.,][]{Gu2009, Miller2009, Hansen2010,
Arras2010, Ibgui2011}. However, the tidal evolution of the interiors of
hot-Jupiters is not well recognized. Moreover, very close-in planets
revolving around their stars with periods of the order of one day, may be
close enough to loose their mass through the Lagrangian point $L_3$ \citep{Li2010}, as they may fill up the Roche lobe.

\Corr{
In the considered model, we assume, that the material is purely viscous, which means that there is no unelasticity nor rigidity in the bodies. The model ignores the so-called $\Lambda$-effect, which leads to differential rotation in the stars \citep[see, e.g.,][]{Kitchatinov1993, Kueker1993}.
However, this effect is well studied for single stars, it is not understood yet how a close planetary companion can tidally affect the differential rotation of the star. And on the other hand, the tidal influence of the star on the planet's differential rotation is not studied well enough. Here, we omit this effect.
}

\Corr{
As it was noted already, the dissipation parameters $\sigma_l$ (and also $\lambda_l$) depend on the tidal frequency, $\omega_{\idm{t}}$. After \citep{Eggleton1998}, we assume, that the energy is dissipated due to turbulent convection in the objects (thus, we consider the so-called \textit{equilibrium tide} model). Therefore, $\sigma_l$ parameter depends on the value of the turbulent viscosity coefficient $\nu_{\idm{t}}$ \citep[see][Eq.~113]{Eggleton1998}. We assumed, that $\nu_{\idm{t}}$ is constant, i.e., does not depend on $\omega_{\idm{t}}$, which is not true. For small $\omega_{\idm{t}}$, $\nu_{\idm{t}}$ is nearly constant, while for larger tidal frequency, it is usually believed to be proportional to the inverse of its square, \citep[see, e.g.,][]{Goodman1997, Terquem1998}.
Moreover, $\nu_{\idm{t}}$ is not isotropic, and its value depends on the Coriolis number \citep{Kueker1994}.
While for the star of the known internal structure, it is possible to calculate $\nu_{\idm{t}}$, for the planet it is usually not possible due to very poor knowledge of it's interior. In our simplified model, we do not take this effect into account.
}

\Corr{
The equilibrium tide is not the only contribution to the energy loss. The other mechanism relies on the damping of the internal waves excited by tidal force (the so-called \textit{dynamical tide}). The waves may be damped by radiative diffusion, convective viscosity as well as non-linear breaking \citep[see, e.g.,][and references therein]{Barker2010}. This mechanism produces different dependence of the dissipative parameter, then the equilibrium tide. In the literature, the so-called quality factor $Q$ (or rather the modified quantity $Q'$) is used to express the energy dissipation rate. If we find formal relation between $\sigma$ and $Q$, the equilibrium tide model leads to $Q \propto \omega_{\idm{t}}$. Dynamical tide model, on the other hand, gives the inverse relation. The dependence is in fact much more complex and is a subject of many studies in the literature \citep{Zahn1975, Goodman1998, Goodman2009, Ogilvie2009, Barker2010, Efroimsky2011}.
}

%
\section{Averaging over the mean anomaly}
%
The orbital equations of motion have the following general form:
\begin{equation}
\ddot{\vec{r}} = -\frac{\mu}{r^3} \, \vec{r} + \vec{f},
\label{eq:translational_general}
\end{equation}
where the first term is for the Keplerian acceleration and $\vec{f}$ is a
perturbation of small magnitude as compared to the Keplerian term.
Therefore, the planet moves on weakly perturbed elliptic orbit. The time
scales of variation of its elements are a few orders of magnitude longer
than the orbital period. Hence, one might perform the averaging of the
equations of motion over fast evolution (i.e., over the mean anomaly) making
use of the {\it averaging principle} \corr{[see, e.g., \cite{Arnold1993}, \S 6.1 or \cite{Arnold1995}, \S 51, \S52]}. Nevertheless, at first we should
verify whether the evolution of $\pmb{\Omega}_0$ and $\pmb{\Omega}_1$ occur
in slow-enough time scale. Because the magnitude of the moment of inertia of
the planet is much smaller than that one of the star, i.e., $\Izero_1 \ll
\Izero_0$, the planetary angular velocity vector varies faster than the
stellar one. With the help of a relatively simple analysis of~(\ref
{eq:final_spins})\corr{\footnote{\corr{The analysis, i.e., an estimation of the order of magnitude of $|\dot{\pmb{\Omega}}_1|$, is done by setting $e=0$, the inclination between $\pmb{\Omega}_1$ and $\vec{r}$ to $\frac{\pi}{4}$ and physical parameters to their typical values.}}}, one finds that the relative variation of $\pmb{\Omega}_1$
during one orbital period has its order of:
\[
\frac{|\dot{\pmb{\Omega}}_1|}{\Omega_1} \,
P_{\idm{orb}} \sim \frac{5 \, \pi}{2} \frac{m_0}{m_1}
\left( \frac{R_1}{r} \right)^3 \frac{\Omega_1}{n},
\]
where $n$ is the mean motion. For a one-day orbit and fast rotating
Jupiter-like planet ($\Omega_1 \sim n$) , the above quantity is of the order
of $6 \times 10^{-2}$. For wider orbits. it obviously decreases. Thus, we
have shown that the assumptions of the averaging principle are fulfilled at
acceptable level even for rather ``extreme'' systems. Nevertheless, we
should keep in mind that this level of approximation is only acceptable when
one would like to compare the results of the mean and exact model.

Having the equations of motion in general form of~(\ref
{eq:translational_general}), one may find the following equations for the
evolution of the orbital angular momentum and Laplace vectors, respectively
$\vec{h}$ and $\vec{e}$:
\begin{equation}
\dot{\vec{h}} = \vec{r} \times \vec{f}, \quad \dot{\vec{e}}
= \frac{1}{\mu} \Big\lbrace \vec{f} \times \vec{h}
+ \dot{\vec{r}} \times \left( \vec{r} \times \vec{f} \right) \Big\rbrace.
\end{equation}
Keplerian orbital elements, such as the semi-major axis, eccentricity,
inclination, longitude of \Corr{ascending} node and argument of pericenter, may be
easily derived from the components of $\vec{h}$ and $\vec{e}$. We choose
this representation instead of the classical Gauss planetary equations,
because it has no singularity with respect to the inclination.

If the motion is considered to as the Keplerian one, all angles $\vec{h}$,
$\vec{e}$, $\pmb{\Omega}_0$ and $\pmb{\Omega}_1$ are constant and the {\it
fast vectors} $\vec{r}$ and $\dot{\vec{r}}$ may be expressed as the following
combinations of $\vec{e}$ and $\vec{h} \times \vec{e}$:

\begin{equation}
\vec{r} = x_{\idm{orb}} \, \frac{\vec{e}}{e} + y_{\idm{orb}}
\, \frac{\vec{h} \times \vec{e}}{h \, e}, \quad
\dot{\vec{r}} = \dot{x}_{\idm{orb}} \, \frac{\vec{e}}{e}
+ \dot{y}_{\idm{orb}} \, \frac{\vec{h} \times \vec{e}}{h \, e},
\end{equation}
where $x_{\idm{orb}}$ and $y_{\idm{orb}}$ are Cartesian $(x,y)$--coordinates
in the Kepler frame, i.e, $x_{\idm{orb}} = r \, \cos\nu$ and $y_{\idm{orb}}
= r \, \sin\nu$, where $\nu$ is the true anomaly. 
Thus, the right-hand sides of the equations for
$\dot{\vec{h}}$, $\dot{\vec{e}}$, $\dot{\pmb{\Omega}}_0$ and
$\dot{\pmb{\Omega}}_1$ are some functions of $\nu$, which are parametrised
by constant vectors $\vec{h}$, $\vec{e}$, $\pmb{\Omega}_0$ and
$\pmb{\Omega}_1$.

According with the principle of averaging\corr{\footnote{\corr{We consider the problem using the averaging principle, although in general this principle is incorrect. 
From a technical point of view and when we don't need to know the relation between mean and actual elements, this principle is equivalent to the first order perturbation theory \citep[for an overview on the classical perturbation theories see][]{FerrazMello2007}.}}}, the equations of motion are averaged out over the mean anomaly
$\mathcal{M}$, under assumption that the orbital elements are fixed
over the averaging period. Formally, the procedure relies on calculating
following integral:
\begin{equation}
\big\langle \Ff \big\rangle \equiv \frac{1}{2 \pi} \int_0^{2 \pi} \Ff \,
d\mathcal{M} = \frac{1}{2 \pi} \int_0^{2 \pi} \left( \Ff \, \mathcal{J} \right)
\, d\nu, \quad
\mathcal{J} \equiv \frac{\left( 1 - e^2 \right)^{\frac{3}{2}}}{\left( 1 + e \,
\cos\nu \right)^2},
\end{equation}
where $\Ff$ is averaged function. In the above formulae, we did a change of
the integration variable, from the mean anomaly to the true anomaly.
Note that we know $\Ff$ as a function of $\nu$ rather then $\mathcal{M}$
(see the discussion in \cite{Migaszewski2008}).

By calculating such integrals over the right-hand sides of the equations of
motion, one finds the secular equations that have the following form:
\begin{eqnarray}
\!\!\!\!\!\!\!\!\!\!\!\!&&\dot{\vec{h}} = -\frac{1}{2\,\beta} \,
\sum_{l=0,1} A_l \, \Ff_{1,l} - \frac{9}{4\,\beta} \sum_{l=0,1}
\sigma_l \, A_l^2 \, \Ff_{2,l},\label{eq:dot_h_sec}\\
\!\!\!\!\!\!\!\!\!\!\!\!&&\dot{\vec{e}} = -\frac{1}{4 \, \beta}
\sum_{l=0,1} A_l \, \Ff_{3,l} - 15 \sum_{l=0,1} \frac{A_l}{m_l} \Ff_4 
- \frac{3 \, \mu}{c^2} \, \Ff_5 - \frac{9}{4 \, \beta} \sum_{l=0,1}
\sigma_l \, A_l^2 \, \Ff_{6,l},\label{eq:dot_e_sec}\\
\!\!\!\!\!\!\!\!\!\!\!\!&&\dot{\pmb{\Omega}}_l = \frac{1}{2 \, \Izero_l}
\, A_l \, \Ff_{1,l} + \frac{9}{4 \, \Izero_l} \, \sigma_l \, A_l^2 \,
\Ff_{2,l}, \quad l=0,1.
\label{eq:dot_Om_sec}
\end{eqnarray}
Formally, variable names of $\dot{\vec{h}}$, $\dot{\vec{e}}$, etc., should be
encompassed by square brackets $\langle ... \rangle$, that denote the averaged
values. However, it will be clear from the context, that after the averaging
all functions should be understood as the {\it secular} or the {\it mean}
quantities. The functions $\Ff$ are given by the following formulae:
\begin{eqnarray}
\Ff_{1,l} &=& \frac{\mu^3 \left( 1 - e^2 \right)^{\frac{3}{2}}}{h^8 \, e^2} \Big\lbrace h^2 \, \left( \vec{e} \cdot \pmb{\Omega}_l \right) \, \left( \vec{e} \times \pmb{\Omega}_l \right) + \big[ \left( \vec{e} \times \vec{h} \right) \cdot \pmb{\Omega}_l \big] \, \left( \vec{e} \times \vec{h} \right) \times \pmb{\Omega}_l \Big\rbrace,\nonumber\\
\Ff_{2,l} &=& \frac{\mu^6 \left( 1 - e^2 \right)^{\frac{3}{2}}}{h^{16} \, e^2} \Big\lbrace 2 \, \mu^2 \, e^2 \, \fE_2 \vec{h} - 2 \, e^2 \, h^4 \, \fE_4 \pmb{\Omega}_l + h^4 \, \fE_3 \left( \vec{e} \cdot \pmb{\Omega}_l \right) \vec{e} \nonumber\\ 
&&+ h^2 \, \fE_5 \big[ \left( \vec{e} \times \vec{h} \right) \cdot \pmb{\Omega}_l \big] \left( \vec{e} \times \vec{h} \right) \Big\rbrace,\nonumber\\
\Ff_{3,l} &=& \frac{\mu^3 \left( 1 - e^2 \right)^{\frac{3}{2}}}{h^{10} \, e^2} \Big\lbrace 2 \, h^2 \big[ \left( \vec{e} \cdot \pmb{\Omega}_l \right) \left( \vec{e} \times \vec{h} \right) \cdot \pmb{\Omega}_l \big] \vec{e} \nonumber\\
&& + \Big[ 2 \, h^2 \, e^2 \, \Omega_l^2 - 7 \, h^2 \left( \vec{e} \cdot \pmb{\Omega}_l \right)^2 - 5 \big[ \left( \vec{e} \times \vec{h} \right) \cdot \pmb{\Omega}_l \big]^2 \Big] \left( \vec{e} \times \vec{h} \right) \nonumber \\ 
&& + 2 \, e^2 \, h^2 \big[ \left( \vec{e} \times \vec{h} \right) \cdot \pmb{\Omega}_l \big] \pmb{\Omega}_l + 4 \, h^2 \, e^2 \left( \vec{e} \cdot \pmb{\Omega}_l \right) \left( \pmb{\Omega}_l \times \vec{h} \right)  \Big\rbrace,\nonumber\\
\Ff_4 &=& \frac{\mu^7 \left( 1 - e^2 \right)^{\frac{3}{2}}}{h^{14}} \, \fE_5 \left( \vec{e} \times \vec{h} \right),\nonumber\\
\Ff_5 &=& \frac{\mu^3 \left( 1 - e^2 \right)^{\frac{3}{2}}}{h^6} \left( \vec{e} \times \vec{h} \right),\nonumber\\
\Ff_{6,l} &=& \frac{\mu^6 \left( 1 - e^2 \right)^{\frac{3}{2}}}{h^{16}} \Big\lbrace 18 \, \mu^2 \, \fE_1 \vec{e} + h^2 \, \fE_5 \big[ \left( \vec{e} \cdot \pmb{\Omega}_l \right) \vec{h} - 11 \left( \vec{h} \cdot \pmb{\Omega}_l \right) \vec{e} \big]\Big\rbrace,\nonumber
\end{eqnarray}
where $\fE_i \equiv \fE_i (e)$ are the following functions of eccentricity

\begin{eqnarray}
&&\fE_1 = 1 + \frac{15}{4} \, e^2 + \frac{15}{8} \, e^4 + \frac{5}{64} \, e^6, 
\quad \fE_2 = 1 + \frac{15}{2} \, e^2 + \frac{45}{8} \, e^4 + \frac{5}{16}
\, e^6,\nonumber\\
&&\fE_3 = 1 + \frac{9}{2} \, e^2 + \frac{5}{8} \, e^4,
\quad \fE_4 = 1 + 3 \, e^2 + \frac{3}{8} \, e^4,
\quad \fE_5 = 1 + \frac{3}{2} \, e^2 + \frac{1}{8} \, e^4.\nonumber
\end{eqnarray}
All these functions are equal to $1$ for circular orbits and are greater than $1$
for elliptic orbits.
\begin{figure}
\centerline{
\hbox{
\includegraphics[width=0.5\textwidth]{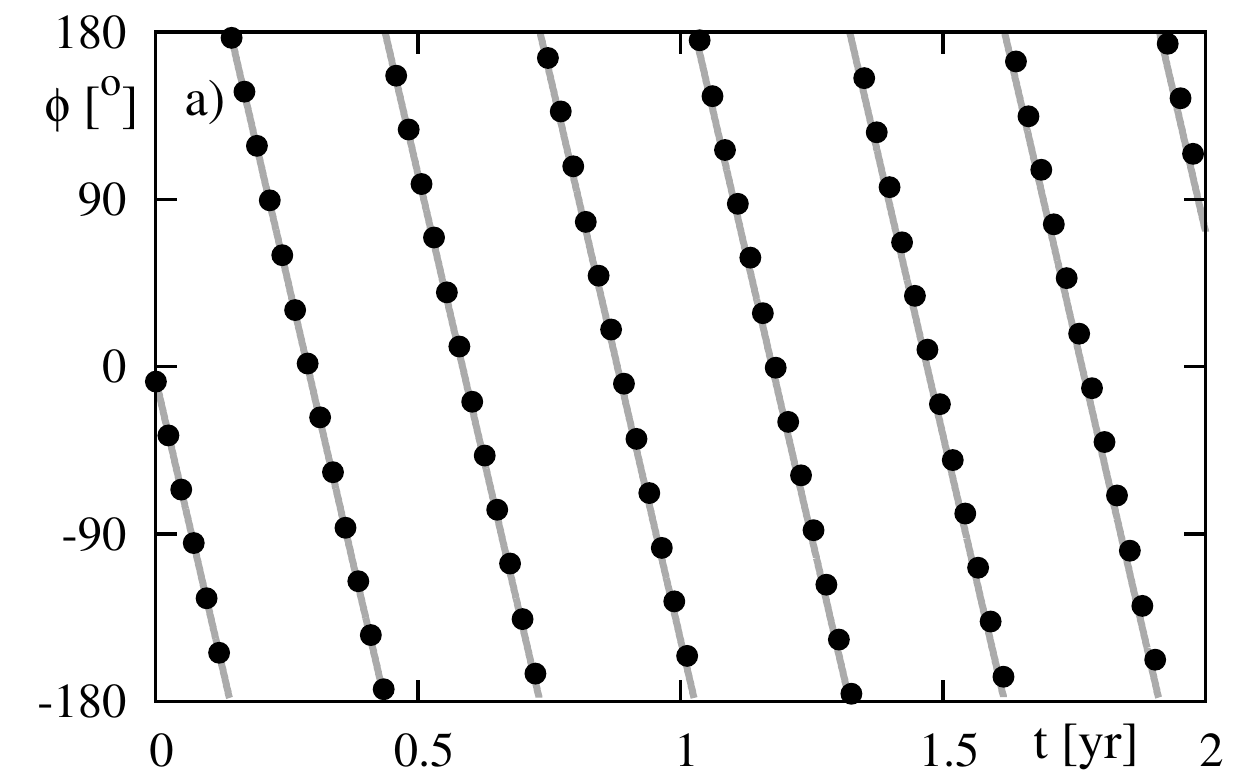}
\includegraphics[width=0.5\textwidth]{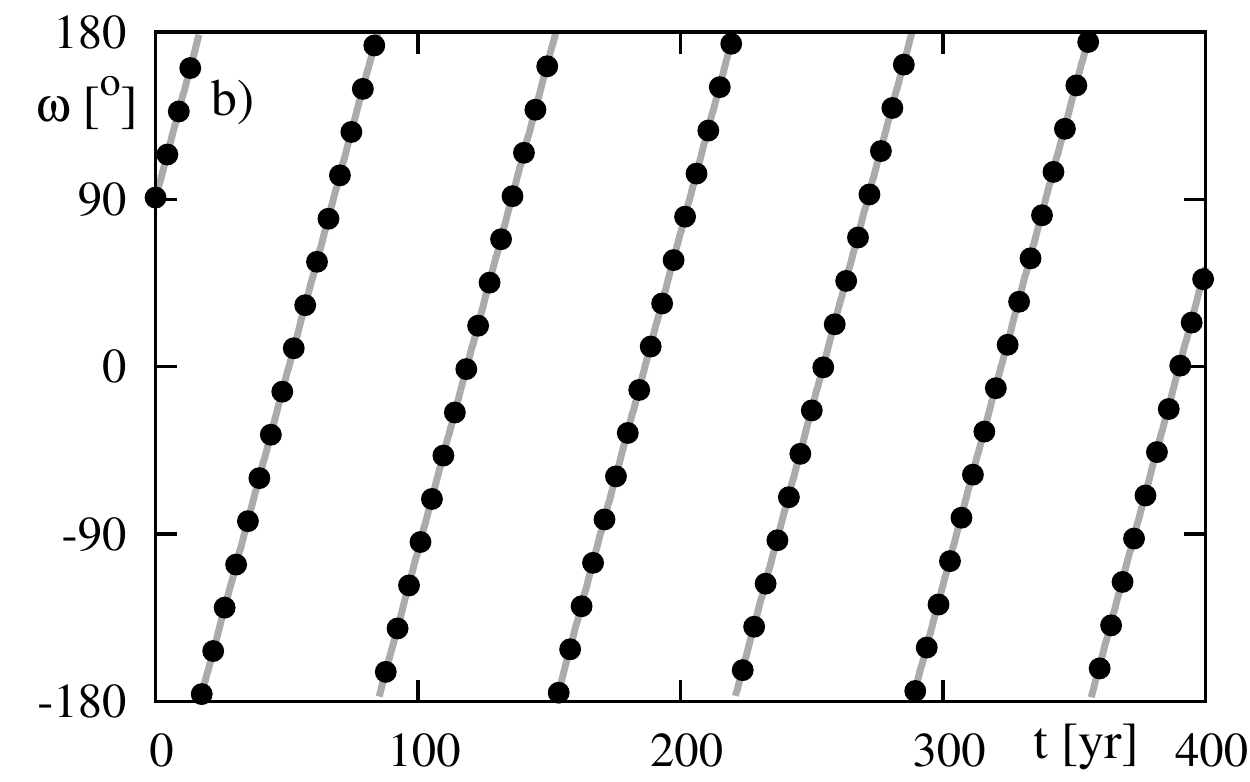}
}
}
\caption{
Temporal evolution of angles $\phi_1(t)$ (the left panel) and $\omega(t)$
(the right panel) derived in terms of the solution to the full equations of
motion (black dots) and to the equations of the first-averaged system (gray
curves). Main parameters of the system are $m_0 = 1\,\msun$, $m_1 = 1\,\mJ$,
$R_0 = 1\,\RS$, $R_1 = 1\,\RJ$, $a = 0.02\,\au$, $e = 0.1$,
$T_{\idm{rot,0}} = 5\,\mbox{d}$ , $T_{\idm{rot,1}} = 1\,\mbox{d}$.
}
\label{fig:evolution_comparison1}
\end{figure}
To verify the correct form of the averaged equations, we compare the
time-evolution of the unaveraged system, Eqs.~(\ref{eq:final_translational})
and~(\ref {eq:final_spins}) with the evolution of the secular system, 
Eq.~(\ref {eq:dot_h_sec})-(\ref{eq:dot_Om_sec}). Figure~\ref
{fig:evolution_comparison1} illustrates the results of that comparison as
plots of the azimuthal angle of $\pmb{\Omega}_1$ in the orbital reference
frame, $\phi_1(t)$ (the left-hand panel), and the argument of pericenter
$\omega(t)$ (the right-hand panel). \corr{The integration of both the non-averaged and the averaged equations of motion were done with the help of the Taylor integrator \citep[software package by][]{Jorba2004}.} As we can see, the results agree very
well, although the precession of the planetary spin is relatively fast, with
the period of about $100$ days. We will \Corr{discuss} further in this paper, that
this time-scale is the fastest one after the orbital period, and the next
one, in terms of the magnitude, is the characteristic time-scale of the
pericenter advance.
\begin{figure}
\centerline{
\hbox{
\includegraphics[width=0.5\textwidth]{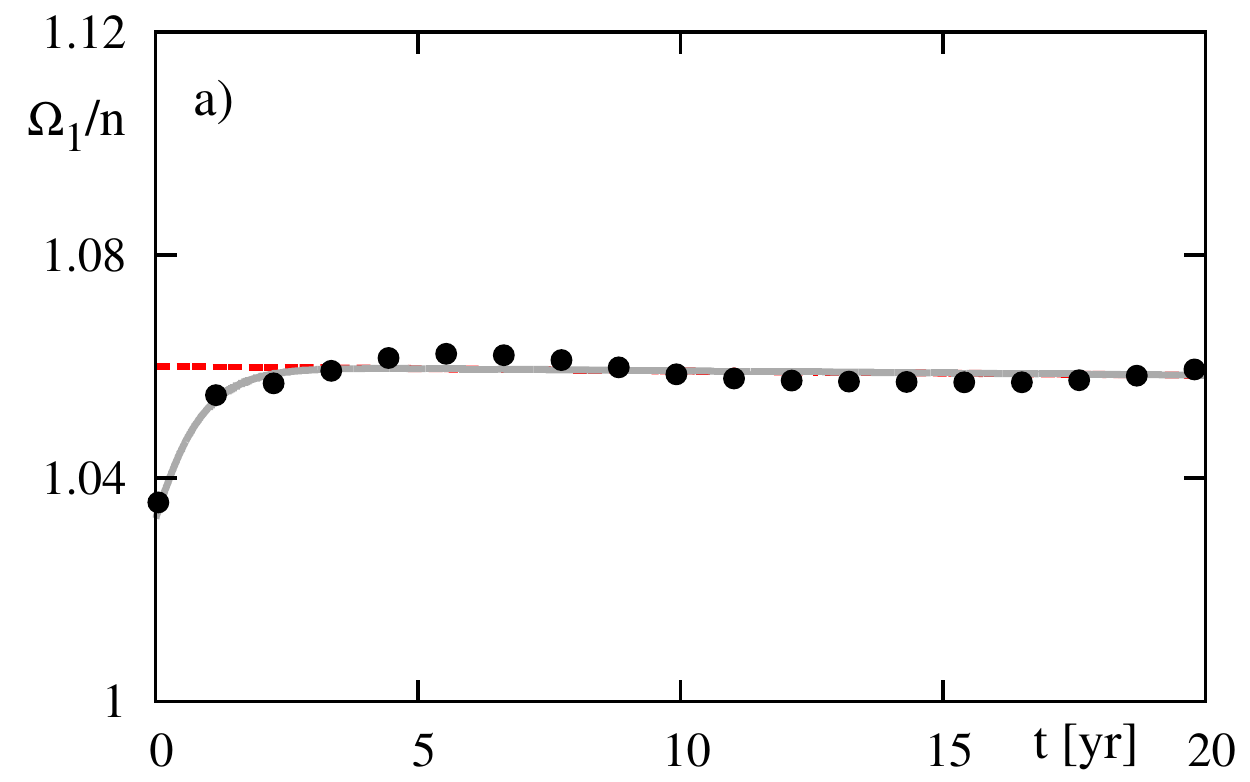}
\includegraphics[width=0.5\textwidth]{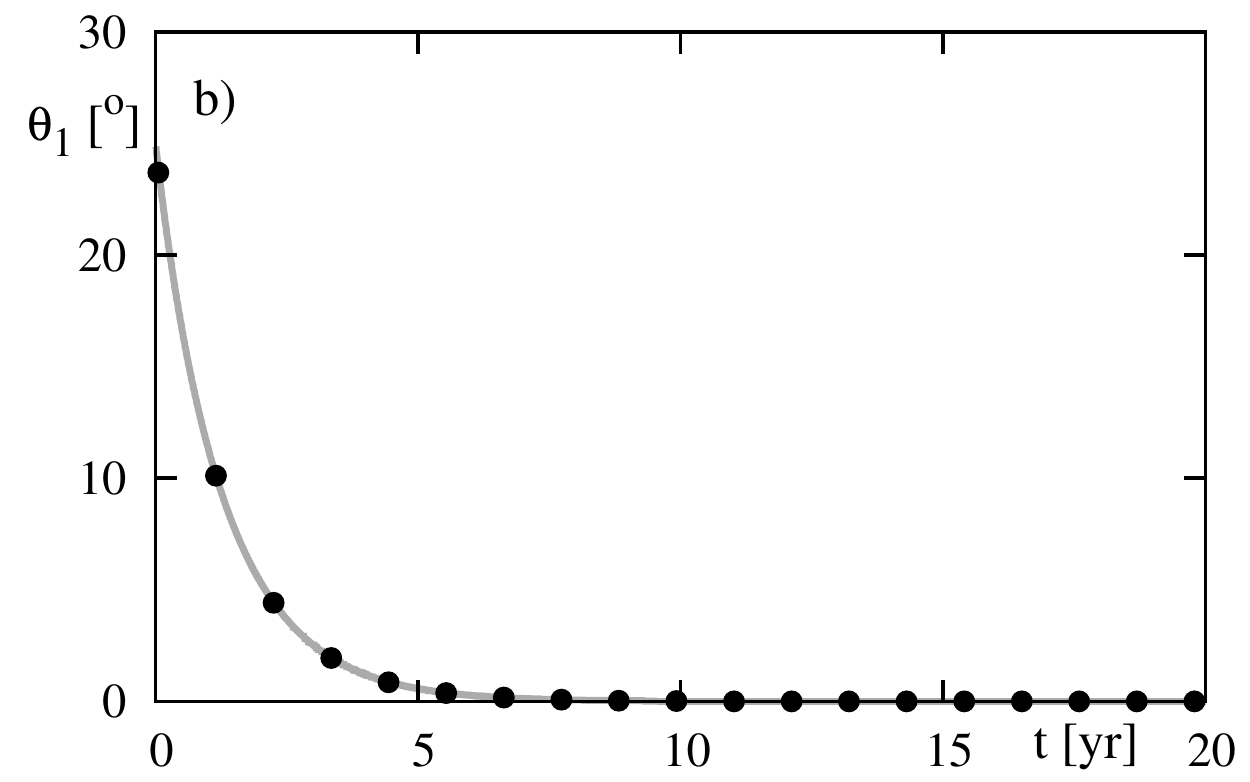}
}
}
\caption{
Evolution of $\Omega_1/n(t)$ (the left panel) and $\theta_1(t)$ (the right
panel) derived by a solution to the full equations of motion (black dots)
and the equations of the first averaged system (gray curves). Parameters of
the system are $m_0 = 1\,\msun$, $m_1 = 1\,\mJ$, $R_0 = 1\,\RS$, $R_1 =
1\,\RJ$, $a = 0.02\,\au$, $e = 0.1$, $T_{\idm{rot,0}} = 5\,\mbox{d}$,
$T_{\idm{rot,1}} = 1\,\mbox{d}$. The dissipative coefficients are $\lambda_0
= 10^{-2}$ and $\lambda_1 = 10^{-2}$. \Corr{Red line is for  \textit{quasi-equilibrium} value of the $\Omega_1/n$ ratio. It is explained further in the text. See subsection~7.4 and Eq.~(89) for details.}
}
\label{fig:evolution_comparison2}
\end{figure}
The experiment  described above concerns the conservative system, with
$\lambda_0 = \lambda_1 = 0$. The next figure~\ref{fig:evolution_comparison2}
shows the results for the non-conservative models, with $\lambda_0 = 10^{-2}$
and $\lambda_1 = 10^{-2}$ (the parameters are unrealistically large, just
to shorten the computations time). The left-hand panel is for the evolution
of the $\Omega_1/n$ ratio. As we will show later on, for an eccentric orbit,
the rotational frequency does not converge exactly to the mean motion $n$,
but rather to some larger value. The left-hand panel illustrates the time
evolution of $\theta_1$ (angle between $\pmb{\Omega}_1$ and $\vec{h}$).
Clearly, it decreases to zero. Both these quantities evolve in relatively
short time-scale; let us recall that  $\lambda_0, \lambda_1$ parameters are
by $3-4$ orders of magnitude too large in this test as compared to likely
values of the real systems. Again, the numerical test shows a perfect
agreement between the results derived from both models.
\begin{figure}
\centerline{
\hbox{
\includegraphics[width=0.5\textwidth]{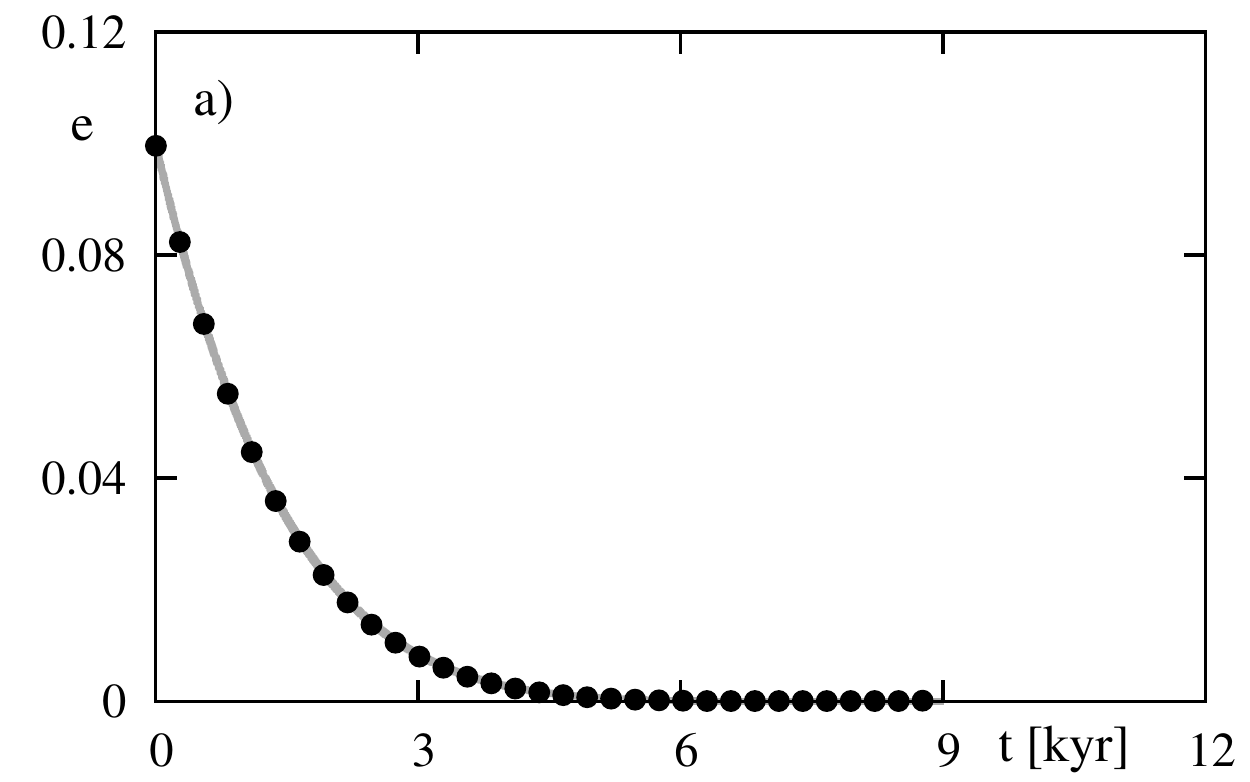}
\includegraphics[width=0.5\textwidth]{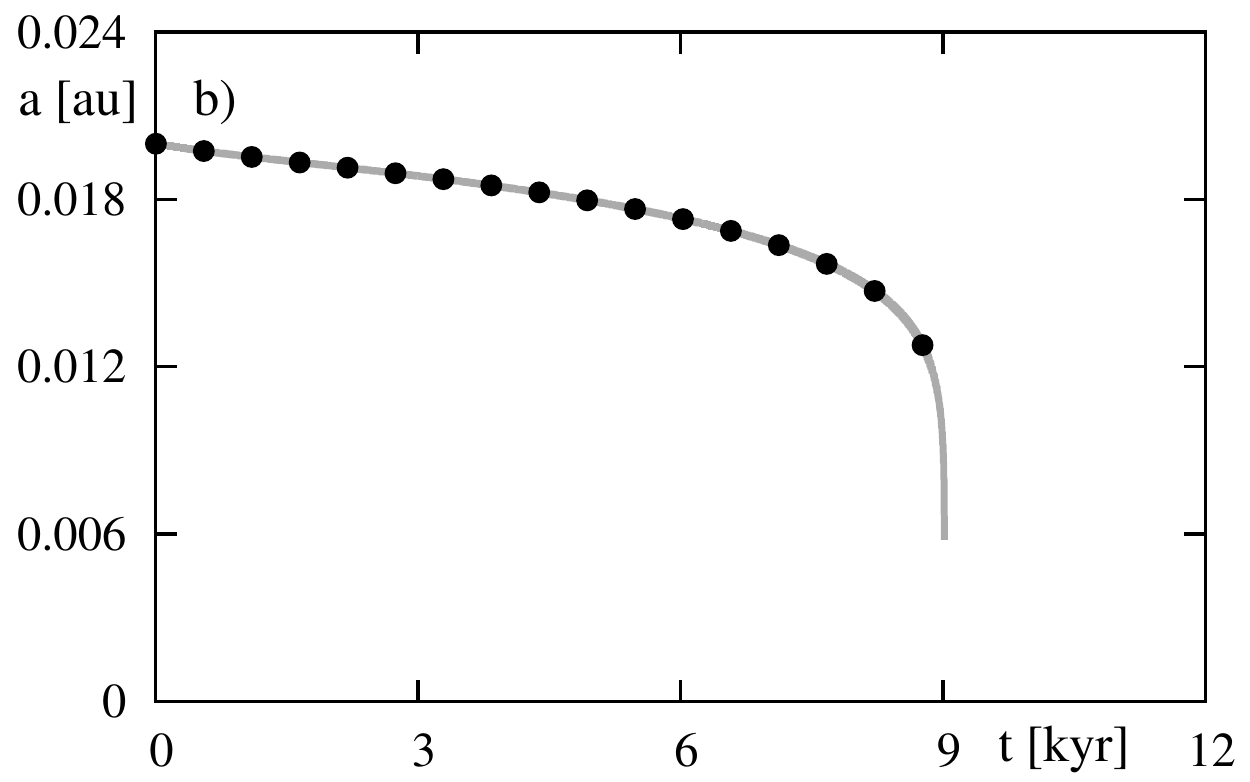}
}
}
\caption{
Evolution of $e(t)$ (the left panel) and $a(t)$ (the right panel) derived by
solving full equations of motion (black dots) and the equations of
first-averaged system (gray curves). Parameters of the system are $m_0 =
1\,\msun$, $m_1 = 1\,\mJ$, $R_0 = 1\,\RS$, $R_1 = 1\,\RJ$, $a = 0.02\,\au$ ,
$e = 0.1$, $T_{\idm{rot,0}} = 5\,\mbox{d}$, $T_{\idm{rot,1}} = 1\,\mbox{d}$.
Dissipative coefficients are $\lambda_0 = 10^{-2}$ and $\lambda_1 = 10^{-2}$.
}
\label{fig:evolution_comparison3}
\end{figure}
Eccentricity $e$ and the semi-major axis $a$ evolve in longer time-scale.
Figure~\ref{fig:evolution_comparison3} shows plots of $e(t)$ (the left-hand
panel) and $a(t)$ (the right-hand panel) for the same parameters as taken in
the previous experiment. Again, the agreement between the exact and the mean
models are excellent.

As we have seen on Fig.~\ref{fig:evolution_comparison1}, the planetary spin
evolves in the time-scale that is only two orders of magnitude longer then
the orbital period. It is still short, as compared to the full time-scale
counted in Gyrs. The integration step would be then still too short to study
the long term dynamics of the system CPU-effectively. In the next section,
we will show that angle $\phi_1$ may be treated as the second fast angle,
and it is possible to perform the second averaging of the secular system.
%
\section{Time-scale of the evolution and the second averaging}
%
Our next goal is to estimate the time-scale of the evolution of the
conservative model. In this section we fix $\sigma_0 = \sigma_1 = 0$. That
implies the following properties of the equations of motion:
\begin{equation}
\vec{h} \cdot \dot{\vec{h}} = 0, \quad \vec{e} \cdot \dot{\vec{e}} = 0,
\quad \pmb{\Omega}_0 \cdot \dot{\pmb{\Omega}}_0 = 0, \quad \pmb{\Omega}_1
\cdot \dot{\pmb{\Omega}}_1 = 0.
\end{equation}
It means that the conservative system possesses at least four first
integrals $e, h, \Omega_0, \Omega_1$. The constant eccentricity and the
magnitude of orbital angular momentum imply also the semi-major axis ($a$)
constant. Moreover, it may be easily shown, that the total angular momentum
$\vec{L} = \beta \vec{h} + \Izero_0 \pmb{\Omega}_0 + \Izero_1 \pmb{\Omega}_1$
is a constant vector. Another constant quantity is $\vec{e} \cdot \vec{h} =
0$. Therefore, the initial set of $12$ scalar equations of motion may be
reduced with the help of the $8$ first integrals, so it should be possible
to transform it to the set of four scalar equations.

Nevertheless, we do not attempt to write them down, but rather to make use of
the properties of the conservative secular system,  in order to simplify
the non-conservative system. At first, let us make a simple estimation of
the time scale of the evolution of vectors $\vec{h}$, $\vec{e}$,
$\pmb{\Omega}_0$, $\pmb{\Omega}_1$ in the conservative system. We choose the
following (say, representative) values of its parameters: $m_0 = 1\,\msun$,
$m_1 = 1\,\mJ$, $a = 0.02\,\au$, $R_0 = 1\,\RS$, $R_1 = 1\,\RJ$, $\Omega_0 =
2\,\pi/(5\,\mbox{d})$, $\Omega_1 = 2\,\pi/(1\,\mbox{d})$, $k_{\idm{L,0}} =
0.03$, $k_{\idm{L,1}} = 0.3$, $\kappa_0 = 0.2$, $\kappa_1 = 0.5$. We also
take into account the dominant term only. We obtain:
\begin{eqnarray}
&&\frac{|\dot{\vec{h}}|}{h} \, P_{\idm{orb}} \sim \pi \, \frac{m_0}{m_1}
\left( \frac{R_1}{a} \right)^5 \left( \frac{\Omega_1}{n} \right)^2 \!\!
\frac{k_{\idm{L,1}}}{\left( 1 - e^2 \right)^2} \sim 8 \times 10^{-6}
\left( 1 - e^2 \right)^{-2},\\
&&\frac{|\dot{\vec{e}}|}{e} \, P_{\idm{orb}} \sim 30 \, \pi \frac{m_0}{m_1}
\left( \frac{R_1}{a} \right)^5 \!\! \frac{k_{\idm{L,1}}}{\left( 1 - e^2 \right)^5}
\sim 2 \times 10^{-4} \left( 1 - e^2 \right)^{-5},\\
&&\frac{|\dot{\pmb{\Omega}}_0|}{\Omega_0} \, P_{\idm{orb}} \sim \frac{5\pi}{2}
\frac{m_1}{m_0} \frac{k_{\idm{L,0}}}{\kappa_0} \frac{\Omega_0}{n} \left( \frac{R_0}{a} \right)^3 \!\! \left( 1 - e^2 \right)^{-\frac{3}{2}} \sim 3 \times 10^{-6} \left( 1 - e^2 \right)^{-\frac{3}{2}},\\
&&\frac{|\dot{\pmb{\Omega}}_1|}{\Omega_1} \, P_{\idm{orb}} \sim \frac{5\pi}{2}
\frac{m_0}{m_1} \frac{k_{\idm{L,1}}}{\kappa_1} \frac{\Omega_1}{n} \left( \frac{R_1}{a} \right)^3 \!\! \left( 1 - e^2 \right)^{-\frac{3}{2}} \sim 6 \times 10^{-2} \left( 1 - e^2 \right)^{-\frac{3}{2}}.
\end{eqnarray}
As we have shown, $\pmb{\Omega}_1$ becomes the fast vector after the first
averaging. Then, we may express this vector in the Keplerian frame, assuming
that $\vec{h}$, $\vec{e}$ are constant vectors. We obtain:
\begin{equation}
\pmb{\Omega}_1 = \Omega_1 \left( \cos\phi_1 \, \sin \theta_1 \,
\frac{\vec{e}}{e} + \sin\phi_1 \, \sin \theta_1 \, \frac{\vec{h}
\times \vec{e}}{h \, e} + \cos \theta_1 \, \frac{\vec{h}}{h}  \right),
\end{equation}
where $\theta_1$ is an angle between $\vec{h}$ and $\pmb{\Omega}_1$, and
$\phi_1$ is azimuthal angle in the orbital plane measured from the direction
of vector $\vec{e}$. Using Eq~(\ref{eq:dot_Om_sec}) for $l=1$, one may find:
\begin{equation}
\dot{\phi_1} = - \frac{A_1}{2 \, \Izero_1} \, \frac{\mu^3 \,
\left(1 - e^2 \right)^{\frac{3}{2}}}{h^6} \Omega_1 \, \cos\theta_1, \quad
\dot{\theta}_1 = 0.
\end{equation}
The characteristic time-scale for $\phi_1$ may be estimated as follows: %
\begin{equation}
\tau_{\phi_1} \equiv \frac{2\,\pi}{\dot{\phi_1}} \approx 90 \, \mbox{d}
\left( \frac{a}{0.02\,\au} \right)^3 \left( \frac{1\,\RJ}{R_1} \right)^3
\frac{m_1}{1\,\mJ} \, \frac{1\,\msun}{m_0} \, \frac{T_{\idm{rot,1}}}{1\,\mbox{d}}
\, \frac{\left( 1 - e^2 \right)^{\frac{3}{2}}}{\cos \theta_1}
\end{equation}
The precession rate is constant in the conservative model. Nevertheless, for
$\theta_1 \approx \pi/2$, $\phi_1$ may not change fast enough to be
considered as the fast angle. That introduces a limitation of the analysis.
To conclude, we stress that in some cases  angle $\phi_1$  may be not slow
enough to make the averaging over the mean anomaly valid (that corresponds
to too large, too close, too fast rotating planet), or it may be not fast
enough to make it possible to perform the second averaging over itself
(Uranus-type orientation of the spin vector). As we will show further on,
the last limitation is weaker than the first one.

After the second averaging, we obtain the following equations of motion:
\begin{eqnarray}
\dot{\vec{h}} &=& -\frac{1}{2 \, \beta} \, A_0 \,
\Ff_{1,0} - \frac{9}{4\,\beta} \left( \sigma_0 \, A_0^2 \, \Ff_{2,0} + \sigma_1
\, A_1^2 \, \Gg_{2,1} \right),\label{eq:dot_h_second}\\
\dot{\vec{e}} &=& -\frac{1}{4\,\beta} \left( A_0 \,
\Ff_{3,0} + A_1 \, \Gg_{3,1} \right) - 15 \left( \frac{A_0}{m_0} +
\frac{A_1}{m_1} \right) \Ff_4 
- \frac{3\,\mu}{c^2} \, \Ff_5 \nonumber \\ 
&&- \frac{9}{4\,\beta} \left( \sigma_0 \,
A_0^2 \, \Ff_{6,0} + \sigma_1 \, A_1^2 \, \Gg_{6,1} \right),\label{eq:dot_e_second}\\
\dot{\pmb{\Omega}}_0 &=& \frac{1}{2 \, \Izero_0}
\, A_0 \, \Ff_{1,0} + \frac{9}{4 \, \Izero_0} \, \sigma_0 \,
A_0^2 \, \Ff_{2,0},\label{eq:dot_Om0_second}
\end{eqnarray}
where
\begin{eqnarray}
&&\Gg_{2,1} = 2 \, \frac{\mu^6 \left( 1 - e^2 \right)^{\frac{3}{2}}}{h^{16}}
\Big\lbrace \mu^2 \, \fE_2 - h^3\,\Omega_1 \cos \theta_1 \, \fE_4 \Big\rbrace \, \vec{h},\\
&&\Gg_{3,1} = \frac{\mu^3 \left( 1 - e^2 \right)^{\frac{3}{2}}}{h^8} \, \Omega_1^2 \left( 5\,\cos^2 \theta_1 - 3 \right) \, \left( \vec{e} \times \vec{h} \right),\\
&&\Gg_{6,1} = \frac{\mu^6 \left( 1 - e^2 \right)^{\frac{3}{2}}}{h^{16}} \Big\lbrace 18\,\mu^2 \, \fE_1 - 11\,h^3\,\Omega_1 \cos \theta_1 \, \fE_5 \Big\rbrace \, \vec{e}.
\end{eqnarray}
\begin{figure}
\centerline{
\hbox{
\includegraphics[width=0.5\textwidth]{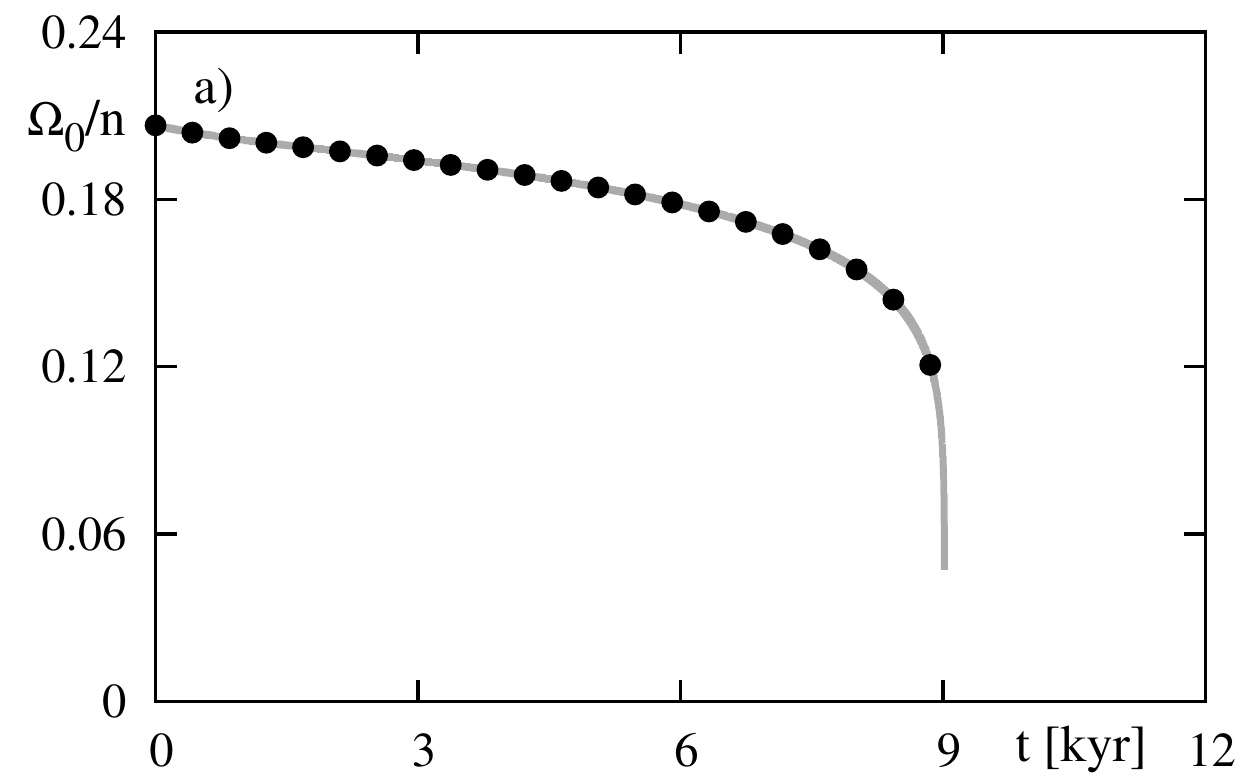}
\includegraphics[width=0.5\textwidth]{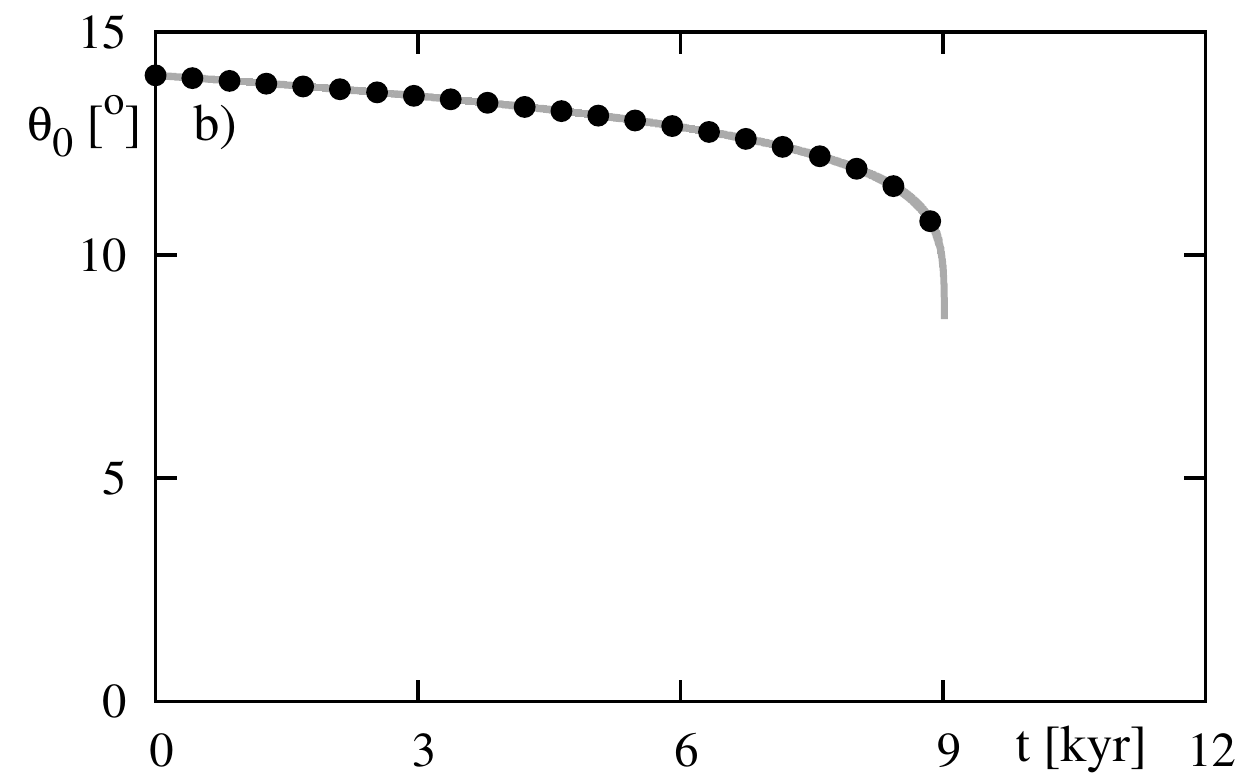}
}
}
\caption{
Evolution of $\Omega_0/n(t)$ (the left-hand panel) and $\theta_0(t)$ (the
right-hand panel) derived from the equations motion of the first-averaged
system (black dots) and the equations of the second-averaged system (gray
curves). Parameters of the configuration are $m_0 = 1\,\msun$, $m_1 = 1\,\mJ$,
$R_0 = 1\,\RS$, $R_1 = 1\,\RJ$, $a = 0.02\,\au$, $e = 0.1$,
$T_{\idm{rot,0}} = 5\,\mbox{d}$, $T_{\idm{rot,1}} = 1\,\mbox{d}$.
Dissipative parameters are $\lambda_0 = 10^{-2}$ and $\lambda_1 = 10^{-2}$.
}
\label{fig:evolution_comparison_second1}
\end{figure}
After the second averaging, the fastest angle is the argument of pericenter.
Thus the time step-size of the integration increases significantly. To
verify the correctness of this step, we perform a new experiments, which
results are illustrated in Fig.~\ref{fig:evolution_comparison_second1}. The
parameters of the system are the same as in the previous tests. At this
time, the plots show the evolution of $\Omega_0/n$ (the left-hand panel) and
$\theta_0$ (the right-hand panel). The agreement between the first averaged
(black dots) and the second-averaged model (grey curve) are basically perfect.

%
\section{The third averaging and the dissipative equations of motion}
%
From equations~(\ref{eq:dot_h_second})-(\ref{eq:dot_Om0_second}) one can
obtain equations governing evolution of $\dot{a}$, $\dot{e}$, $\dot{\Omega}_0$,
$\dot{\theta}_0$ and also $\dot{\Omega}_1$, $\dot{\theta}_1$. They read as
follows:
\begin{eqnarray}
&& \dot{a} = - \frac{9}{\beta} \frac{1}{a^7 \left( 1 - e^2 \right)^{\frac{15}{2}}} \sum_{l=0,1} \sigma_l \, A_l^2 \bigg[ \fE_6 - \fE_2 \left( 1 - e^2 \right)^{\frac{3}{2}} \frac{\Omega_l}{n} \, \cos \theta_l \bigg],\label{eq:dot_a_final}\\
&& \dot{e} = -\frac{9}{4 \, \beta} \frac{e}{a^8 \left( 1 - e^2 \right)^{\frac{13}{2}}} \sum_{l=0,1} \sigma_l \, A_l^2 \bigg[ 18 \, \fE_1 - 11 \, \fE_5 \left( 1 - e^2 \right)^{\frac{3}{2}} \frac{\Omega_l}{n} \, \cos \theta_l \bigg],\label{eq:dot_e_final}
\end{eqnarray}

\begin{eqnarray}
&& \dot{\Omega}_0 = \frac{9}{4 \, \Izero_0} \frac{\sigma_0 \, A_0^2 \, n}{a^6 \left( 1 - e^2 \right)^6} \bigg[ 2\,\fE_2 \cos \theta_0 - \left( 1 - e^2 \right)^{\frac{3}{2}} \frac{\Omega_0}{n} \, \left( 2 \, \fE_4 - \mathcal{P} \sin \theta_0 \right) \bigg],\\
&& \dot{\theta}_0 = - \frac{9}{4 \, \Izero_0} \frac{\sigma_0 \, A_0^2}{a^6 \left( 1 - e^2 \right)^6} \, \frac{n}{\Omega_0} \bigg[ 2\,\fE_2 \sin \theta_0 - \frac{\Omega_0}{n} \left( 1 - e^2 \right)^{\frac{3}{2}} \times \nonumber\\
&& \quad\quad\quad\quad\quad\quad\quad\quad\quad\quad\quad 
\left( \mathcal{P} \cos \theta_0 - \frac{1}{\alpha} \Big\lbrace 2 \, \fE_4 \sin \theta_0 - \mathcal{P} \Big\rbrace \right) \bigg],\\
&& \dot{\Omega}_1 = \frac{9}{4 \, \Izero_1} \frac{\sigma_1 \, A_1^2 \, n}{a^6 \left( 1 - e^2 \right)^6} \bigg[ 2\,\fE_2 \cos \theta_1 - \fE_4 \left( 1 - e^2 \right)^{\frac{3}{2}} \frac{\Omega_1}{n} \left( 1 + \cos^2{\theta_1} \right) \bigg],\label{eq:dot_Om1_final}\\
&& \dot{\theta}_1 = - \frac{9}{4 \, \Izero_1} \frac{\sigma_1 \, A_1^2}{a^6 \left( 1 - e^2 \right)^6} \, \frac{n}{\Omega_1} \sin \theta_1 \bigg[ 2\,\fE_2 - \fE_4 \left( 1 - e^2 \right)^{\frac{3}{2}} \frac{\Omega_1}{n} \cos \theta_1 \bigg].\label{eq:dot_i1_final}
\end{eqnarray}
where
\[
\fE_6 = 1 + \frac{31}{2} \, e^2 + \frac{255}{8} \, e^4 + \frac{185}{16} \, e^6 + \frac{25}{64} \, e^8,
\]
\begin{equation}
\mathcal{P} \equiv \fE_3 \, \frac{\left( \vec{e} \cdot \pmb{\Omega}_0 \right)^2}{e^2 \, \Omega_0^2 \sin \theta_0} + \fE_5 \, \frac{\big[ \left( \vec{e} \times \vec{h} \right) \cdot \pmb{\Omega}_0 \big]^2}{h^2 \, e^2 \, \Omega_0^2 \sin \theta_0}.
\end{equation}
These equations do not depend on the longitude of the ascending node nor
on the azimuthal angle of $\pmb{\Omega}_0$.

A simple estimation of the time-scale of the equations of motion derived in
the previous section shows that, after the second averaging, the argument of
pericenter $\omega$ becomes the fastest angle in the system. It varies
typically in a time-scale of $10^4$ years. To be more specific, and to find
limitations to the choice of $\omega$ as the fast angle, we obtain a direct
form of $\dot{\omega}$ in the second averaged system (the conservative
system). It is a sum of five terms, i.e., $\dot{\omega} = \dot{\omega}_{3,0}
+ \dot{\omega}_{3,1} + \dot{\omega}_{4,0} + \dot{\omega}_{4,1} +
\dot{\omega}_5$, where particular terms of the sum correspond to functions
$\Ff_{3,0}$, $\Ff_{3,1}$, $\Ff_4$ (these are contributions form the star and
the planet), $\Ff_5$, respectively. We obtain:
\begin{eqnarray}
&& \dot{\omega}_{3,0} = \frac{A_0 \, \Omega_0^2}{4 \, \beta \, a^{\frac{7}{2}} \, \mu^{\frac{1}{2}} \left( 1 - e^2\right)^2} \left( 2 \, \alpha \cos \theta_0 + 5 \cos^2{\theta_0} - 1  \right), \\
&& \dot{\omega}_{3,1} = \frac{A_1 \, \Omega_1^2}{4 \beta \, a^{\frac{7}{2}} \, \mu^{\frac{1}{2}} \left( 1 - e^2\right)^2} \left( 5 \cos^2{\theta_1} - 3 \right),\\
&& \dot{\omega}_{4,0} = \frac{15 \, \mu^{\frac{1}{2}} \, \fE_5 }{a^{\frac{13}{2}} \left( 1 - e^2 \right)^5} \, \frac{A_0}{m_0},\\
&& \dot{\omega}_{4,1} = \frac{15 \, \mu^{\frac{1}{2}} \, \fE_5 }{a^{\frac{13}{2}} \left( 1 - e^2 \right)^5} \, \frac{A_1}{m_1},\\
&& \dot{\omega}_5 = \frac{3 \, \mu^{\frac{3}{2}}}{c^2 \, a^{\frac{5}{2}} \, \left( 1 - e^2 \right)}.
\end{eqnarray}
We introduce a new quantity:
\[
\alpha \equiv \frac{\beta \, h}{\Izero_0 \Omega_0}, 
\]
which is the ratio of the magnitude of the orbital angular momentum to the
magnitude of the stellar rotational angular momentum. For typical parameters
considered in this work, $\alpha$ is of the order of unity. Contributions
form the tidal deformation of the objects $\dot{\omega}_{4,0}$,
$\dot{\omega}_{4,1}$, as well as relativistic term $\dot{\omega}_5$ are
always positive. On contrary, the first two terms, stemming form the
rotational deformation of the star and the planet, have some critical values
of $\imut$ at which the sign of the rate frequency changes its sign (from
positive to negative).
%
\subsection{The critical inclinations}
%
We may now introduce the critical inclinations $\Icrit$ and $\thetacrit$.
The first quantity is a solution to the equation
$\dot{\omega}_{3,0}(\theta_0) = 0$, the second  one is a solution to the
equation $\dot{\omega}_{3,1}(\theta_1) = 0$. One can easily obtain the
following expressions:
\begin{equation}
\cos \Icrit = -\frac{1}{5} \left( \alpha \pm \sqrt{\alpha^2 + 5} \right),
\quad \cos \thetacrit = \pm \sqrt{\frac{3}{5}}.
\end{equation}
The critical inclination $\Icrit$ has the following limits:
\begin{equation}
\lim_{\alpha \to 0} \cos \Icrit = \pm \sqrt{\frac{1}{5}},
\quad \lim_{\alpha \to \infty} \cos \Icrit = 0.
\end{equation}

\begin{figure}
\centerline{
\includegraphics[width=0.8\textwidth]{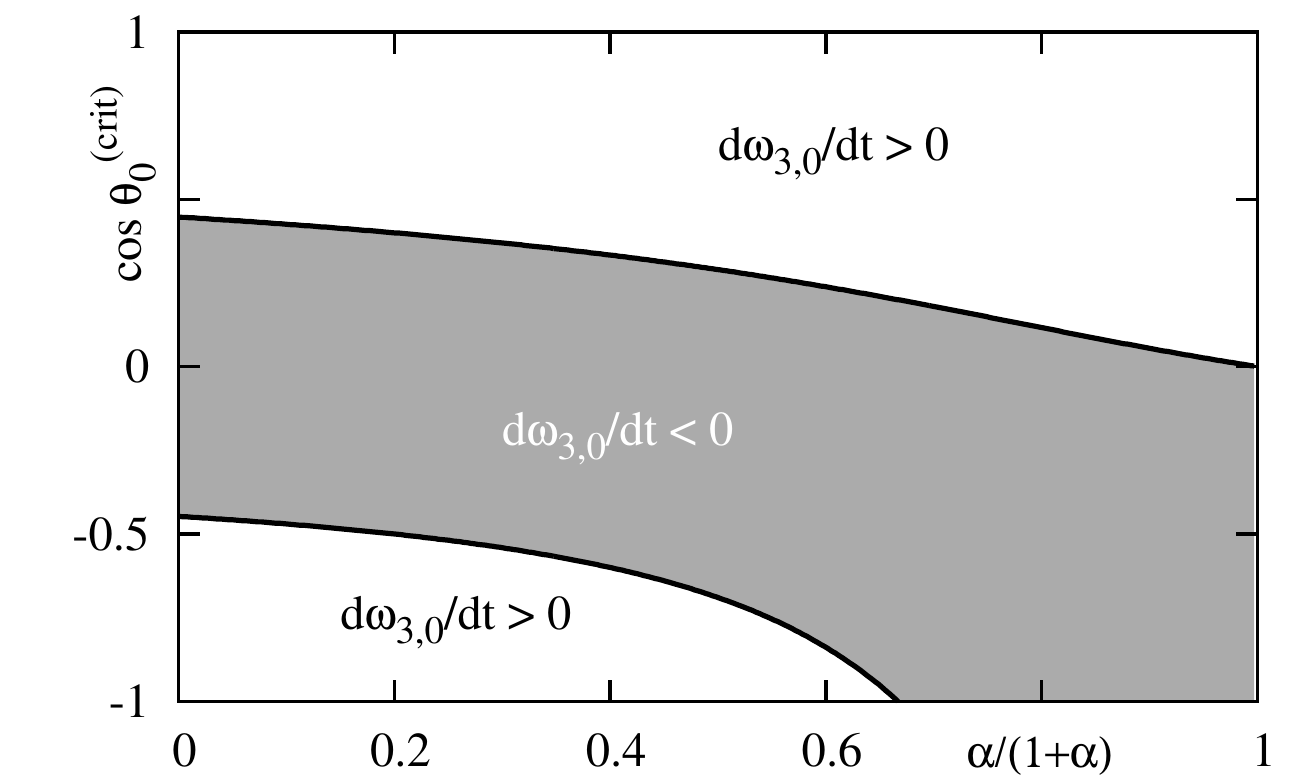}
}
\caption{
Critical inclination $\Icrit$ as a function of $\bar{\alpha} \equiv
\alpha/(1 + \alpha)$. Shaded areas correspond to retrograde rotation of the
pericenter, while the white colored regions are for the prograde rotation of
the pericenter.
}
\label{fig:critical_inclinations}
\end{figure}
Figure~\ref{fig:critical_inclinations} shows graphs of $\cos \Icrit$ as a
function of $\bar{\alpha} \equiv \alpha/(1 + \alpha)$. In this paper we
consider systems with $\alpha \sim 1$ (or $\bar{\alpha} \sim 0.5$), thus
magnitudes of the stellar and orbital angular momenta are of the same order.
Note that for $\theta_0 \sim \Icrit$, a contribution to $\dot{\omega}_{3,0}$
emerging due to the rotational deformation of the star is very small. If it
was the only contribution, $\omega$ could not be considered as the fast
angle. Similarly, if $\theta_1 \sim \thetacrit$, the contribution of the
rotational deformation of the planet is negligible.
%
\subsection{The time-scale of the rotation of pericenter}
%
To justify $\omega$ as the fast angle in the second-averaged
system, we estimate the time-scale $\tau_{3,l} \equiv 2 \, \pi
/ \dot{\omega}_{3,l}$, $\tau_{4,l} \equiv 2 \, \pi / \dot{\omega}_{4,l}$ (
$l=0,1$), $\tau_5 \equiv 2 \, \pi / \dot{\omega}_5$. We obtain:
\begin{eqnarray}
&& \tau_{3,0} \approx 10^5 \, \mbox{yr} \left( \frac{a}{0.02\,\au} \right)^{\frac{7}{2}} \left( \frac{m_0}{1\,\msun} \right)^{\frac{1}{2}} \left( \frac{1\,\RS}{R_0} \right)^5 \left( \frac{T_{\idm{rot,0}}}{5\,\mbox{d}} \right)^2  \frac{6 \left( 1 - e^2 \right)^{\frac{3}{2}}}{2\, \alpha\, C_0 + 5 \,C_0^2 - 1}, \nonumber\\
&& \tau_{3,1} \approx 10^5 \, \mbox{yr} \left( \frac{a}{0.02 \, \au} \right)^{\frac{7}{2}} \frac{m_1}{1\,\mJ} \left( \frac{1\,\msun}{m_0} \right)^{\frac{1}{2}} \left( \frac{1\,\RJ}{R_1} \right)^5 \left( \frac{T_{\idm{rot,1}}}{1\,\mbox{d}} \right)^2 \frac{2 \left( 1 - e^2 \right)^2}{5 \, C_1^2 - 3},\nonumber\\
&& \tau_{4,0} \approx 10^4 \, \mbox{yr} \left( \frac{a}{0.02\,\au} \right)^{\frac{13}{2}} \left( \frac{m_0}{1\,\msun} \right)^{\frac{1}{2}} \frac{1\,\mJ}{m_1} \left( \frac{1\,\RS}{R_0} \right)^5 \frac{\left( 1 - e^2 \right)^5}{\fE_5},\nonumber\\
&& \tau_{4,1} \approx 80 \, \mbox{yr} \left( \frac{a}{0.02\,\au} \right)^{\frac{13}{2}} \left( \frac{1\,\msun}{m_0} \right)^{\frac{3}{2}} \frac{m_1}{1\,\mJ} \left( \frac{1\,\RJ}{R_1} \right)^5 \frac{\left( 1 - e^2 \right)^5}{\fE_5},\nonumber\\
&& \tau_5 \approx 2 \times 10^3 \, \mbox{yr} \left( \frac{a}{0.02\,\au} \right)^{\frac{5}{2}} \left( \frac{1\,\msun}{m_0} \right)^{\frac{3}{2}} \left( 1 - e^2 \right),
\end{eqnarray}
where $C_l \equiv \cos \theta_l$ and $T_{\idm{rot,0}}, T_{\idm{rot,1}}$
denote the rotation periods of the star and the planet, respectively. The
typical (characteristic) time-scales were calculated for $k_{\idm{L},0} = 0.03$ and
$k_{\idm{L},1} = 0.3$.

The effect of the tidal deformation of the planet dominates over remaining
perturbations  in the range of the typical parameters. For a wider orbit,
$\tau_{4,1}$ increases faster then $\tau_{3,0}$ or $\tau_{3,1}$.
Nevertheless, for a Sun-like star and a Jupiter-like planet, the effect of
the rotational deformation of the bodies may be never the most important
contribution to $\dot{\omega}$. For a wider orbit, the dominating
perturbation is the relativistic term. We have shown, that in the
interesting range of physical and orbital parameters of the system, the
argument of pericenter always increases (i.e., $\dot{\omega}>0$), and may be
treated as the fast angle in the third averaging.

\begin{figure}
\centerline{
\includegraphics[width=0.8\textwidth]{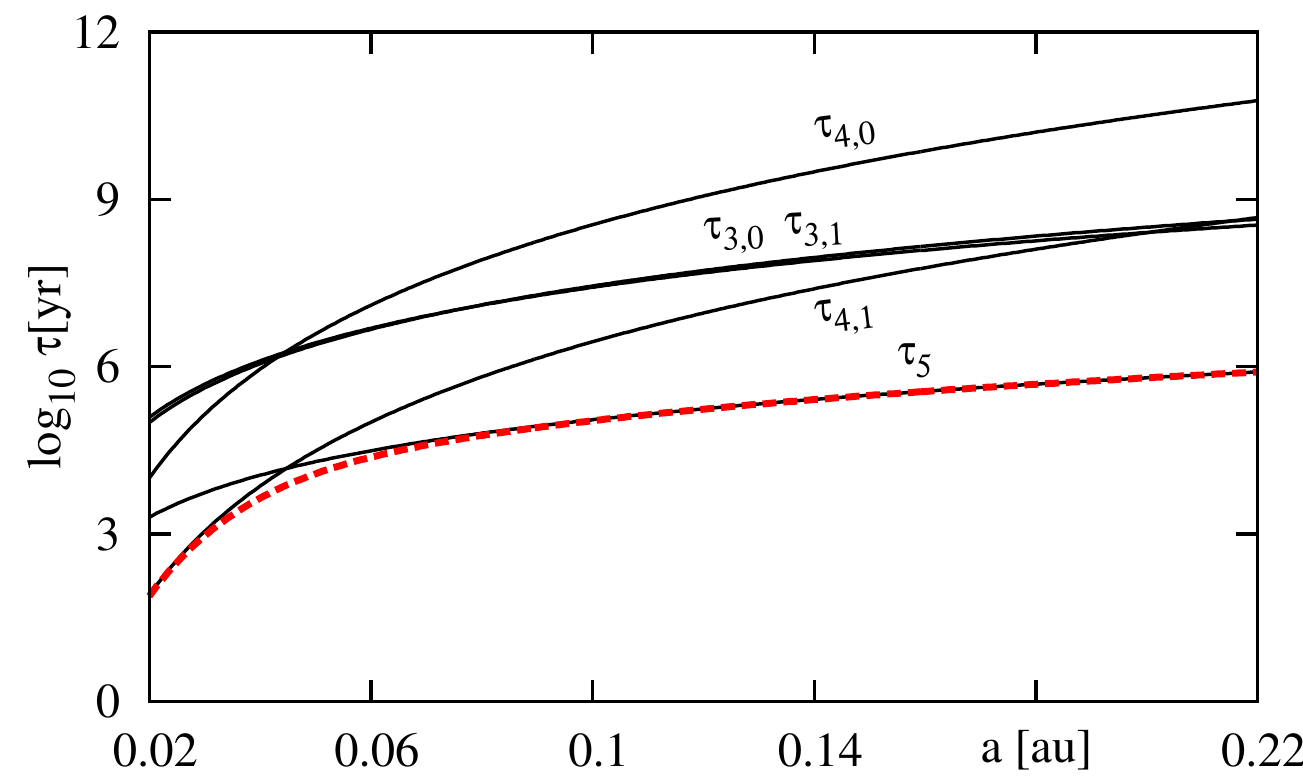}
}
\caption{A time-scale of the rotation of periapses.}
\label{fig:time_scales_conservative}
\end{figure}
Figure~\ref{fig:time_scales_conservative} illustrates graphs of
characteristic time-scale of the evolution of the conservative system. Thin,
black curves are for the time-scale $\tau_{3,l}$, $\tau_{4,l}$ and $\tau_5$
discussed above. The dashed, red curve is for the join effect of two
dominating perturbations, i.e., the general relativity and the tidal
deformation of the planet. The time-scale due to the rotational deformation
of the object are always longer than the former one, thus $\dot{\omega}$ is
always positive. This figure ensures us, that the third averaging (over
$\omega$), is valid in the considered range of parameters. Nevertheless, for
fast rotating (periods of a one day) and large stars (with $R_0 \sim 2\,\RS$), 
the time-scale $\tau_{3,0}$ may be shorter for some orbits (e.g., $a =
0.1\,\au$) than $\tau_{4,1}$ and $\tau_5$.
%
\subsection{The time-scale of the evolution of the system}
%
Because of the small magnitude of the moment of inertia of the planet, it
has to be verified whether the dissipative time-scale for $\dot{\Omega}_1$
and $\dot{\theta}_1$ may be of the same order, as the time-scale of the
rotation of periastron. For the completeness of the analysis, we estimate
the time-scales of all variables, i.e., $a, e, \Omega_0, \theta_0, \Omega_1,
\theta_1$. Because the right-hand sides of the equations of motion have
rather complex form, we limit this estimation to the case of small $e$,
$\theta$ and $\Omega_l/n \ll 1$. For the typical parameters of the system,
one finds the following characteristic time-scales:
\begin{eqnarray}
&&\tau_{a,*} \approx 6.5 \times 10^7 \mbox{yr} \, \frac{10^{-5}}{\lambda_0} \left( \frac{a}{0.02\,\au} \right)^8 \left( \frac{m_0}{1\,\msun} \right)^{\frac{1}{2}} \frac{1\,\mJ}{m_1} \left( \frac{1\,\RS}{R_0} \right)^{\frac{13}{2}}, \nonumber\\
&&\tau_{a,p} \approx 5.5 \times 10^7 \mbox{yr} \, \frac{10^{-6}}{\lambda_1} \left( \frac{a}{0.02\,\au} \right)^8 \left( \frac{1\,\msun}{m_0} \right)^2 \left( \frac{m_1}{1\,\mJ} \right)^{\frac{3}{2}} \left( \frac{1\,\RJ}{R_1} \right)^{\frac{13}{2}}, \nonumber\\
&&\tau_{e,*} \approx 1.5 \times 10^7 \mbox{yr} \, \frac{10^{-5}}{\lambda_0} \left( \frac{a}{0.02\,\au} \right)^8 \left( \frac{m_0}{1\,\msun} \right)^{\frac{1}{2}} \frac{1\,\mJ}{m_1} \left( \frac{1\,\RS}{R_0} \right)^{\frac{13}{2}}, \nonumber\\
&&\tau_{e,p} \approx 1.2 \times 10^7 \mbox{yr} \, \frac{10^{-6}}{\lambda_1} \left( \frac{a}{0.02\,\au} \right)^8 \left( \frac{1\,\msun}{m_0} \right)^2 \left( \frac{m_1}{1\,\mJ} \right)^{\frac{3}{2}} \left( \frac{1\,\RJ}{R_1} \right)^{\frac{13}{2}}, \nonumber
\end{eqnarray}
\begin{eqnarray}
&&\tau_{\Omega_0} \approx \tau_{\theta_0} \approx 1.2 \times 10^8 \mbox{yr} \, \frac{10^{-5}}{\lambda_0} \left( \frac{a}{0.02\,\au} \right)^{\frac{15}{2}} \frac{m_0}{1\,\msun} \left( \frac{1\,\mJ}{m_1} \right)^2 \left( \frac{1\,\RS}{R_0} \right)^{\frac{9}{2}} \frac{5\,\mbox{d}}{T_{\idm{rot},0}}, \nonumber\\
&&\tau_{\Omega_1} \approx \tau_{\theta_1} \approx 1.3 \times 10^4 \mbox{yr} \, \frac{10^{-6}}{\lambda_1} \left( \frac{a}{0.02\,\au} \right)^{\frac{15}{2}} \left( \frac{1\,\msun}{m_0} \right)^{\frac{5}{2}} \left( \frac{m_1}{1\,\mJ} \right)^{\frac{3}{2}} \left( \frac{1\,\RJ}{R_1} \right)^{\frac{9}{2}} \frac{1\,\mbox{d}}{T_{\idm{rot},1}}, \nonumber
\end{eqnarray}
where for any quantity denoted as $X$, the time-scale is defined as $\tau_X
\equiv X/|\dot{X}|$. The equations of motion for $a$ and $e$ are sums of
terms representing contributions due to the energy dissipation in the star
and in the planet, i.e, $\dot{a} = \dot{a}^{(*)} + \dot{a}^{(\idm{p})}$,
$\dot{e} = \dot{e}^{(*)} + \dot{e}^{(\idm{p})}$. We calculated these
time-scales for contributions coming from each body separately.
\begin{figure}
\centerline{
\includegraphics[width=0.8\textwidth]{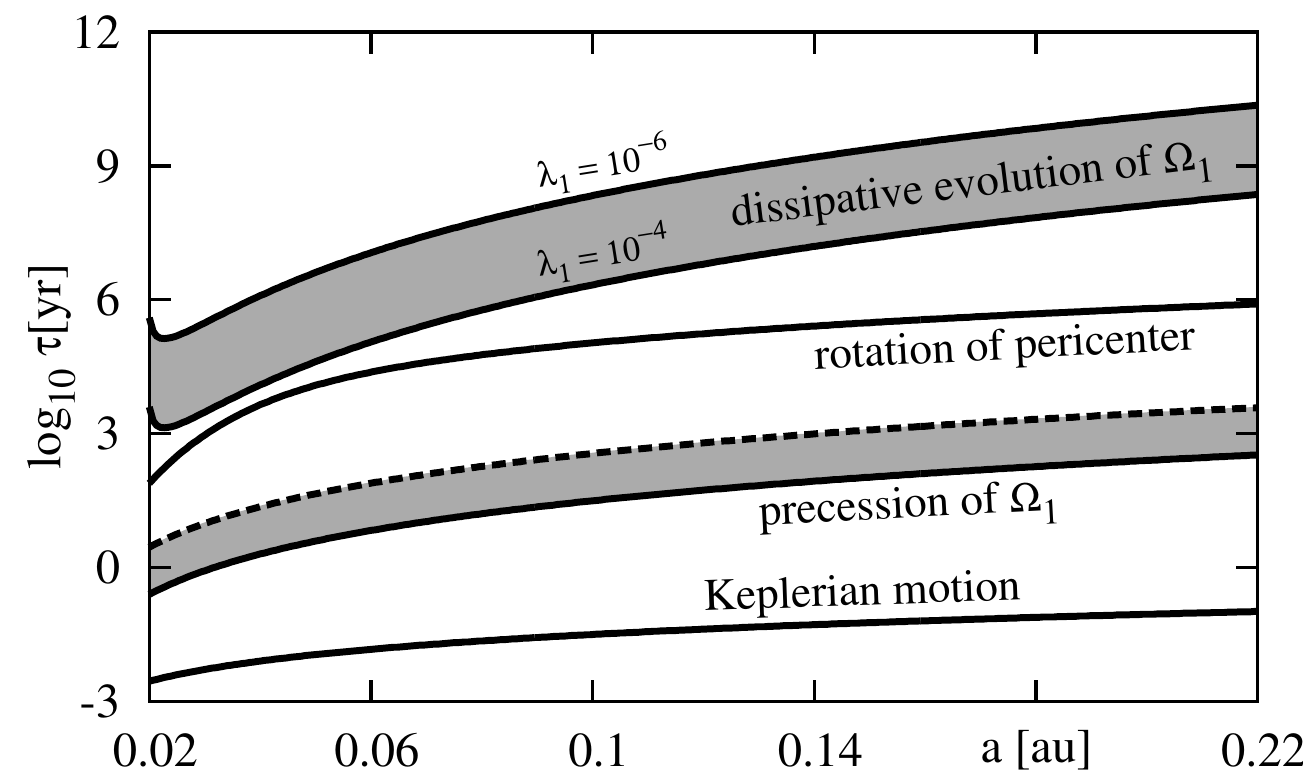}
}
\caption{
Time-scales of the Keplerian motion, precession of the planetary spin,
rotation of the pericenter, and the dissipative evolution of the planetary
spin. Parameters have typical values, $T_{\idm{rot},1} = 1\,\mbox{d}$.
}
\label{fig:time_scales}
\end{figure}
The dissipative evolution of the planetary spin occurs relatively fast.
However, one should keep in mind that these estimates were done under the
assumption of $\Omega_1/n \ll 1$. This means that we neglect terms with
$\Omega_1/n$ in the equations of motion. This is not correct in general, and
that simplification has been used only to derive the time-scales. Moreover,
the magnitude of $\lambda_1$ is not well known. A concise discussion of the
time-scales present in the system may be summarized graphically in Figure~
\ref{fig:time_scales}. It illustrates a separation of particular time-scales
for Keplerian motion, precession of the planetary spin (or frequency
associated with the angle $\phi_1$), rotation of the pericenter (a
frequency of variations of $\omega$), and finally, the time-scale of the
dissipative evolution of planetary spin. According with formulae derived for
$\tau_{\Omega_1}$, we included also the dependency of the mean motion, see
Eq.~(\ref{eq:dot_Om1_final}) with $e=0, \theta=0$. For extreme systems, with
a one-day orbital period, the mean motion may be comparable with the period
of the rotation of the pericenter, and $\Omega_1, \theta_1$ may be
considered as the fast quantities, similarly to $\omega$.

Nevertheless, the equations for $\dot{\Omega}_1$ and $\dot{\theta}_1$ do not
depend on $\omega$, and the equations for $\dot{\Omega}_0$ and
$\dot{\theta}_0$, which depend on $\omega$, do not depend of $\Omega_1,
\theta_1$. Therefore, we can average out the equations for the evolution of
the stellar spin, without considering the evolution of the planetary spin,
even if it varies in comparable time-scale.

\subsection{The third averaging and the quasi-synchronization of the planet's spin}
%
As we have shown, the time derivatives of $\Omega_0$ and $\theta_0$ depend on the
fast vector $\vec{e}$, i.e., they depend on the fast angle $\omega$. One can
easily find that $\langle \mathcal{P} \rangle = \fE_4 \sin \theta_0$, thus,
after the third averaging, the equations for $\dot{\Omega}_0$ and
$\dot{\theta}_0$ read as follows:
\begin{eqnarray}
&& \dot{\Omega}_0 = \frac{9}{4 \, \Izero_0} \frac{\sigma_0 \, A_0^2 \, n}{a^6 \left( 1 - e^2 \right)^6} \bigg[ 2\,\fE_2 \cos \theta_0 - \fE_4 \left( 1 - e^2 \right)^{\frac{3}{2}} \frac{\Omega_0}{n} \left( 1 + \cos^2{\theta_0} \right) \bigg],\label{eq:dot_Om0_final}\\
&& \dot{\theta}_0 = - \frac{9}{4 \, \Izero_0} \frac{\sigma_0 \, A_0^2}{a^6 \left( 1 - e^2 \right)^6} \, \frac{n}{\Omega_0} \sin \theta_0 \bigg[ 2\,\fE_2 - \fE_4 \left( 1 - e^2 \right)^{\frac{3}{2}} \frac{\Omega_0}{n} \left( \cos \theta_0 - \frac{1}{\alpha} \right) \bigg].\label{eq:dot_i0_final}
\end{eqnarray}
The spin of the planet evolves much faster than the other quantities, i.e.,
$a, e, \Omega_0, \theta_0$. Assuming that these parameters are constant, one may
find that $\Omega_1$ and $\theta_1$ tend to an equilibrium. There exists
only one such an equilibrium, which is linearly stable, i.e.,
\begin{equation}
\Omega_1\Big|_{\idm{eq}} = n \, \frac{\fE_2}{\fE_4 \left( 1 - e^2\right)^{\frac{3}{2}}} \geq n, \quad \theta_1\Big|_{\idm{eq}} = 0.
\end{equation}
Thus, the rotational velocity of the planet tends to $n$ only for circular
orbits, while for $e>0$, the equilibrium value is larger than the mean
motion. That is why we call this state as {\em quasi-synchronous}, rather
than the synchronous one. For $e>0$, the energy is still dissipated in the
planetary interior, even if the spin has reached the equilibrium orientation.

Time-scales $\tau_{a,p}$, $\tau_{e,p}$ are relatively short, $\sim 10^6$
years for a one-day orbit. Yet they were obtained under the assumption that
the planetary rotational velocity remains separated from the
quasi-synchronous state, which is true only for $t < \tau_{\Omega_1}$. When
the planet is located at this state, its contribution to $\dot{a}$ and
$\dot{e}$ may be smaller. Putting the equilibrium values for $\Omega_1$ and
$\theta_1$ to Eq.~(\ref{eq:dot_a_final}) and (\ref{eq:dot_e_final}), we find
the following expressions:
\begin{eqnarray}
&& \dot{a} = - \frac{9}{2 \, \beta} \frac{1}{a^7 \left( 1 - e^2 \right)^{\frac{15}{2}}} \bigg\lbrace 2\,\sigma_0 \, A_0^2 \bigg[ \fE_6 - \fE_2 \left( 1 - e^2 \right)^{\frac{3}{2}} \frac{\Omega_0}{n} \, \cos \theta_0 \bigg] \nonumber \\
&& \quad\quad\quad\quad\quad\quad\quad\quad\quad\quad\quad\quad
+ 7\,\sigma_1 \, A_1^2 \, e^2 \, \frac{\fE_7}{\fE_4} \bigg\rbrace ,\label{eq:dot_a_final2}\\
&& \dot{e} = -\frac{9}{4 \, \beta} \frac{e}{a^8 \left( 1 - e^2 \right)^{\frac{13}{2}}} \bigg\lbrace \sigma_0 \, A_0^2 \bigg[ 18 \, \fE_1 - 11 \, \fE_5 \left( 1 - e^2 \right)^{\frac{3}{2}} \frac{\Omega_0}{n} \, \cos \theta_0 \bigg] \nonumber \\
&& \quad\quad\quad\quad\quad\quad\quad\quad\quad\quad\quad\quad
+ 7\,\sigma_1 \, A_1^2 \, \frac{\fE_7}{\fE_4} \bigg\rbrace, \label{eq:dot_e_final2}
\end{eqnarray}
where
\[\fE_7 = 1 + \frac{45}{14} \, e^2  + 8 \, e^4  + \frac{685}{224} \, e^6  + \frac{255}{448} \, e^8  + \frac{25}{1792} \, e^{10}.
\]
However, the tides in the planet cause a fast orbital decay; it acts in
relatively short time-scale of the order of $\tau_{e,p}$. After
circularization of the orbit, the tides in the planet do not contribute to
$\dot{a}$. We notice here, that in multi-planet systems the eccentricity as
well as inclination of the orbit vary in the conservative time-scale and the
spin of the planet may unlikely reach the equilibrium. Therefore, only in
single-planet systems, it is possible to use the simplified equations of
motion (i.e., the equations with fixed $\theta_1 = 0$ and $\Omega_1 =
\Omega_1|_{\idm{eq}}$).

\section{The dissipative evolution of the system}
%
The final set of the equations of motion, i.e., Eq.~(\ref
{eq:dot_Om0_final}), (\ref{eq:dot_i0_final}), (\ref{eq:dot_a_final2}), (\ref
{eq:dot_e_final2}) has an integral of the magnitude of the total angular
momentum, $L \equiv |\vec{L}|$, where
\begin{equation}
\vec{L} = \beta \, \vec{h} + \Izero_0 \, \pmb{\Omega}_0.
\end{equation}
To describe  the evolution of the system, one has to solve three equations
of motion, i.e., $\dot{a}, \dot{e}$ and $\dot{\Omega}_0$, for a fixed value
of $L$. Still, we cannot write down the solution to these equations, and we
have to integrate them numerically. Moreover, we may study some particular
solutions, like the equilibria, analytically.
%
\subsection{The stability of the equilibrium}
%
The final approximation of the evolutionary equations of the  system
possesses one type of equilibrium. It corresponds to a state, in which the
mechanical energy is not lost, i.e., $e=0, \Omega_0 = n_* =
\sqrt{\mu/a_*^3}, \theta_0 = 0$, where $a_*$ (or $n_*$) may be treated as a
parameter of {\em a family of such equilibria}. For some $a_*$, the
equilibrium may be stable, while for some other value -- unstable.  To study
the stability, we write down the linear variational equations in the
vicinity of the stationary solution as follows:
\begin{eqnarray}
\dot{\delta}_a &=& \frac{27 \, \sigma_0 \, A_0^2}{2 \, \beta \, a_*^8} \, \delta_a + \frac{9 \, \sigma_0 \, A_0^2}{\beta \, a_*^7 \, n_*} \, \delta_{\Omega_0}, \nonumber\\
\dot{\delta}_e &=& - \frac{63}{4 \, \beta \, a_*^8}  \left( \sigma_0 \, A_0^2 + \sigma_1 \, A_1^2 \right) \delta_e,\nonumber\\
\dot{\delta}_{\Omega_0} &=& - \frac{27 \, \sigma_0 \, A_0^2 \, n_*}{4 \, \Izero_0 \, a_*^7} \, \delta_a - \frac{9 \, \sigma_0 \, A_0^2}{2 \, \Izero_0 \, a_*^6 \, n_*} \, \delta_{\Omega_0}, \nonumber\\
\dot{\delta}_{\theta_0} &=& -\frac{9 \, \sigma_0 \, A_0^2}{4 \, \beta \, a_*^8} \left( \frac{\beta \, a_*^2}{\Izero_0} + 1 \right) \delta_{\theta_0},
\end{eqnarray}
where $\delta_a, \delta_e, \delta_{\Omega_0}$ and $\delta_{\theta_0}$ are
the variations of $a, e, \Omega_0, \theta_0$, respectively. Equations
for $\delta_e$ and $\delta_{\theta_0}$ are separable, and they admit simple
solutions, i.e., both these quantities decrease exponentially to $0$:
\begin{equation}
\delta_e(t) = \delta_e(0) \exp(\lambda_e \, t), \quad
\delta_{\theta_0}(t) = \delta_{\theta_0}(0) \exp(\lambda_{\theta_0} \, t), \quad \lambda_e < 0, \quad \lambda_{\theta_0} < 0.
\end{equation}
The equations for $\delta_a$ and $\delta_{\Omega_0}$ are conjugate  each
to other. To solve them, we define a new quantity $x \equiv \Omega_0 / n$ and
write down the relevant variational equation:
\begin{equation}
\dot{\delta}_x = - \frac{9 \, \sigma_0 \, A_0^2}{\beta \, a_*^8} \left( \frac{\beta \, a_*^2}{\Izero_0} - 3 \right) \delta_x.
\end{equation}
This equation also admits a simple solution: $\delta_x(t) = \delta_x(0)
\exp(\lambda_x t)$, where $\lambda_x$ may be either positive or negative.
The stability condition reads as follows:
\begin{equation}
\lambda_x < 0 \quad \Leftrightarrow \quad
\frac{\beta \, a^2}{\Izero_0} > 3.
\end{equation}
For typical values of the physical parameters, the equilibrium is stable if
\begin{equation}
a > 0.074\,\au \left( \frac{m_0}{1\,\msun} \right)^{\frac{1}{2}}
\left( \frac{1\,\mJ}{m_1} \right)^{\frac{1}{2}} \left( \frac{\kappa_0}{0.2}
\right)^{\frac{1}{2}} \frac{R_0}{1\,\RS}.
\end{equation}
The orbital period corresponds to $a=0.074\,\au$ is $\sim 7.4\,\mbox{d}$.
Thus, for  a Jupiter-like planet orbiting a Sun-like star, with orbital
period shorter then $\sim 7.4\,\mbox{d}$, the planet does not tend to the
equilibrium. When the equilibrium is unstable, the initial condition
$\Omega_0 < n$ forces the planet to fall down onto its host star.

Considering the time-scale of the tidal evolution of orbits with semi-major
axes larger than this critical value, one may conclude that the equilibrium
may be unlikely attained by a Jovian planet. Moreover, in a more realistic
model, this equilibrium does not exist at all \citep{Barker2009}, because
additional effects, like the angular momentum dissipation due to stellar
winds, may decrease the rotational velocity of the star.

\begin{figure}
\centerline{
\hbox{
\includegraphics[width=0.5\textwidth]{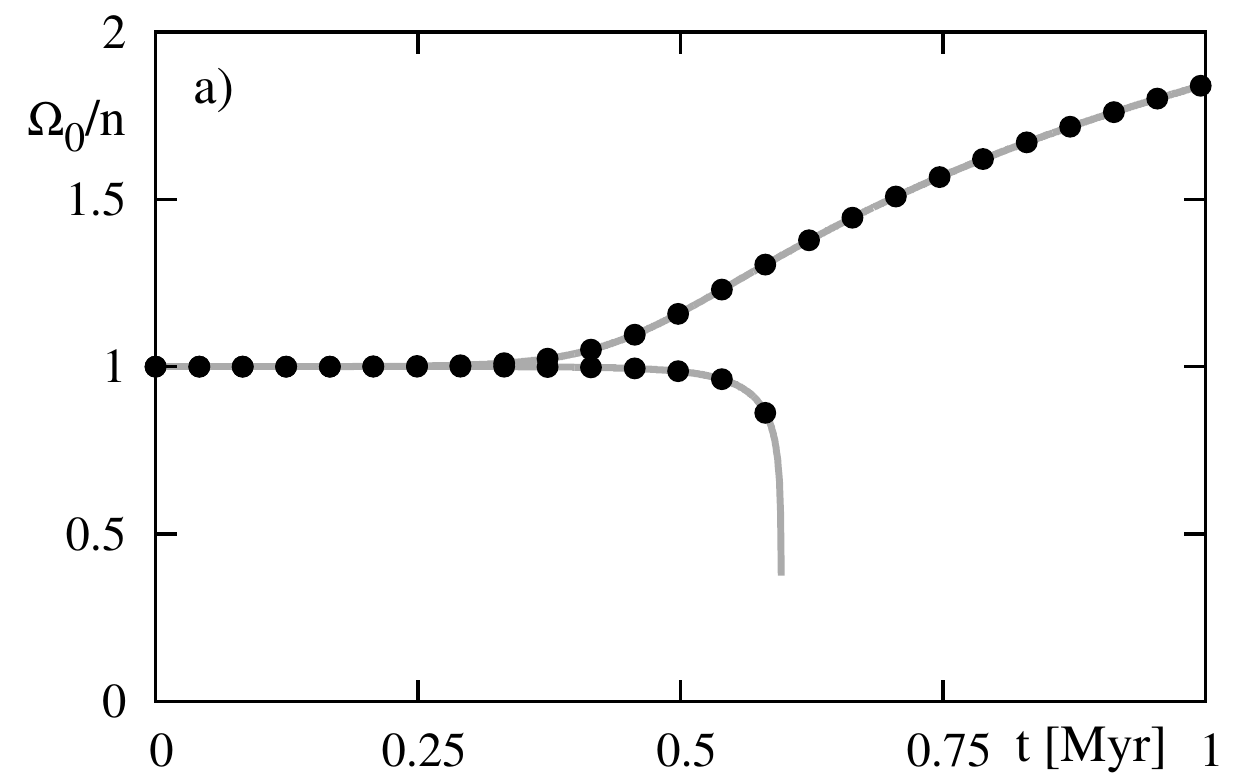}
\includegraphics[width=0.5\textwidth]{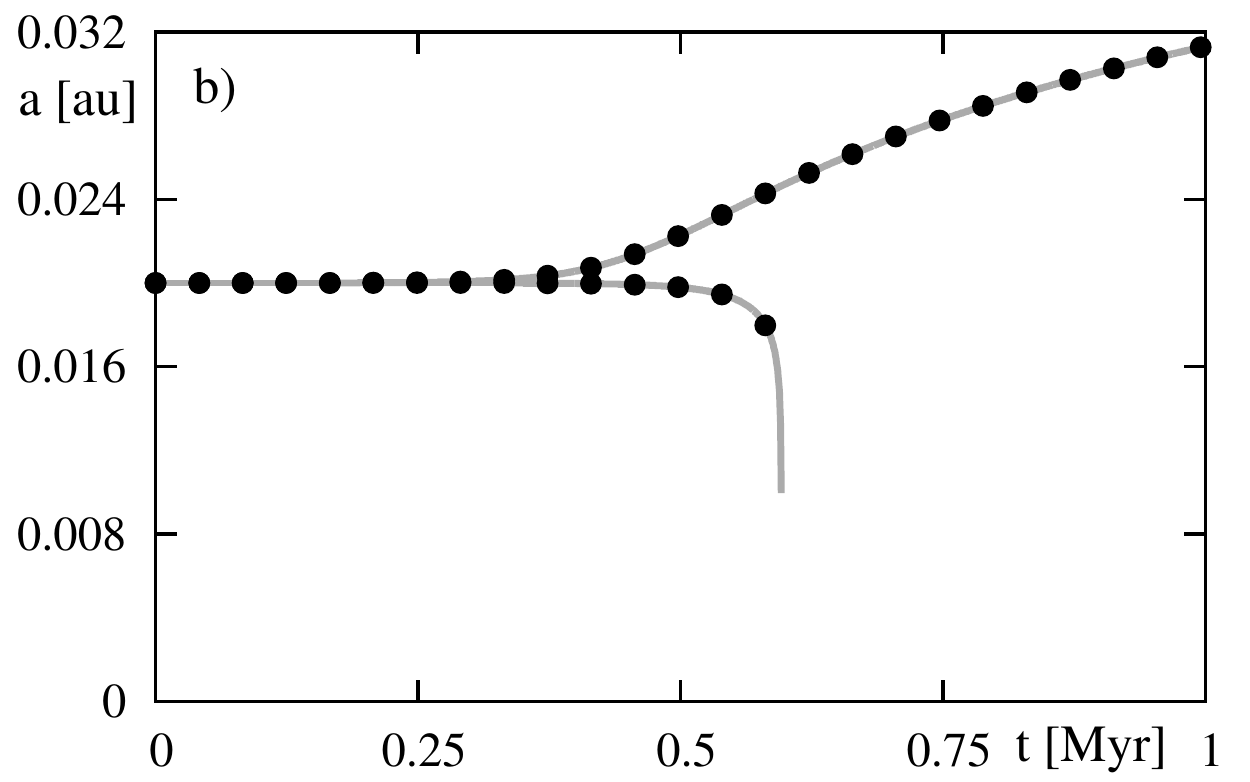}
}
}
\caption{Evolution of $\Omega_0/n(t)$ (the left-hand panel) and $a(t)$ (the
right-hand panel) derived through the solution of the equations of motion of
the second averaged system (black dots) and the equations of third-averaged
system (gray curves). Parameters of the system are $m_0 = 1\,\msun$,
$m_1 = 1\,\mJ$, $R_0 = 1\,\RS$, $R_1 = 1\,\RJ$, $a = 0.02\,\au$.
Dissipative coefficients are $\lambda_0 = 10^{-2}$ and $\lambda_1 = 10^{-2}$.
The initial system is close to the equilibrium state.
}
\label{fig:evolution_comparison_third1}
\end{figure}
Figure~\ref{fig:evolution_comparison_third1} illustrates the evolution of
the system initially located in the vicinity of the equilibrium. Black dots
are for the solutions to the equations of motion of the second averaged
system, the grey curve is for the third-averaged system with the spin of
the planet in the synchronous state. This test  shows that the model is
correct. It also illustrates the behavior of the system near the unstable
equilibrium. The system with initially $\Omega_0 \gtrsim n$ excites both the
semi-major axis and the $\Omega_0/n$ ratio. Still, the rates of variations
of $a$ and $\Omega_0/n$ do not suppress dissipation of the total energy. On
the other hand, for initial $\Omega_0 \lesssim n$, the planetary destiny is
to fall down onto the parent star.
%
\subsection{Parametric study of the dissipative evolution}
\begin{figure}
\centerline{
\vbox{
\hbox{
\includegraphics[width=0.5\textwidth]{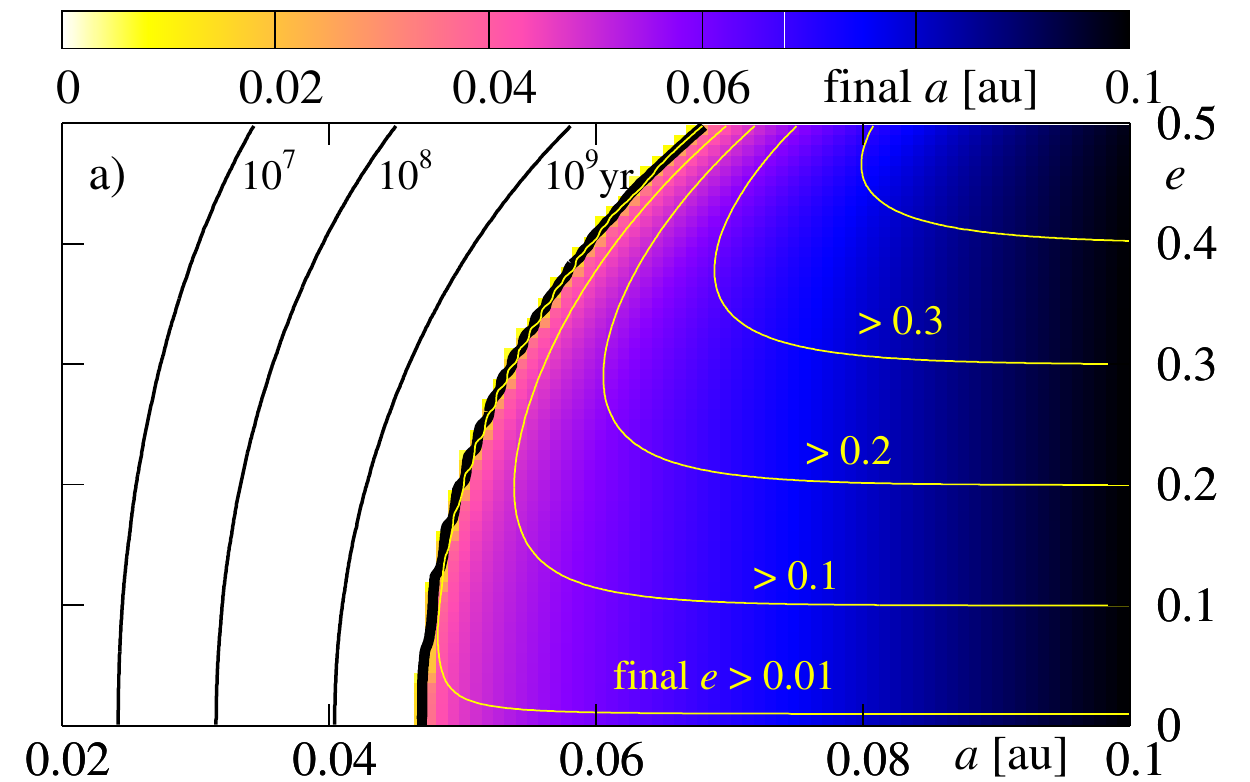}
\includegraphics[width=0.5\textwidth]{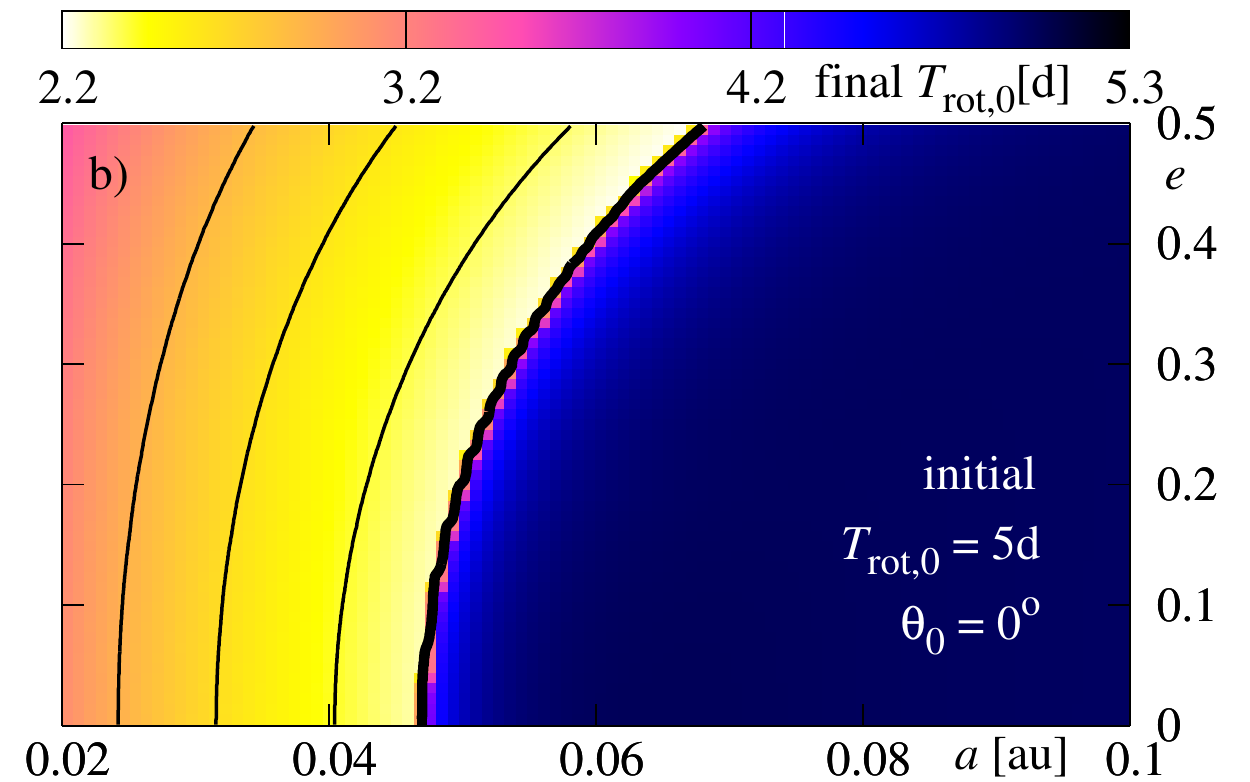}
}
\hbox{
\includegraphics[width=0.5\textwidth]{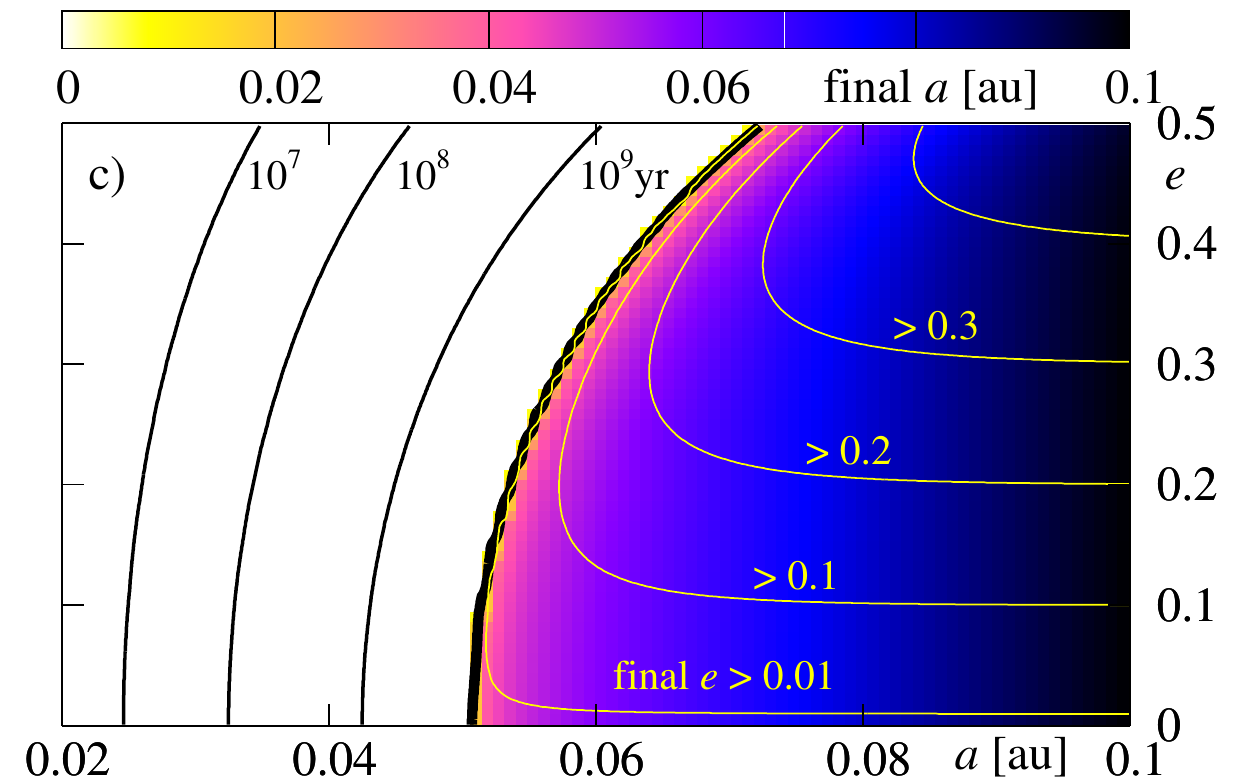}
\includegraphics[width=0.5\textwidth]{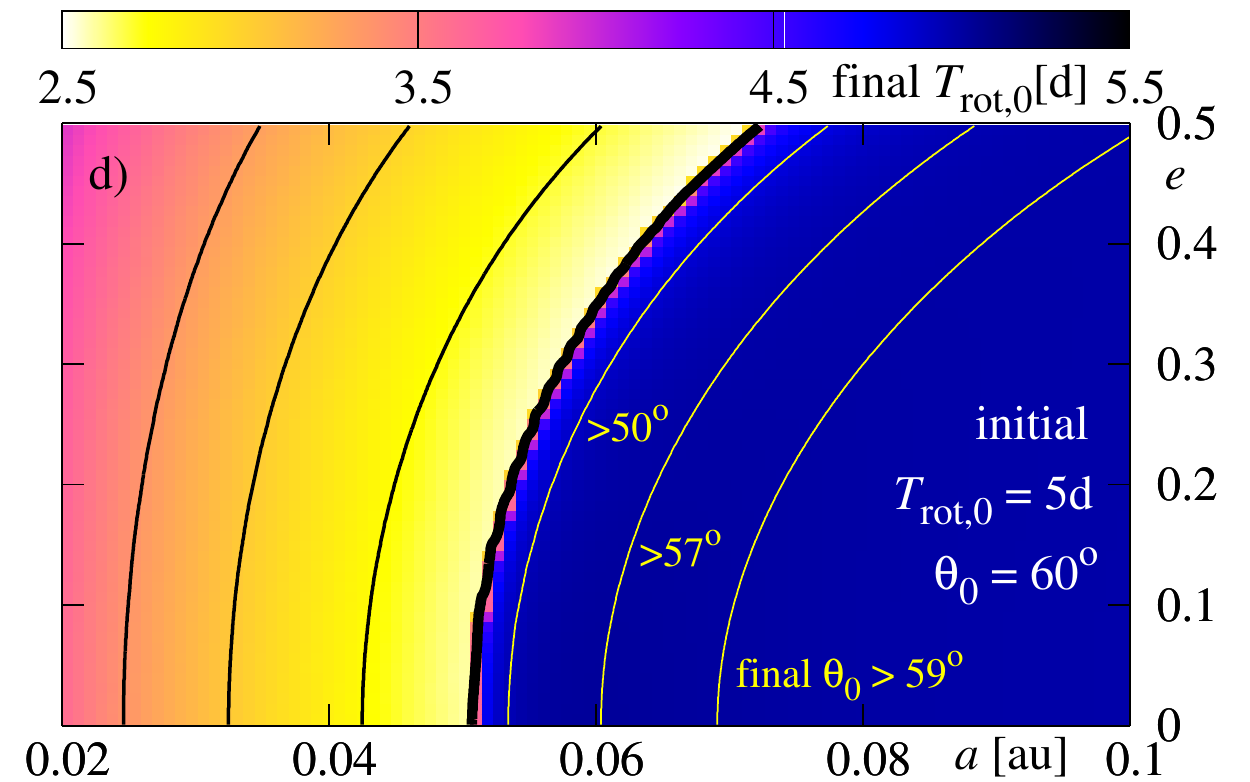}
}
\hbox{
\includegraphics[width=0.5\textwidth]{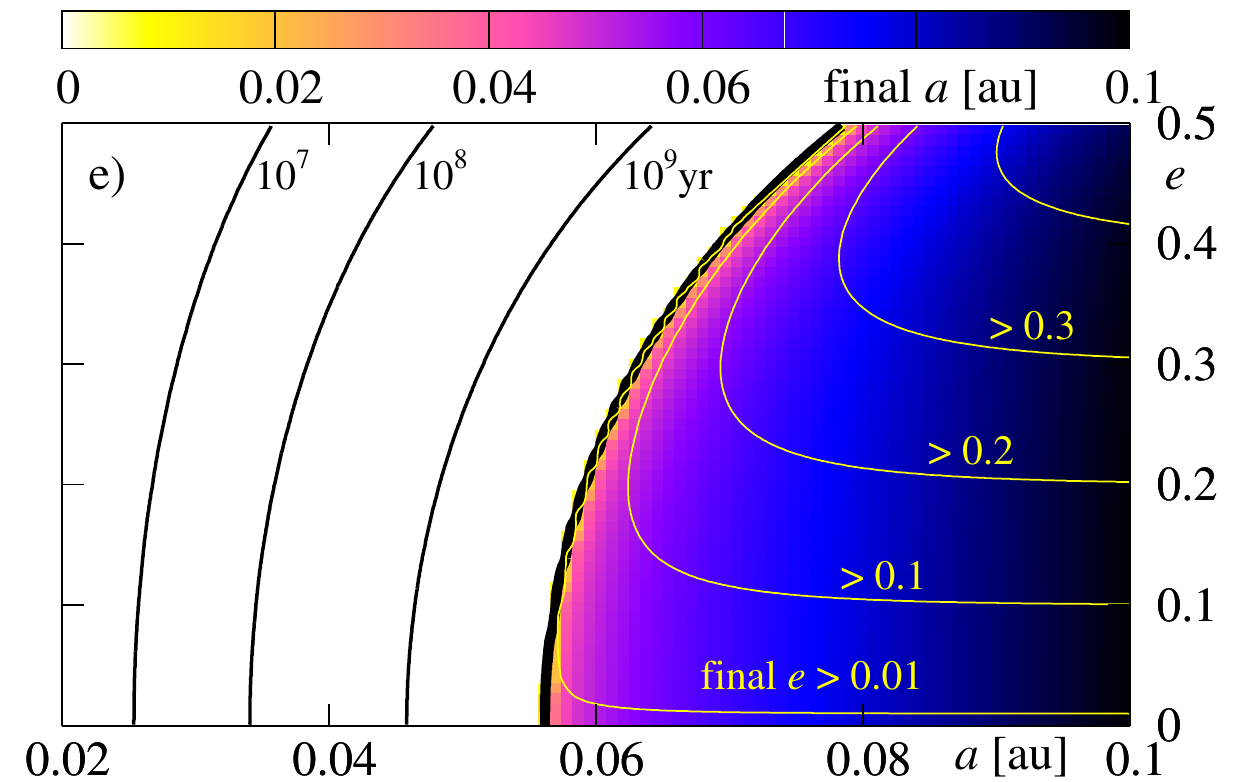}
\includegraphics[width=0.5\textwidth]{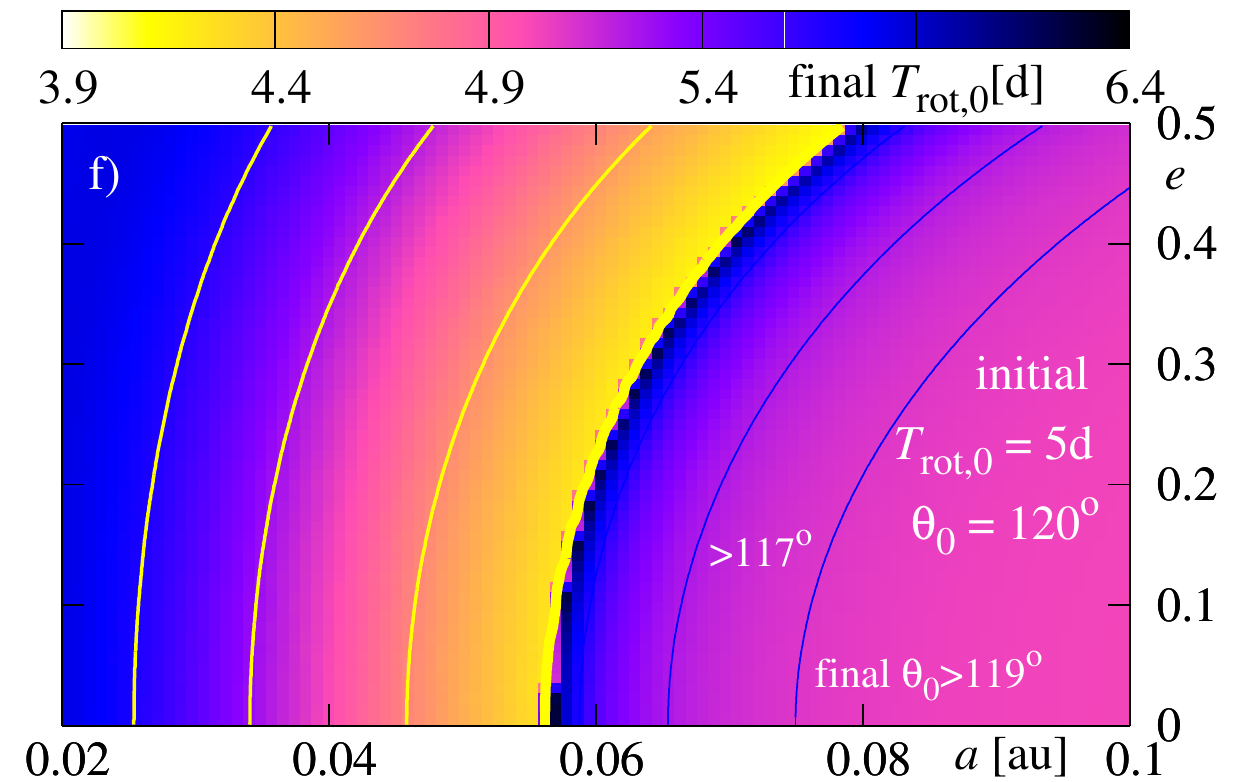}
}
}
}
\caption{
Maps of the final state of the system as a function of the initial
semi-major axis and eccentricity. Panels in the right-hand column are for
the final $a$, the right-hand column is for the final rotational period of
the star. Parameters of the system are the following: $m_1 = 1\,\msun$, $m_1
= 1\,\mJ$, $R_0 = 1\,\RS$, $R_1 = 1\,\RJ$. Polytropic indices of the star
and the planet are equal to $3$ and $1.5$, respectively. The energy
dissipation constants are $\lambda_0 = 5 \times 10^{-5}, \lambda_1 = 5
\times 10^{-7}$. The initial $T_{\idm{rot},0} = 5\,\mbox{d}$. The initial
angle $\theta_0$ is $0^{\circ}, 60^{\circ}, 120^{\circ}$ for the top, the
middle and the bottom rows, respectively. The integration time is $5$~Gyr.
}
\label{fig:scan1}
\end{figure}

\begin{figure}
\centerline{
\vbox{
\hbox{
\includegraphics[width=0.5\textwidth]{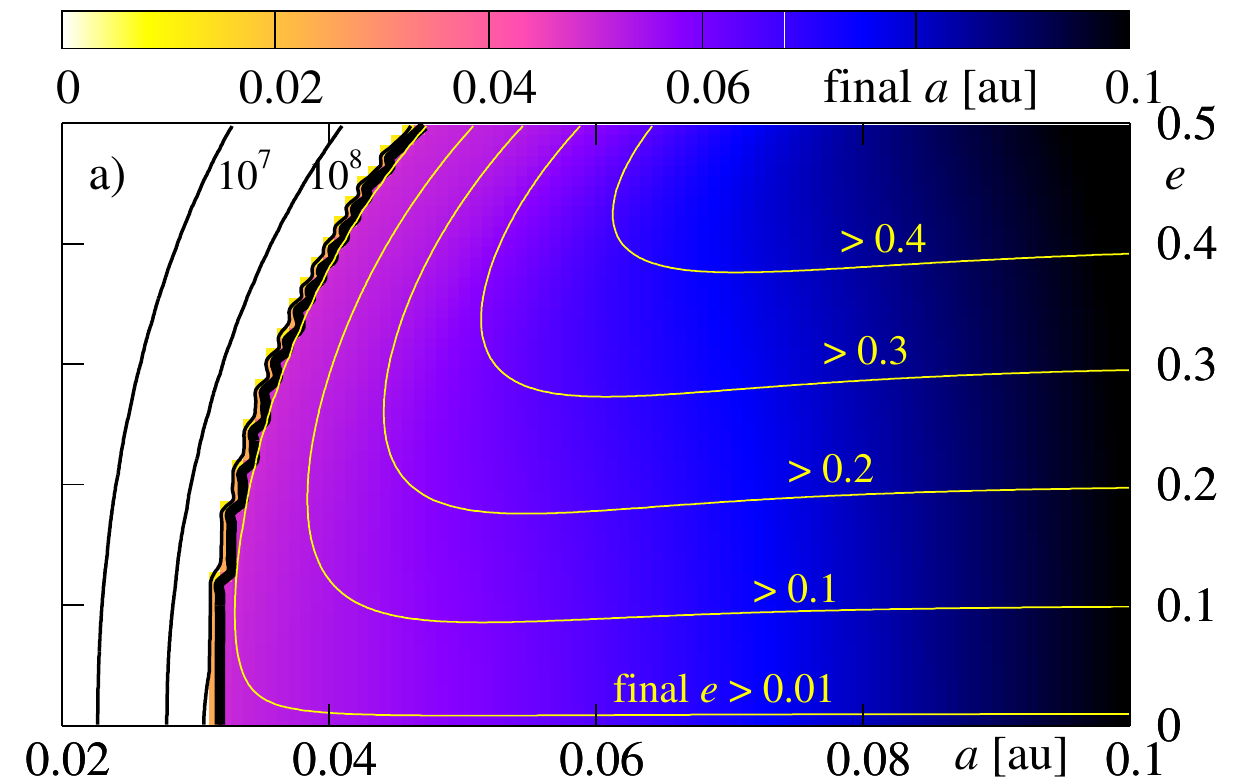}
\includegraphics[width=0.5\textwidth]{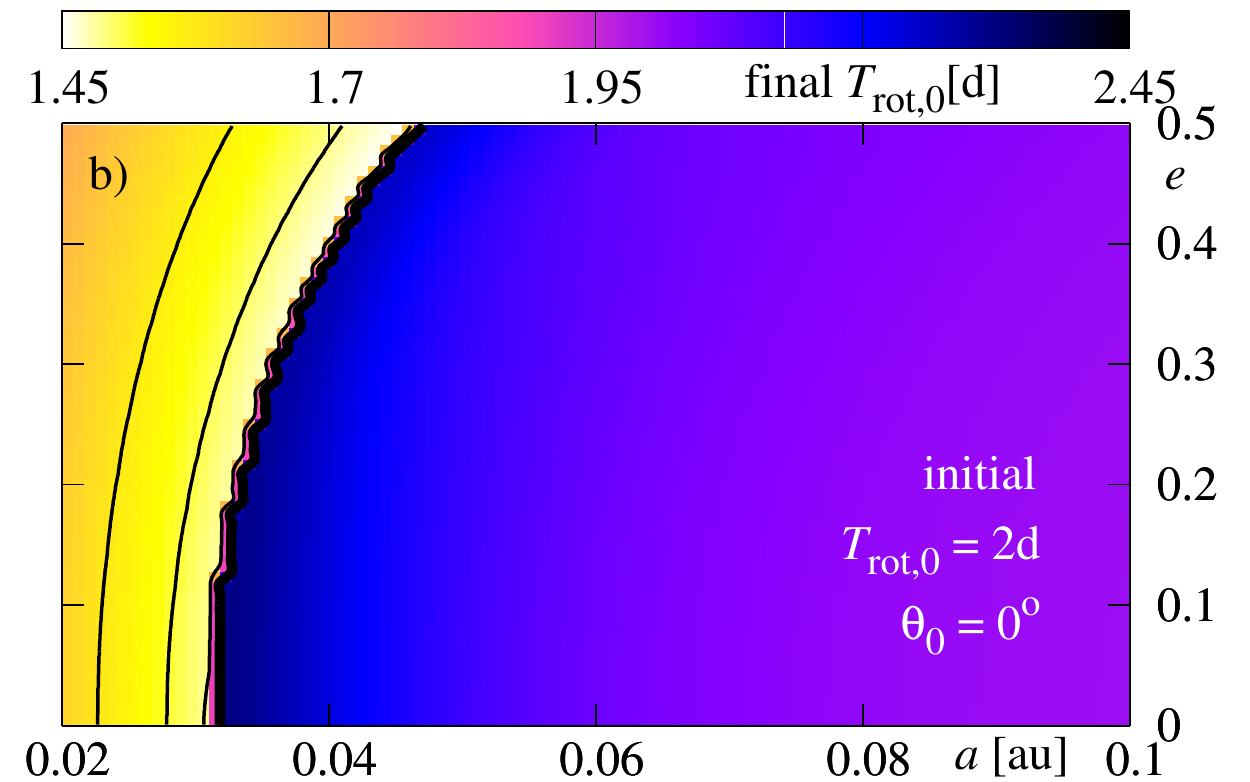}
}
\hbox{
\includegraphics[width=0.5\textwidth]{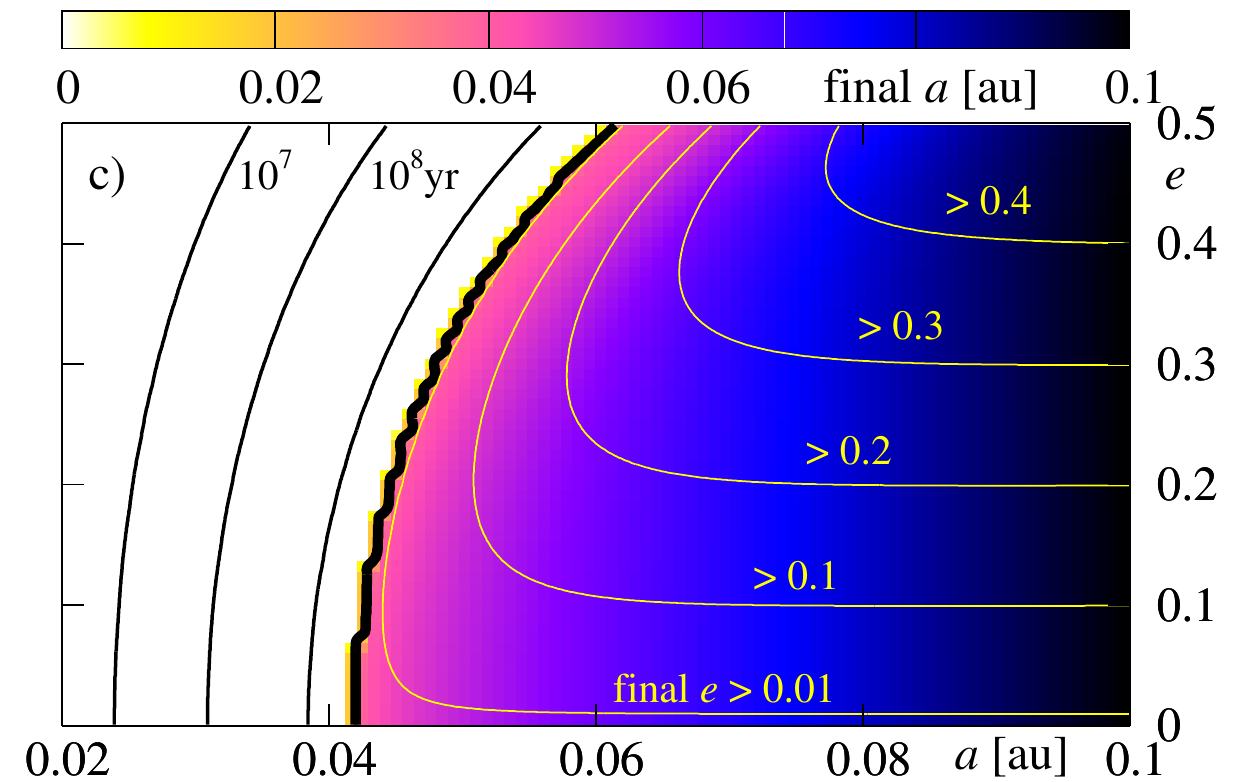}
\includegraphics[width=0.5\textwidth]{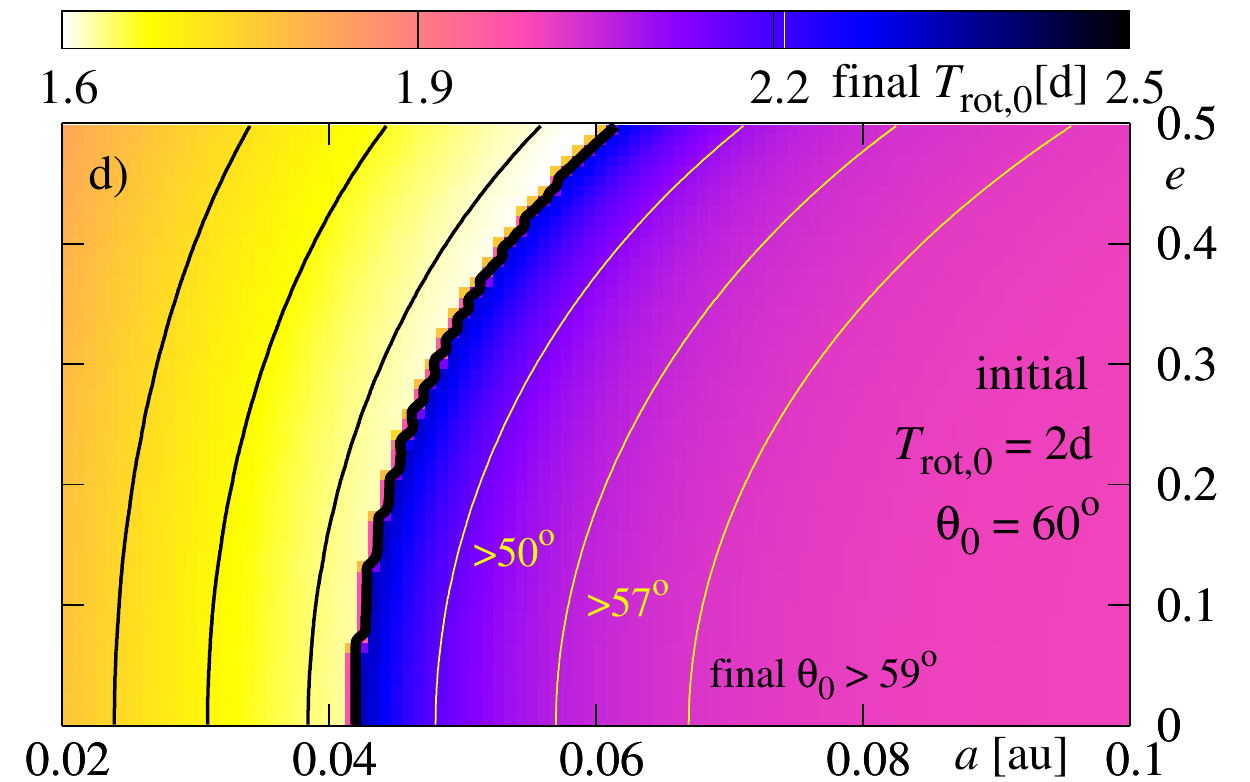}
}
\hbox{
\includegraphics[width=0.5\textwidth]{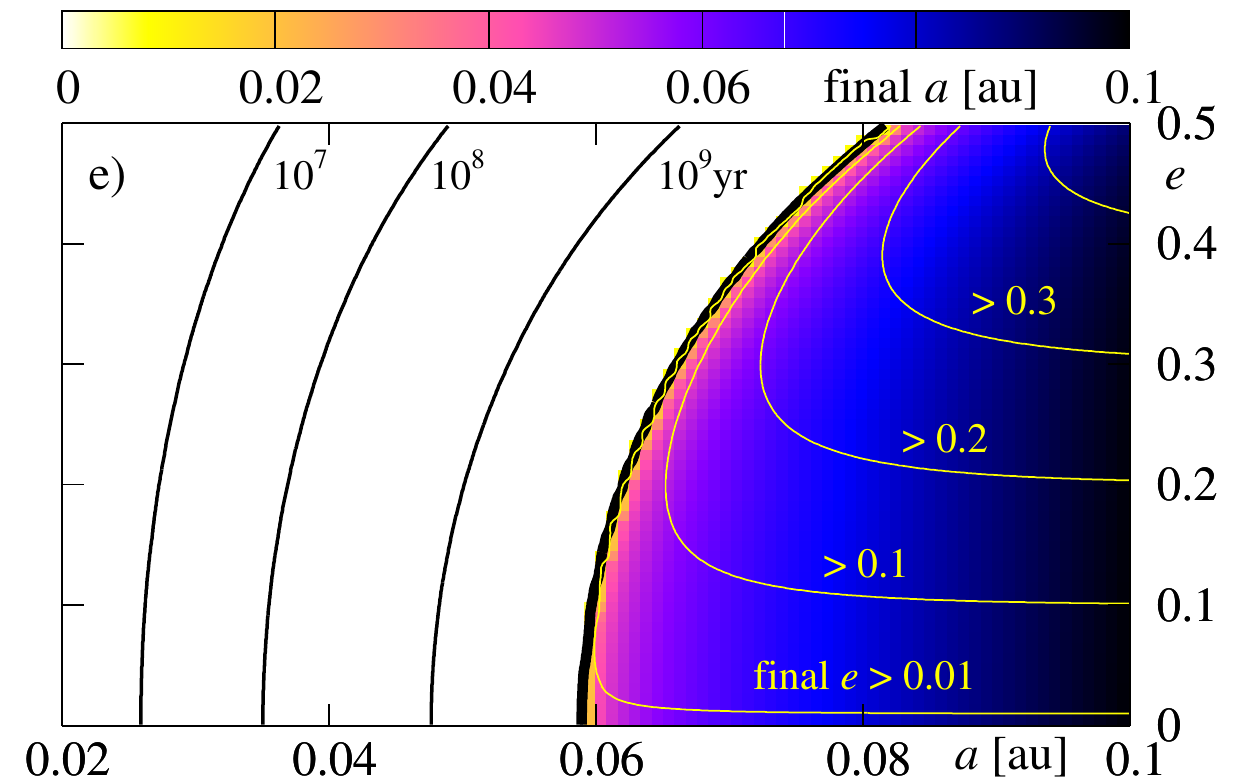}
\includegraphics[width=0.5\textwidth]{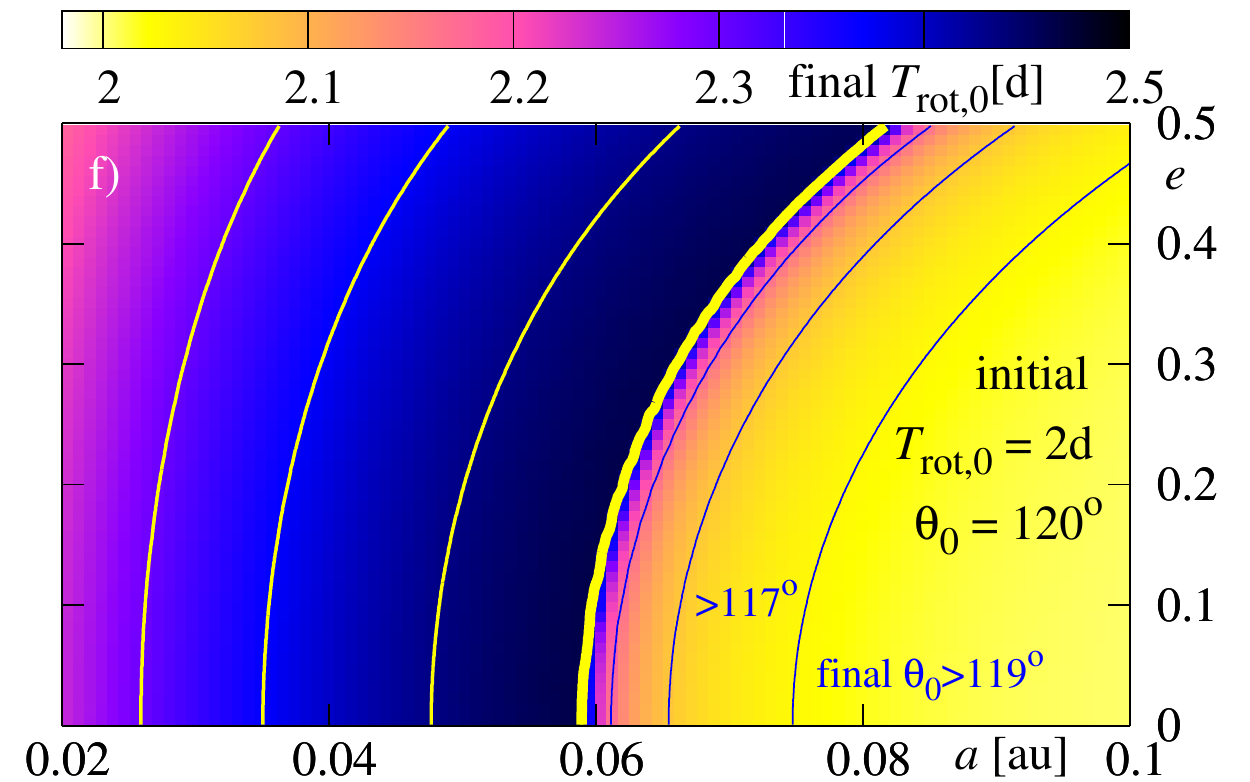}
}
}
}
\caption{The same as shown on Fig.~\ref{fig:scan1},
but for initial $T_{\idm{rot},0} = 2\,\mbox{d}$.}
\label{fig:scan2}
\end{figure}

\begin{figure}
\centerline{
\vbox{
\hbox{
\includegraphics[width=0.5\textwidth]{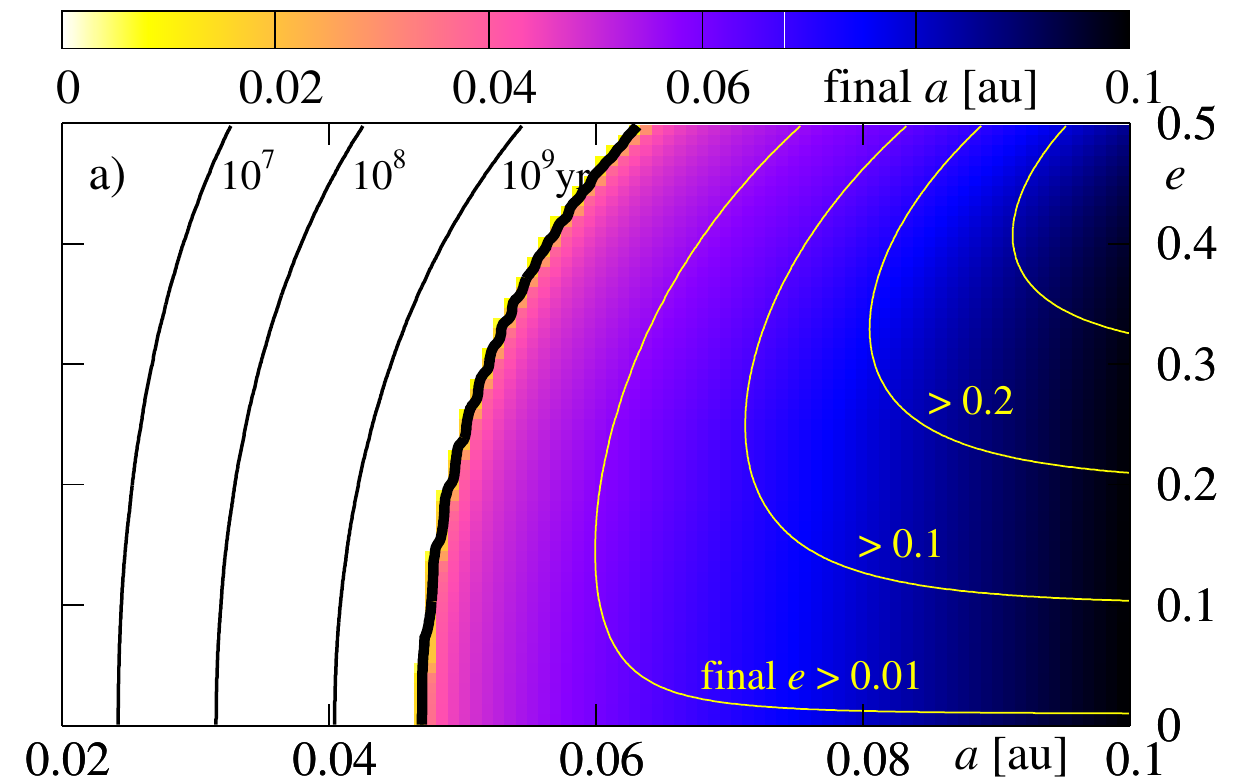}
\includegraphics[width=0.5\textwidth]{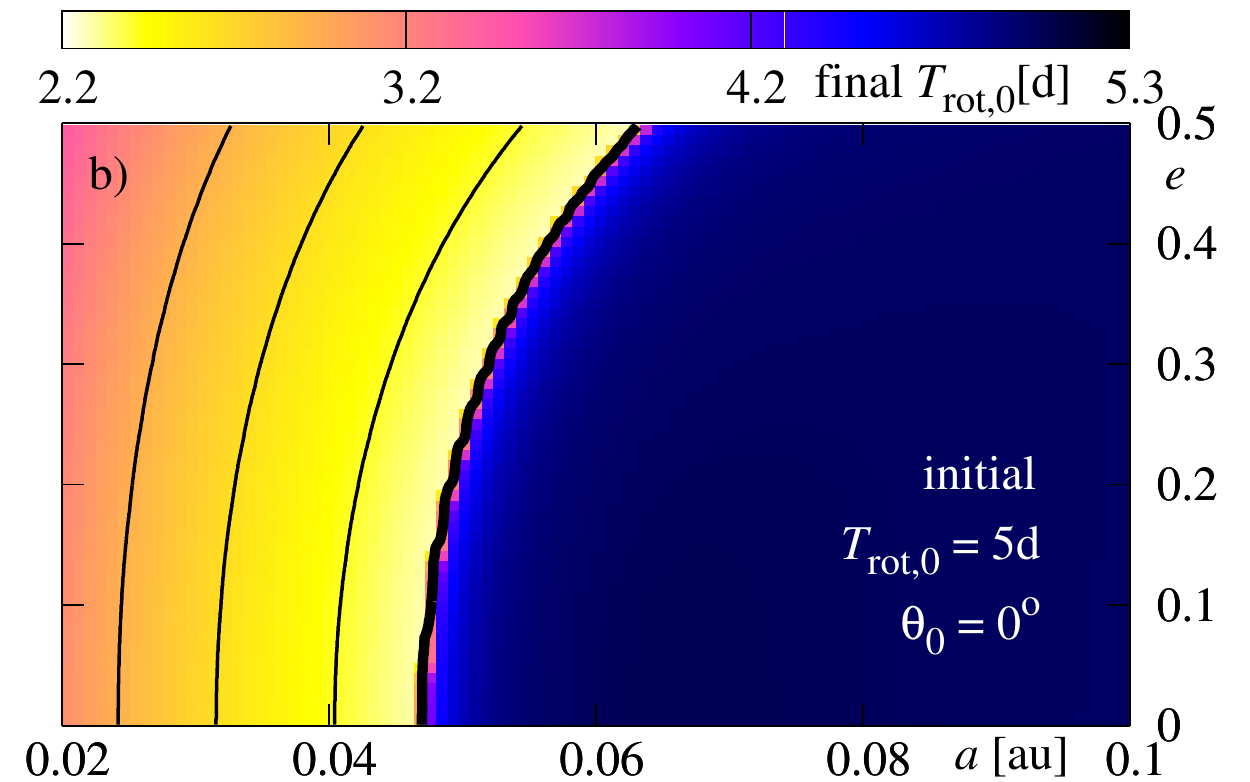}
}
\hbox{
\includegraphics[width=0.5\textwidth]{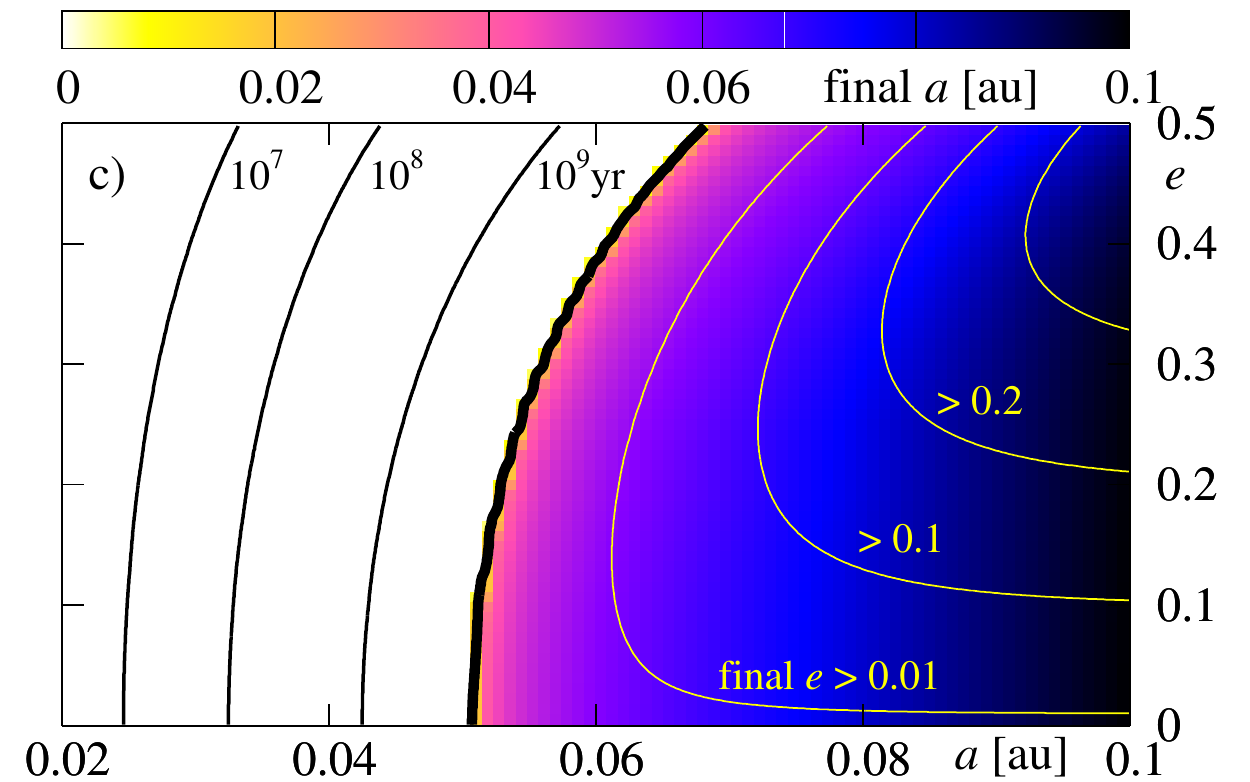}
\includegraphics[width=0.5\textwidth]{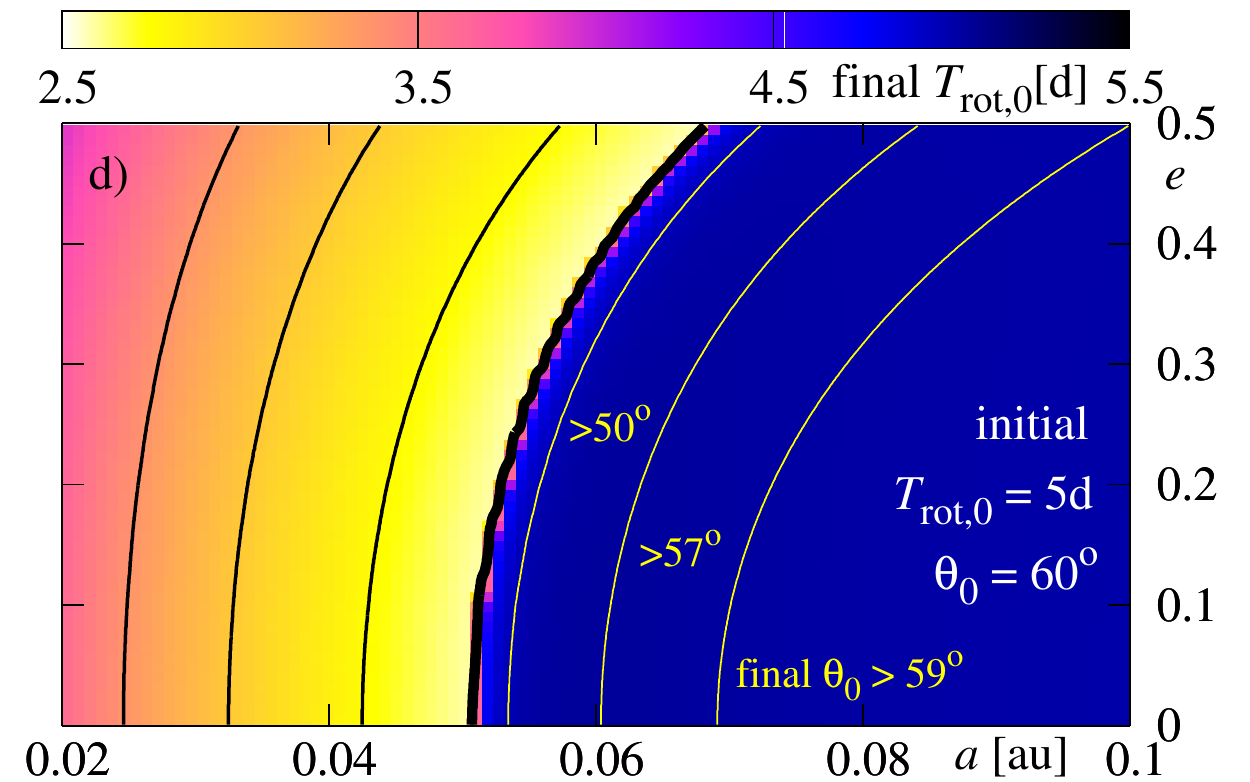}
}
\hbox{
\includegraphics[width=0.5\textwidth]{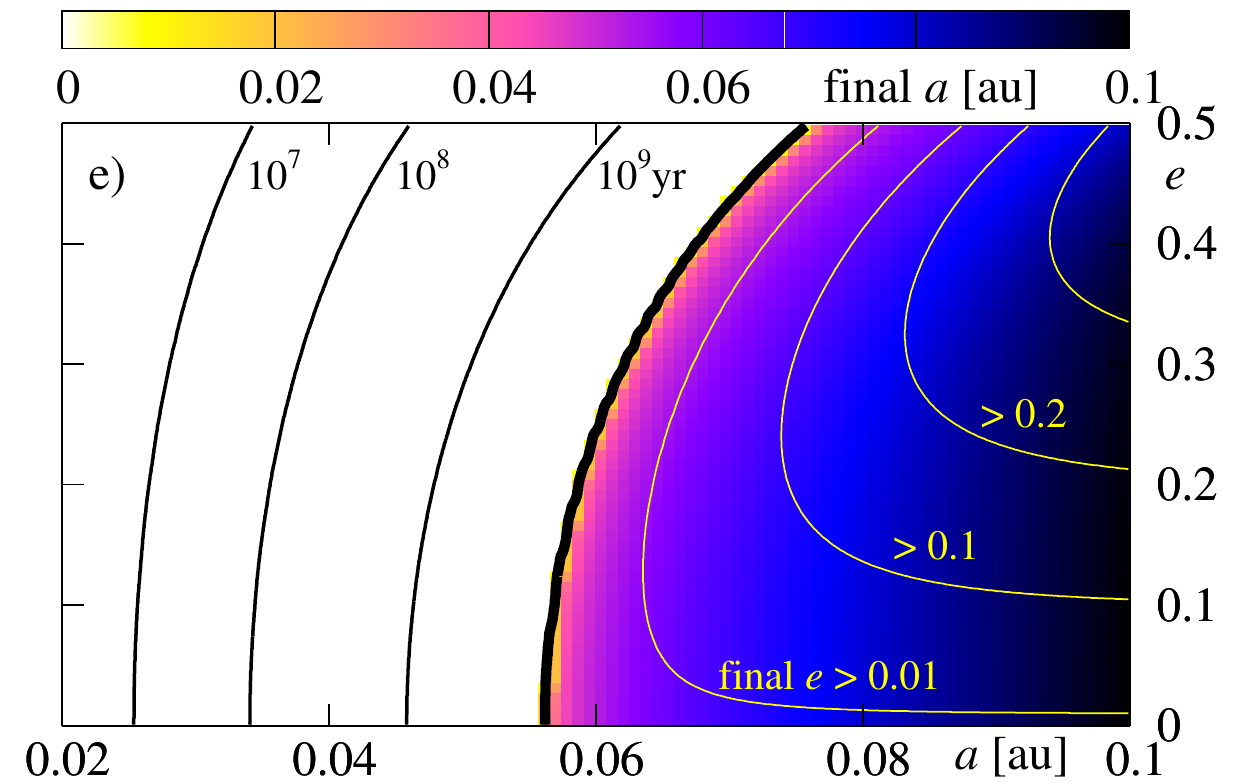}
\includegraphics[width=0.5\textwidth]{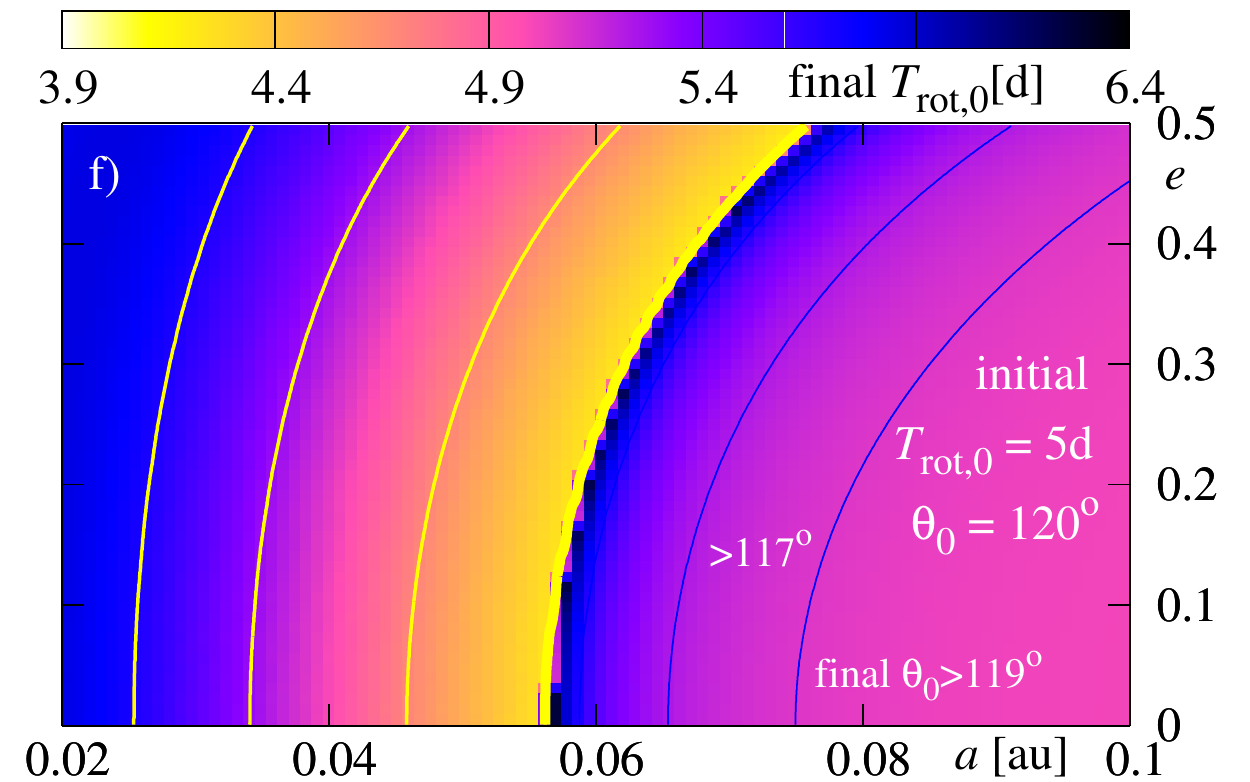}
}
}
}
\caption{The same as in Fig.~\ref{fig:scan1}, but for $\lambda_0 = 5 \times
10^{-5}, \lambda_1 = 5 \times 10^{-5}$. The initial rotational period of the
star is $T_{\idm{rot},0} = 5\,\mbox{d}$.}
\label{fig:scan3}
\end{figure}

\begin{figure}
\centerline{
\vbox{
\hbox{
\includegraphics[width=0.5\textwidth]{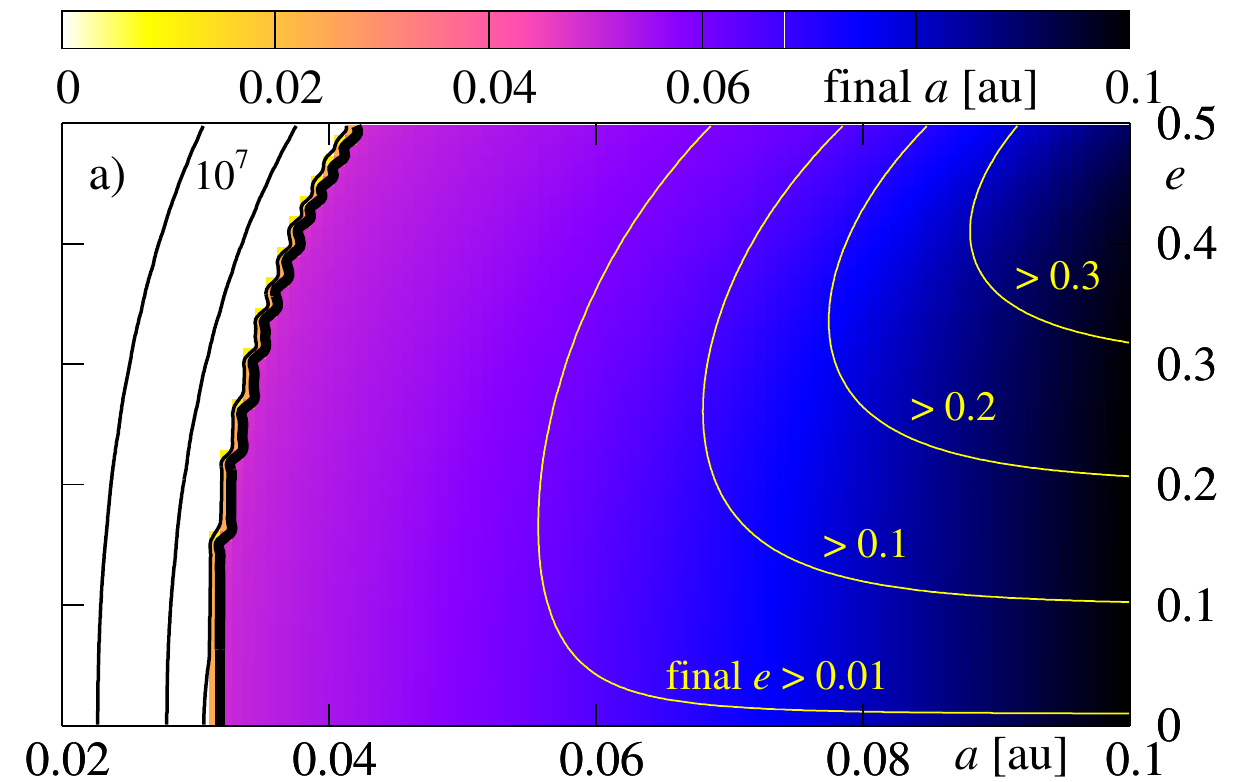}
\includegraphics[width=0.5\textwidth]{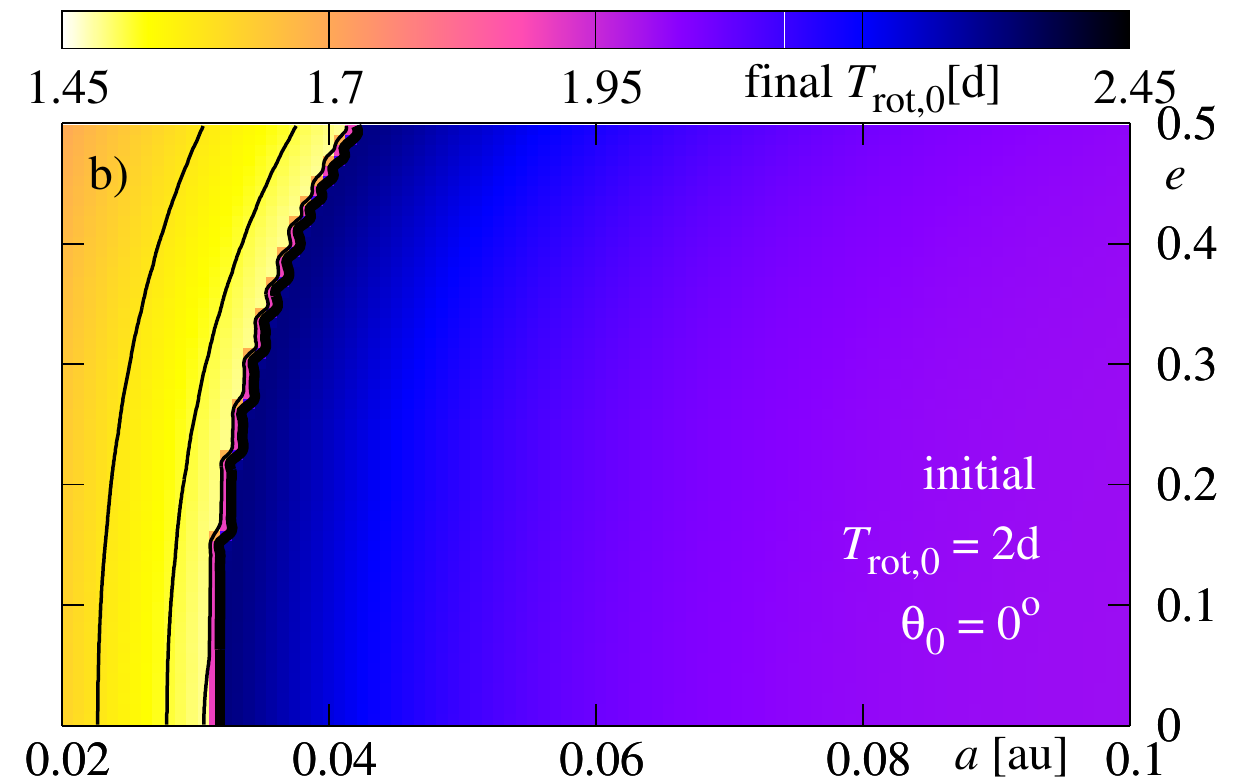}
}
\hbox{
\includegraphics[width=0.5\textwidth]{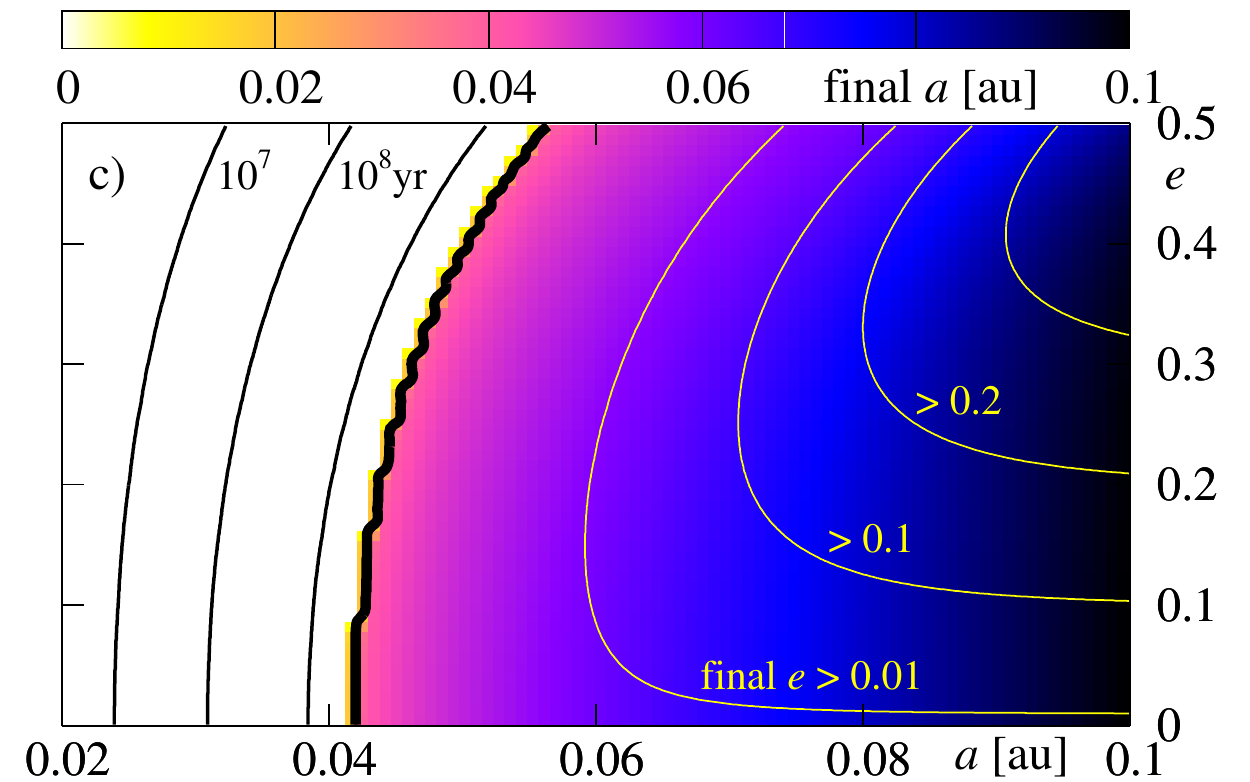}
\includegraphics[width=0.5\textwidth]{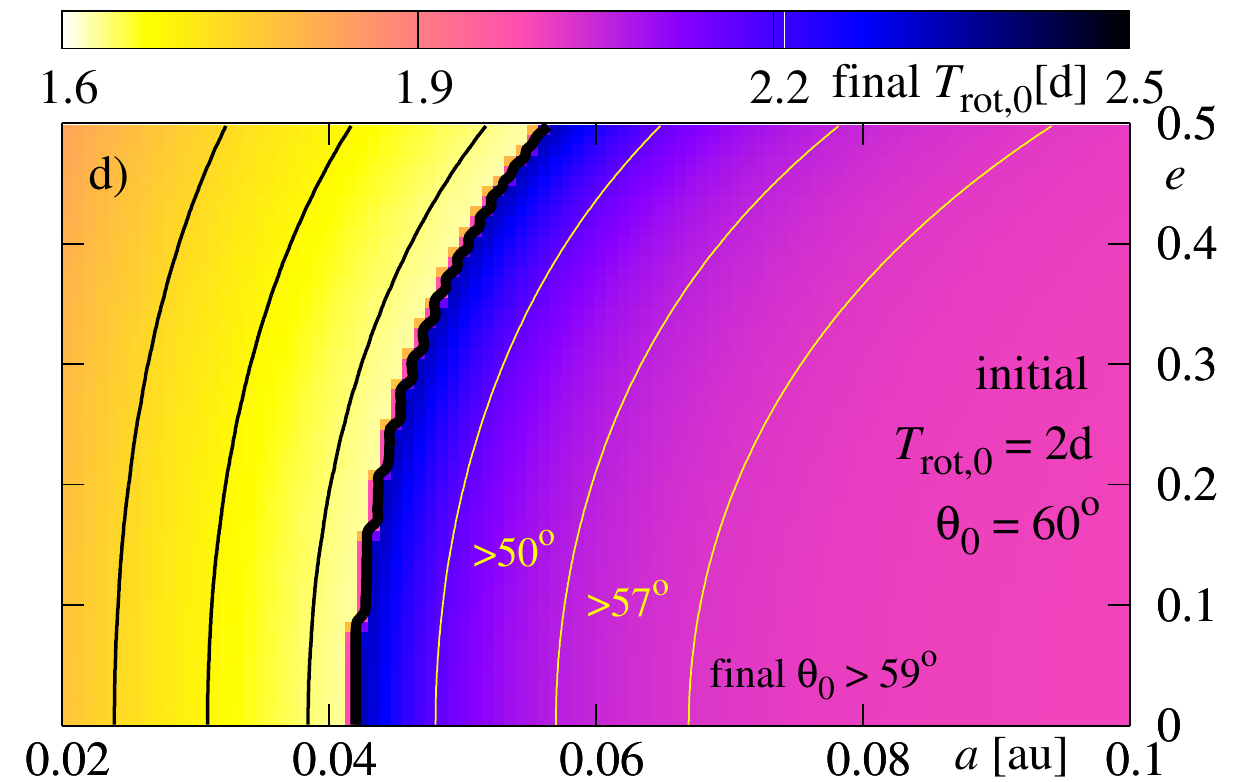}
}
\hbox{
\includegraphics[width=0.5\textwidth]{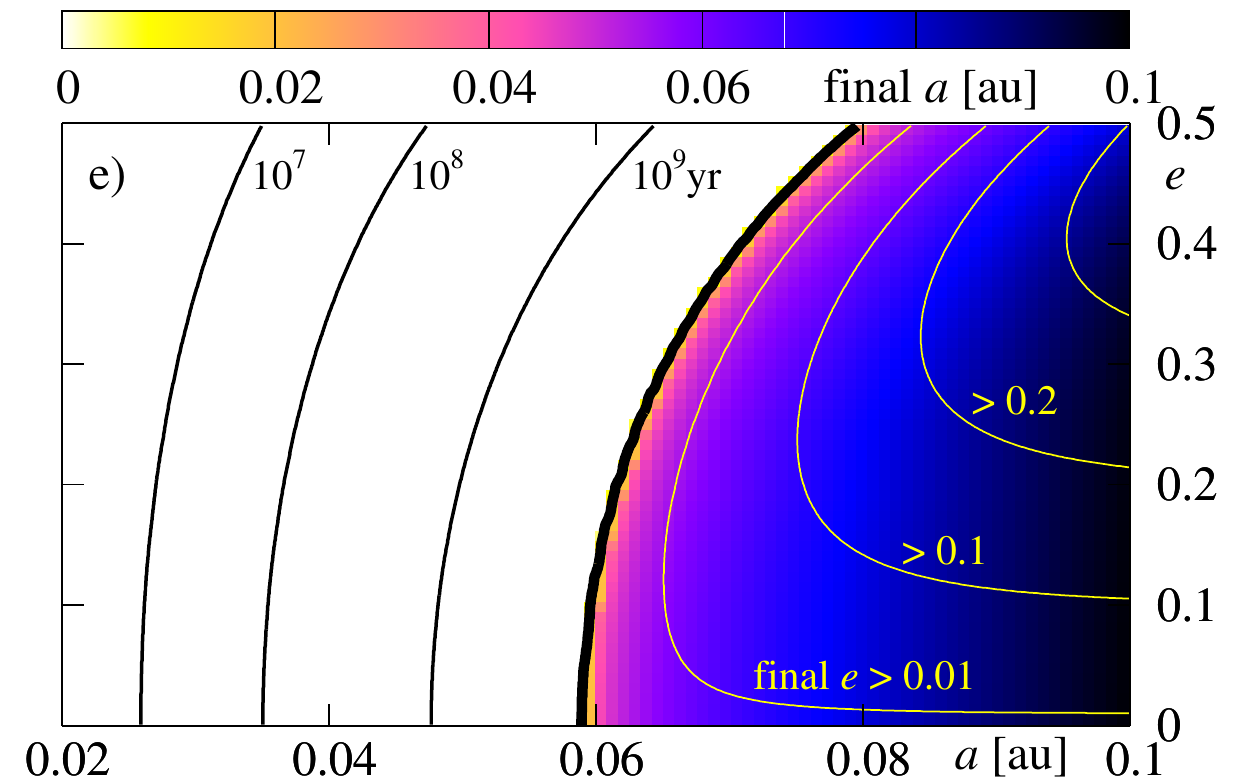}
\includegraphics[width=0.5\textwidth]{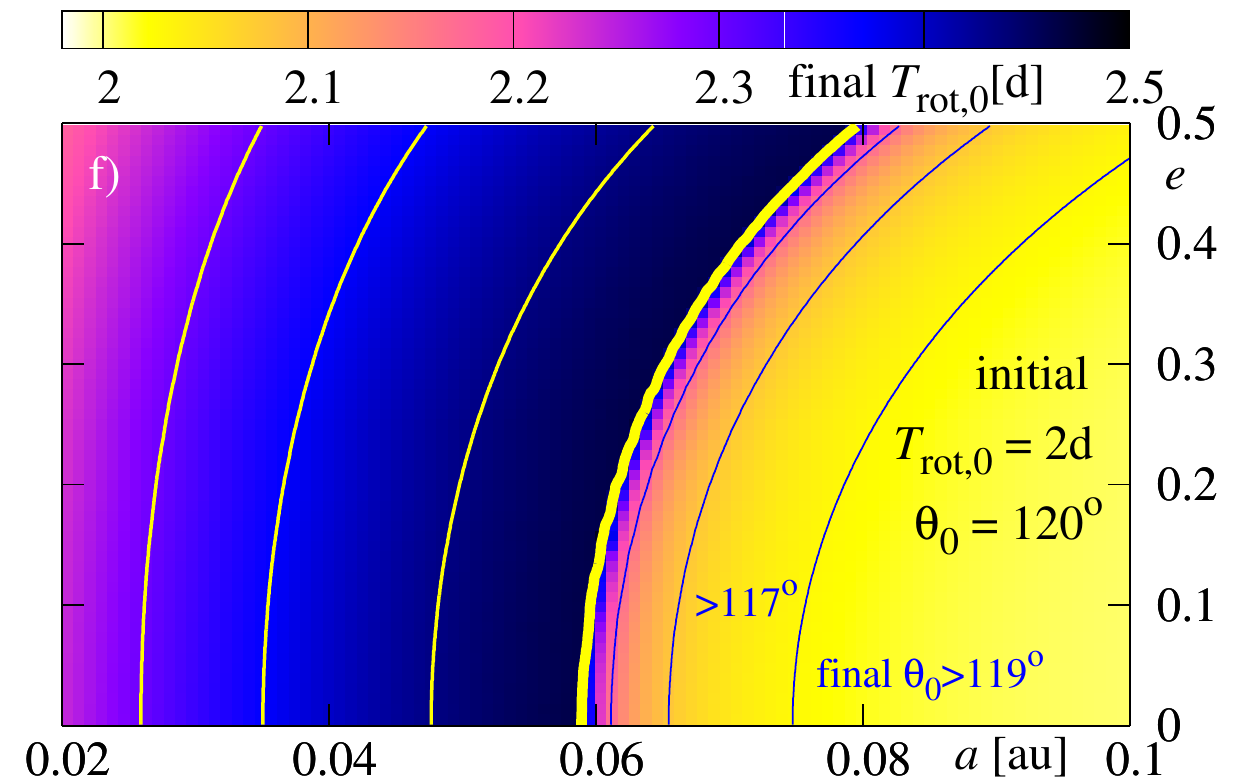}
}
}
}
\caption{The same as in Fig.~\ref{fig:scan1}, but for $\lambda_0 = 5 \times
10^{-5}, \lambda_1 = 5 \times 10^{-5}$. The initial rotational period of the
star is $T_{\idm{rot},0} = 2\,\mbox{d}$.}
\label{fig:scan4}
\end{figure}

%
Here, we present the results of the analysis of the evolution of the system
for various initial conditions. We adopt typical physical parameters of the
system as follows. The masses are $m_0 = 1\,\msun$ and $m_1 = 1\,\mJ$, for
the star and the planet, respectively. Their radii are $R_0 = 1\,\RS$ and
$R_1 = 1\,\RJ$. The mass distribution is described by polytropic models with
indices $n_0 = 3$ and $n_1 = 1.5$. Parameters of the energy dissipation
rates are $\lambda_0 = 5 \times 10^{-5}$ and $\lambda_1 = 5 \times 10^{-7}$
in the first case (results are presented on Figs.~\ref {fig:scan1} and \ref
{fig:scan2}) and $\lambda_0 = \lambda_1 = 5 \times 10^{-5}$ in the second
case (Figs.~\ref{fig:scan3} and \ref{fig:scan4}). We also consider a few
sets of initial $T_{\idm{rot},0}$ and $\theta_0$.
Figure~\ref{fig:scan1} illustrates the results derived for the initial
rotational period of the star $T_{\idm{rot},0} = 5\,\mbox{d}$, and for three
initial values of the inclination between the stellar equator and the orbit
(from the top to the bottom row, it is $0^{\circ}, 60^{\circ}, 120^{\circ}$,
respectively). Colors encode the final $a$ (the left-hand column) and final
$T_{\idm{rot},0}$ (the right-hand column).

At first, lets us study panel (a), obtained for initial $\theta_0 = 0$. For
the initial $(a,e)$ in the white region, the planet falls down onto the star
during time which is shorter than $5$~Gyr. The black, thick curve is for the
border of survival. For the initial conditions located at the $(a, e)$-plane, 
on the right side of this curve, the planet does not fall onto the
star during the whole time of integration $5$~Gyr. Other black (thin) curves
are for initial conditions for which the planet survives during $10^7$~yr,
$10^8$~yr and $10^9$~yr, respectively. As we can see, for parameters adopted
in this experiment, the border is placed at $\sim 0.046\,\au$ for initially
circular orbits and at $\sim 0.068\,\au$ for initial $e=0.5$. Clearly,
planets with initially larger $e$ migrate towards the star faster. As we
already noticed, the contribution to $\dot{a}$ stemming from the energy
dissipation in the interior of the planet is typically larger or similar to
that one due to the dissipation in the star. Nevertheless, this term is
proportional to $e^2$ and vanishes for circular orbit. The eccentricity is
damped in similar time-scale than $a$ for corresponding values of $\lambda_0,
\lambda_1$, thus the dissipation in the planet is important not only at the
early stages of the dissipative evolution.
The evolution of the eccentricity may be also read from the discussed figure. 
Contours plotted with yellow thin curves are for the levels of
constant final $e$, which are $0.001$, $0.1$, $0.2$, $0.3$ and $0.4$,
respectively. As we can see, for initial conditions located close to $\sim
0.01\,\au$ on the right-hand side of ``the survival curve'', the
eccentricity is damped significantly, while for larger initial $a$, its
variability remains small. For $a \gtrsim 0.074\,\au$, the dissipation
time-scales $\tau_e$ are longer than the integration time, and the
eccentricity is not altered significantly. Also the decay of the orbit is
much slower.

The second panel (b) shows the color-map of the final $T_{\idm{rot},0}$. The
black curves are again for the time of planetary survival. The border of $5$
~Gyr is also a border which divides the map of the final state of the star
onto two parts. In cases when the planet falls down onto the star (on the
left-hand side of the border curve), the rotation of the star becomes
faster. For the initial conditions near the border, the rotational period
decreases from $5$ days to $\sim 2.2$ days. It may be understood, because
the initial orbital angular momentum of the planet increases for larger
initial $a$. For initial $a \gtrsim 0.074\,\au$, for which the planet
survives, and a variation of $a$ is not large, also a change of $\Omega_0$
is not large.

The next two rows show the results for initial $\theta_0 = 60^{\circ}$ (the
middle row) and $120^{\circ}$ (the bottom row). In the case of panels (c)
and (d), the differences between these results and the previous ones are not
significant. The border of planetary survival shifts slightly towards larger
initial $a$. The minimal values of the final $T_{\idm{rot},0}$ for $\theta_0
= 60^{\circ}$ increase also slightly,  $\sim 3\,\mbox{d}$ near the border.
This is an effect of addition of two vectors, i.e. $\beta \, \vec{h}$ and
$\Izero_0 \, \pmb{\Omega}_0$, which are mutually inclined. The angular
momentum exchange that may manifest itself in the decrease of
$T_{\idm{rot},0}$ is smaller here than in a coplanar system. The inclination
$\theta_0$ does not change significantly during the evolution. The yellow
contours shown at panel (d) indicate constant levels of the final $\theta_0
= 50^{\circ}, 57^{\circ}, 59^{\circ}$. The systems near the survival border
change their $\theta_0$ by about of $10$~degrees, while for systems located
close to $\sim 0.02\,\au$, on the right-hand side of this border, the
inclination is almost constant. The observed slow rate of inclination
dumping agrees  with the determined inclination in hot-Jupiter systems \citep[e.g.,][]{Triaud2010, Pont2010}.

For the retrograde orbits, i.e., with initial $\theta_0 > \frac{\pi}{2}$,
here $\theta_0 = 120^{\circ}$, the final state of the system is different.
The survival border shifts again. But most significant differences may be
visible in the final rotational period of the star. The picture is somehow a
negative with respect to those obtained for $\theta_0 = 0^{\circ}$ and
$\theta_0 = 60^{\circ}$. The larger final $T_{\idm{rot},0}$ are for those
initial conditions for which the planet falls down onto the star or has
survived but stays close to the border. The difference may be explained due
to the $z$ -components of initial $\beta\,\vec{h}$ and
$\Izero_0\,\pmb{\Omega}_0$ have opposite signs.

Again, the inclination of configurations which end with $a>0$ does not
change significantly. The evolution of the eccentricity in all three cases
looks like very similar. The main difference between the studied cases
concerns the final $T_{\idm{rot},0}$.
The next figure~\ref{fig:scan2} illustrates the results derived for 
parameters used in the previous experiment; only the initial rotational
period of the star is now $T_{\idm{rot},0} = 2\,\mbox{d}$. The
view of the final state of the system is similar to that one presented in
Fig.~\ref{fig:scan1}. There is a difference, however, that relies in a shift
of the survival border towards smaller initial semi-major axes for $\theta_0
< \frac{\pi}{2}$, and towards larger $a$ for $\theta_0 > \frac{\pi}{2}$. It
may be understood rather easily. For the faster rotation of the star, the
border $\Omega_0 = n$ shifts at the $(a, e)$-plane towards larger initial $a$. 
A contribution proportional to the ratio $\Omega_0/n$ implies an increase
of $a, e, \theta_0$ and a decrease of $\Omega_0$; it acts opposite to the
remaining terms. Larger $\Omega_0/n$, for some selected initial condition,
means that this term is more significant from these terms. It manifests
itself also in the fact that for initially coplanar configurations, there
exist regions in $(a, e)$-plane, in which the eccentricity is excited to
larger values. For systems with $\theta_0 > \pi/2$ the survival border
shifts towards smaller $a$. It is quite clear, because the term $\Omega_0/n$
is multiplied by $\cos\theta_0$ which is negative for the retrograde orbit.
This term causes a decrease of both $a$ and $e$, and has larger magnitude
for greater $\Omega_0$ (shorter $T_{\idm{rot},0}$).

In the next two figures~\ref{fig:scan3}~and~\ref{fig:scan4}, we present a
similar study to the previous experiments, but derived for larger values of the
energy dissipation constant $\lambda_1 = 5 \times 10^{-5}$. The results are
illustrated in the same manner. The only difference relies now in 
significantly smaller final eccentricity than it was derived in the previous
experiments. Again, it may be explained if we recall that the contribution
$\dot{e}_{\idm{p}}$ increases by two orders of magnitude with respect to the
previous tests. For large $\lambda_1$ it dominates over $\dot{e}_*$ and
forces a fast dumping the eccentricity. On the other hand, the term
$\dot{a}_{\idm{p}}$ is important only at the early stages of the evolution,
when $e$ is significantly different from $0$. When $e$ becomes very small,
the term $\dot{a}_{\idm{p}}$ does not accelerate the orbit decay. The
borders of survival as well as maps of the final $T_{\idm{rot},0}$ are very
similar to those ones obtained for smaller $\lambda_1$.

\begin{figure}
\centerline{
\hbox{
\includegraphics[width=0.5\textwidth]{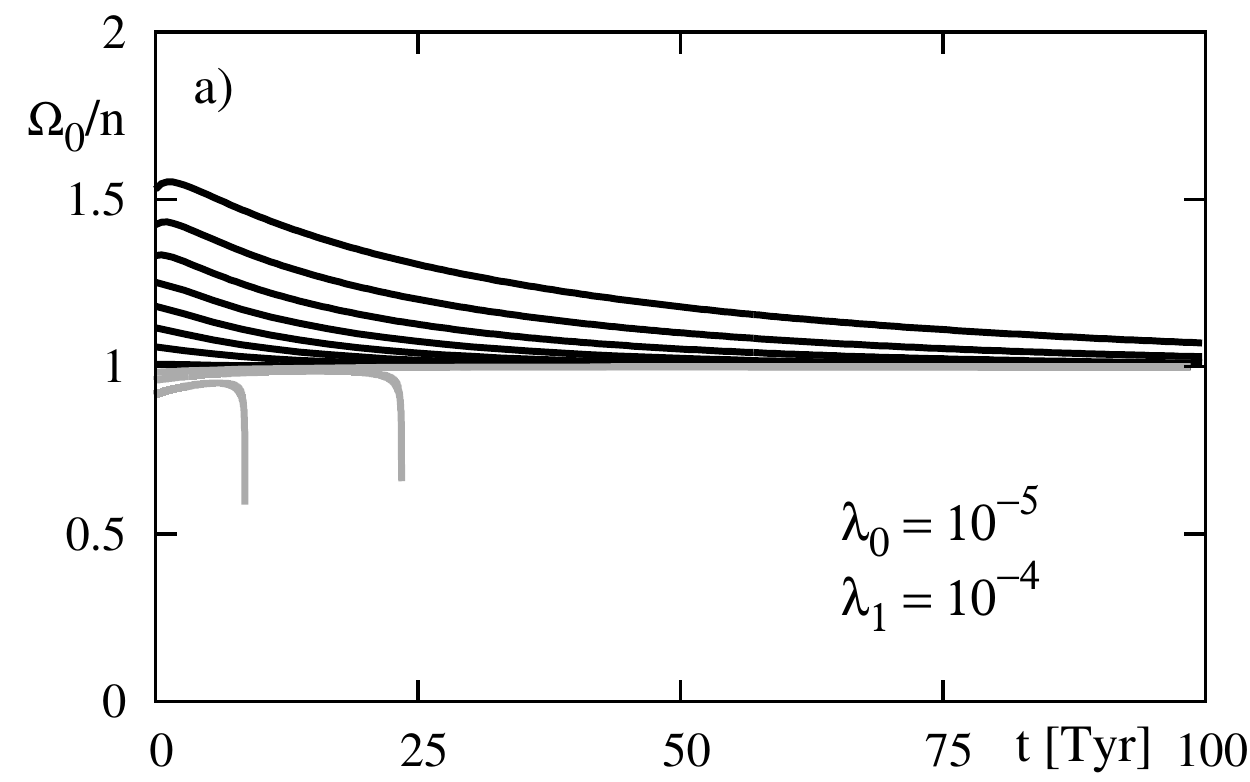}
\includegraphics[width=0.5\textwidth]{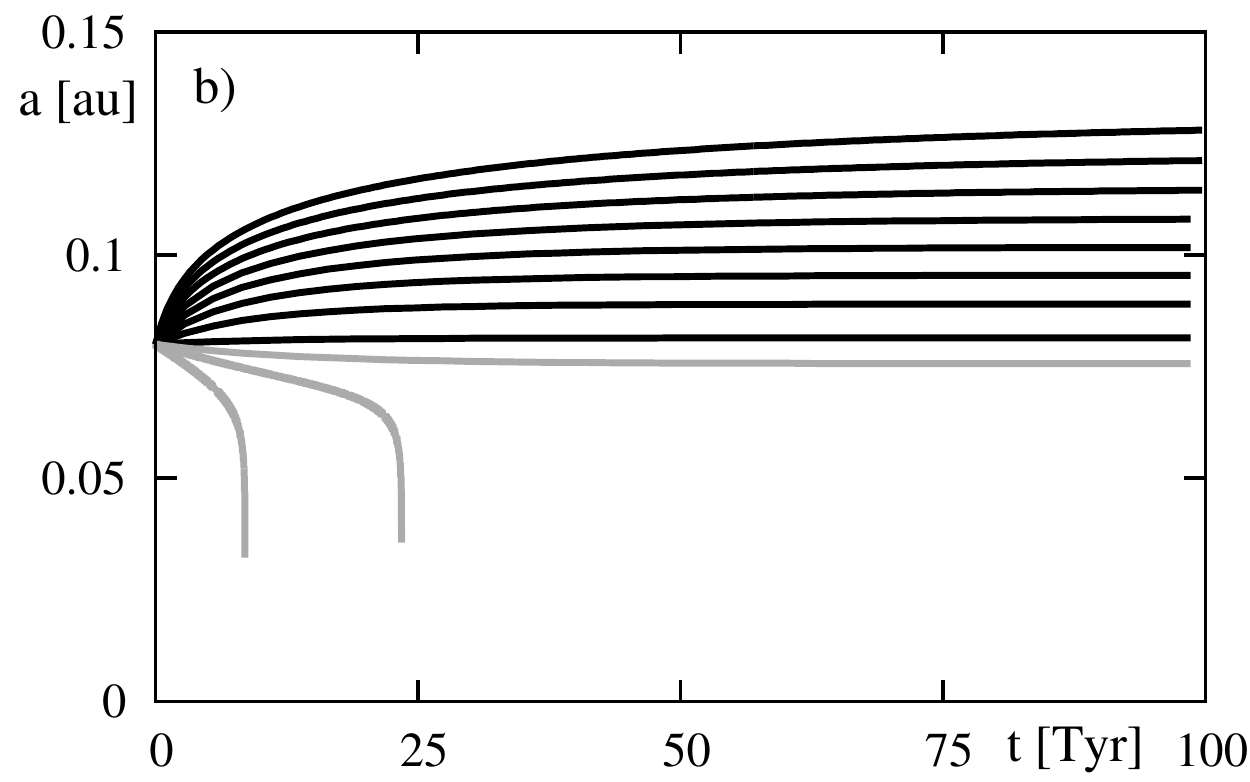}
}
}
\caption{
Time evolution of $\Omega_0/n(t)$ (the left-hand panel) and $a(t)$ (the
right-hand panel) of the third-averaged system. Parameters of the
system are $m_0 = 1\,\msun$, $m_1 = 1\,\mJ$, $R_0 = 1\,\RS$, $R_1 = 1\,\RJ$,
$a = 0.08\,\au$. Dissipative parameters are $\lambda_0 = 10^{-5}$ and
$\lambda_1 = 10^{-4}$. The initial system is close to the equilibrium (i.e.,
$e \approx 0, \theta_0 \approx 0, \Omega_0 \sim 1$). Grey curves are for
the initial $\Omega_0/n < 1$, the black curves are for $\Omega_0/n > 1$.
}
\label{fig:stable_equilibrium}
\end{figure}

It should be noted here, that in none cases illustrated on Fig.~\ref
{fig:scan1}-\ref{fig:scan4}, the system ended in the synchronous state. For
$a \lesssim 0.074\,\au$ this equilibrium is unstable, while for $a \gtrsim
0.074\,\au$ the time-scales of the evolution are too long to fix the system
in this state. Figure~\ref{fig:stable_equilibrium} illustrates the evolution
of the system which is initially close to the equilibrium for $a = 0.08\,\au$
(the rotational period of the star $T_{\idm{rot},0} \in [5.4,
9.4]\,\mbox{d}$). Although the equilibrium is stable, the system tends
towards this state very slowly; the time on the $x$-axis is counted in {\em
terayears}! Moreover, for initial $\Omega_0/n < 1$ (the grey curves), the
planet  likely will fall down onto the star, even if at the beginning the
system seems tend to the equilibrium. It may be explained relatively easily.
If the initial $\Omega_0/n < 1$, then for $a \gtrsim 0.074\,\au$ this ratio
increases, tending to unity, and simultaneously $a$ decrease. When $a$
decreases below the limit of stability, the equilibrium becomes unstable and
the system evolves outwards the equilibrium.

%

\section{Summary and conclusions}
%
In this work we revisited the problem of the generalized model of a
single-planet system including the mechanical energy dissipation. We derived
the equations of motion from the very basic formulation of the Lagrange
equations of the second kind. In that approach, the potential is a function
of the angular velocities of the bodies. Our derivation is different from
the standard approach used to derive the rotational equations of motion of
the rigid body, in which the potential depends only on the Euler angles
describing the orientation of the rigid body, and does not depend on their
time derivatives. We obtained the equations possessing a different general
form, but in the particular case considered here they remain very similar to
the Eulerian equations. Moreover, as we show, an additional term appearing
in the right-hand side of the equation for $\dot{\pmb{\Omega}}_l$ does not
introduce any secular contribution.

The derived equations of motion were averaged out over time-scales 
corresponding to the conservative evolution of the system. We analyzed all 
characteristic time-scales, and we showed that they may be ordered in terms 
of a hierarchical set of variables, which then may be treated as fast 
variables in a recursive averaging process. The fastest component of the 
evolution is the Keplerian mean motion of the planet. After the averaging 
over the mean anomaly, we obtained the first-averaged system. This set of 
equations describes the evolution in which the fastest variable is the 
precessing angular momentum vector of the planet. Then the second averaging 
was performed over the azimuthal angle of $\pmb{\Omega}_1$ in the orbital 
reference frame. In this way we obtained the second averaged system. 
Finally, the third averaging was performed over the argument of pericenter. 
Thus, to eliminate secular variability that occurs in the conservative 
time-scale, and to obtain the dissipative equations of motion, we had to 
average out the system over three fast angles, i.e, $\mathcal{M}, \phi_1, 
\omega$. The precision of the averaging has been tested at all stages of 
the procedure, and we demonstrate  that this approach leads to correct 
results.

Next, we have shown that the dissipative evolution of the angular momentum 
of the planet occurs much faster than the orbital evolution as well as a 
variability of the stellar angular momentum. The inclination of 
$\pmb{\Omega}_1$ with respect to $\vec{h}$ decrease to $0$ in the 
time-scale only slightly longer than the characteristic time-scale of the 
pericenter rotation, for a close-in planet. Simultaneously, $\Omega_1$ 
tends to an equilibrium value which is $\geq n$, where the frequencies are 
equal only for the circular orbit. The equilibrium is stable and we can fix 
$\theta_1$ and $\Omega_1$ at their equilibrium values. In this way we 
derived the final set of equations governing the long term evolution of the 
system that admits energy dissipation, Eq.~(\ref {eq:dot_Om0_final}),~(\ref
{eq:dot_i0_final}),~(\ref{eq:dot_a_final2}) and~(\ref{eq:dot_e_final2}).

The obtained equations of motion describe a dynamical system with one 
equilibrium corresponding to the circular, coplanar and synchronized 
orbit, which is unstable for orbits inside certain critical distance from 
the star,  and which is stable for orbits outside this border. For a 
Sun-Jupiter system that border is on $a \sim 0.074\,\au$. Studying the 
evolution of the system for generic values of physical parameters, we have 
shown however, that this equilibrium is unlikely to be reached. As we 
demonstrated, the final state of the system is a complex function of the 
initial conditions and physical parameters of the model. The derived 
equations make it possible to study this problem very effectively, both 
analytically and  in terms of the CPU overhead, because all the evolution 
of dynamical variables occurring in the intermediate time-scale related to 
the conservative corrections have been averaged out.

\begin{acknowledgements}
I would like to thank Beno{\^\i}t Noyelles, Michael Efroimsky and the anonymous reviewer for the informative reviews that improved the manuscript and Sylvio Ferraz-Mello for comments on the averaging theory.
Many thanks to Krzysztof Go{\'z}dziewski for a discussion and corrections of
the manuscript. This work was supported by the Polish Ministry of Science
and Higher Education grant N/N203/402739. The author is a recipient of the
stipend of the Foundation for Polish Science (programme START, editions
2010 and 2011). This research was carried out with the support of the ''HPC Infrastructure for Grand Challenges of Science and Engineering'' Project, co-financed by the European Regional Development Fund under the Innovative Economy Operational Programme.
\end{acknowledgements}

\section*{Appendices}

\appendix
\corr{\section{A proof that $\dot{E}$ given by formulae (\ref{eq:Edot}) fulfills the condition (\ref{eq:condition_Edot})}
\label{appB}
In the considered case, the left-hand side of condition~(\ref{eq:condition_Edot}) reads as follows:
\[
\sum_{i=1}^9 \dot{q}_i \, \frac{\partial \, \dot{E}}{\partial \, \dot{q}_i} = \dot{\vec{r}} \cdot \frac{\partial \, \dot{E}}{\partial \, \dot{\vec{r}}} \,+\, \dot{\vec{s}} \cdot \frac{\partial \, \dot{E}}{\partial \, \dot{\vec{s}}} \,+\, 
\dot{\vec{p}} \cdot \frac{\partial \, \dot{E}}{\partial \, \dot{\vec{p}}}.
\]
It is quite easy to show, that the following equality occurs, when $\dot{E}$ is given by Eq.~(\ref{eq:Edot}):
\[
\dot{\vec{r}} \cdot \frac{\partial \, \dot{E}}{\partial \, \dot{\vec{r}}} + \pmb{\Omega}_0 \cdot \frac{\partial \, \dot{E}}{\partial \, \pmb{\Omega}_0} + \pmb{\Omega}_1 \cdot \frac{\partial \, \dot{E}}{\partial \, \pmb{\Omega}_1}= 2 \, \dot{E}.
\]
Now, to prove a consistency of the Lagrange equations with a dissipative term in the considered problem, we have to show that
\[
\dot{\vec{s}} \cdot \frac{\partial \, \dot{E}}{\partial \, \dot{\vec{s}}} = \pmb{\Omega}_0 \cdot \frac{\partial \, \dot{E}}{\partial \, \pmb{\Omega}_0}
\]
(and similarly for $\dot{\vec{p}}$ and $\pmb{\Omega}_1$ terms). It is again quite elementary and may be checked by using the above formulae for $\partial \, \dot{E} / \partial \, \dot{\vec{s}}$ and the relation between $\pmb{\Omega}$ and $\dot{\vec{s}}$ (equation~\ref{eq:Omega_quaternions}) as well as the fact, that the following equality occurs for unit quaternion $\mathfrak{s}$ (i.e., $s_0^2 + \vec{s}^2 = 1$):
\[
\frac{\partial \, \dot{s}_0}{\partial \, \dot{\vec{s}}} = - \frac{\vec{s}}{s_0}.
\]
{\em Q.E.D.}
}

\corr{\section{Obtaining the equation (\ref{eq:Lagrange_matrix})}
\label{appA}
The equation in the middle row of Eq.~(\ref{eq:Lagrange_vec_s}) may be transformed into equation~(\ref{eq:Lagrange_matrix}) by using the fact that $\pmb{\Omega}_0 = \pmb{\Omega}_0(\vec{s}, \dot{\vec{s}})$. For the component $x$ we have (index $0$ in $\pmb{\Omega}_0$ is omitted):
\[
\frac{\partial\,f}{\partial\,s_x} = \frac{\partial\,\pmb{\Omega}}{\partial\,s_x} \cdot \frac{\partial\,f}{\partial\,\pmb{\Omega}}, \quad
\frac{\partial\,f}{\partial\,\dot{s}_x} = \frac{\partial\,\pmb{\Omega}}{\partial\,\dot{s}_x} \cdot \frac{\partial\,f}{\partial\,\pmb{\Omega}},
\]
where $f$ is for $\LL$ or $\dot{E}$. For the time derivative appearing in the Lagrange equation we obtain:
\[
\frac{d}{dt} \left( \frac{\partial\,\LL}{\partial\,\dot{s}_x} \right) = \frac{d}{dt} \left( \frac{\partial\,\pmb{\Omega}}{\partial\,\dot{s}_x} \right) \cdot \frac{\partial\,\LL}{\partial\,\pmb{\Omega}} \,+\, \frac{\partial\,\pmb{\Omega}}{\partial\,\dot{s}_x} \cdot \frac{d}{dt} \left( \frac{\partial\,\LL}{\partial\,\pmb{\Omega}} \right).
\]
Similar expressions may be obtained for $y$ and $z$ components and then we write the following:
\[
\frac{d}{dt} \left( \frac{\partial\,\LL}{\partial\,\dot{\vec{s}}} \right) = 
\left(
\begin{array}{c c c}
\displaystyle{\frac{d}{dt} \frac{\partial\,\Omega_x}{\partial\,\dot{s}_x}} & \displaystyle{\frac{d}{dt} \frac{\partial\,\Omega_y}{\partial\,\dot{s}_x}} & \displaystyle{\frac{d}{dt} \frac{\partial\,\Omega_z}{\partial\,\dot{s}_x}} \\
\displaystyle{\frac{d}{dt} \frac{\partial\,\Omega_x}{\partial\,\dot{s}_y}} & \displaystyle{\frac{d}{dt} \frac{\partial\,\Omega_y}{\partial\,\dot{s}_y}} & \displaystyle{\frac{d}{dt} \frac{\partial\,\Omega_z}{\partial\,\dot{s}_y}} \\
\displaystyle{\frac{d}{dt} \frac{\partial\,\Omega_x}{\partial\,\dot{s}_z}} & \displaystyle{\frac{d}{dt} \frac{\partial\,\Omega_y}{\partial\,\dot{s}_z}} & \displaystyle{\frac{d}{dt} \frac{\partial\,\Omega_z}{\partial\,\dot{s}_z}} \\
\end{array}
\right) \,
\left(
\begin{array}{c}
\displaystyle{\frac{\partial\,\LL}{\partial\,\Omega_x}}\\
\displaystyle{\frac{\partial\,\LL}{\partial\,\Omega_y}}\\
\displaystyle{\frac{\partial\,\LL}{\partial\,\Omega_z}}\\
\end{array}
\right) \, + \,
\left(
\begin{array}{c c c}
\displaystyle{\frac{\partial\,\Omega_x}{\partial\,\dot{s}_x}} & \displaystyle{\frac{\partial\,\Omega_y}{\partial\,\dot{s}_x}} & \displaystyle{\frac{\partial\,\Omega_z}{\partial\,\dot{s}_x}} \\
\displaystyle{\frac{\partial\,\Omega_x}{\partial\,\dot{s}_y}} & \displaystyle{\frac{\partial\,\Omega_y}{\partial\,\dot{s}_y}} & \displaystyle{\frac{\partial\,\Omega_z}{\partial\,\dot{s}_y}} \\
\displaystyle{\frac{\partial\,\Omega_x}{\partial\,\dot{s}_z}} & \displaystyle{\frac{\partial\,\Omega_y}{\partial\,\dot{s}_z}} & \displaystyle{\frac{\partial\,\Omega_z}{\partial\,\dot{s}_z}} \\
\end{array}
\right) \,
\left(
\begin{array}{c}
\displaystyle{\frac{d}{dt} \frac{\partial\,\LL}{\partial\,\Omega_x}}\\
\displaystyle{\frac{d}{dt} \frac{\partial\,\LL}{\partial\,\Omega_y}}\\
\displaystyle{\frac{d}{dt} \frac{\partial\,\LL}{\partial\,\Omega_z}}\\
\end{array}
\right),
\]
\[
\frac{\partial\,\LL}{\partial\,\vec{s}} = 
\left(
\begin{array}{c c c}
\displaystyle{\frac{\partial\,\Omega_x}{\partial\,s_x}} & \displaystyle{\frac{\partial\,\Omega_y}{\partial\,s_x}} & \displaystyle{\frac{\partial\,\Omega_z}{\partial\,s_x}} \\
\displaystyle{\frac{\partial\,\Omega_x}{\partial\,s_y}} & \displaystyle{\frac{\partial\,\Omega_y}{\partial\,s_y}} & \displaystyle{\frac{\partial\,\Omega_z}{\partial\,s_y}} \\
\displaystyle{\frac{\partial\,\Omega_x}{\partial\,s_z}} & \displaystyle{\frac{\partial\,\Omega_y}{\partial\,s_z}} & \displaystyle{\frac{\partial\,\Omega_z}{\partial\,s_z}} \\
\end{array}
\right) \,
\left(
\begin{array}{c}
\displaystyle{\frac{\partial\,\LL}{\partial\,\Omega_x}}\\
\displaystyle{\frac{\partial\,\LL}{\partial\,\Omega_y}}\\
\displaystyle{\frac{\partial\,\LL}{\partial\,\Omega_z}}\\
\end{array}
\right),
\]
\[
\frac{1}{2} \, \frac{\partial\,\dot{E}}{\partial\,\dot{\vec{s}}} = 
\left(
\begin{array}{c c c}
\displaystyle{\frac{d}{dt} \frac{\partial\,\Omega_x}{\partial\,\dot{s}_x}} & \displaystyle{\frac{d}{dt} \frac{\partial\,\Omega_y}{\partial\,\dot{s}_x}} & \displaystyle{\frac{d}{dt} \frac{\partial\,\Omega_z}{\partial\,\dot{s}_x}} \\
\displaystyle{\frac{d}{dt} \frac{\partial\,\Omega_x}{\partial\,\dot{s}_y}} & \displaystyle{\frac{d}{dt} \frac{\partial\,\Omega_y}{\partial\,\dot{s}_y}} & \displaystyle{\frac{d}{dt} \frac{\partial\,\Omega_z}{\partial\,\dot{s}_y}} \\
\displaystyle{\frac{d}{dt} \frac{\partial\,\Omega_x}{\partial\,\dot{s}_z}} & \displaystyle{\frac{d}{dt} \frac{\partial\,\Omega_y}{\partial\,\dot{s}_z}} & \displaystyle{\frac{d}{dt} \frac{\partial\,\Omega_z}{\partial\,\dot{s}_z}} \\
\end{array}
\right) \,
\left(
\begin{array}{c}
\displaystyle{\frac{1}{2} \, \frac{\partial\,\dot{E}}{\partial\,\Omega_x}}\\
\displaystyle{\frac{1}{2} \, \frac{\partial\,\dot{E}}{\partial\,\Omega_y}}\\
\displaystyle{\frac{1}{2} \, \frac{\partial\,\dot{E}}{\partial\,\Omega_z}}\\
\end{array}
\right).
\]
Collecting these terms according to middle row of Eq.~(\ref{eq:Lagrange_vec_s}), we obtain Eq.~(\ref{eq:Lagrange_matrix}).
}

\bibliographystyle{spbasic}      

\bibliography{ms}   

\end{document}